\documentclass[journal]{IEEEtran}
\usepackage{amsmath}
\usepackage{color}

\usepackage{graphicx}
\usepackage{epstopdf}
\usepackage[caption=false,font=normalsize,labelfont=sf,textfont=sf]{subfig}
\usepackage[numbers,sort&compress]{natbib}
\usepackage[ruled,linesnumbered,noend]{algorithm2e}
\usepackage[numbers,sort&compress]{natbib}
\usepackage{bm}
\usepackage{amssymb}
\usepackage{enumerate}
\usepackage{graphicx}
\usepackage{float}
\usepackage{amsmath}

\usepackage{amsfonts} 

\usepackage{url}
\usepackage[justification=centering]{caption}
\usepackage{color,xcolor}

\begin{document}

\title{6G-Enabled Smart Railways}
%
\author{Bo Ai,~\IEEEmembership{Fellow,~IEEE},
	Yunlong~Lu,~\IEEEmembership{Member,~IEEE},
        Yuguang~Fang,~\IEEEmembership{Fellow,~IEEE},
        Dusit~Niyato,~\IEEEmembership{Fellow,~IEEE},
        Ruisi~He,~\IEEEmembership{Member,~IEEE},
        Wei~Chen,~\IEEEmembership{Member,~IEEE},
        Jiayi~Zhang,~\IEEEmembership{Member,~IEEE},
        Guoyu~Ma,~\IEEEmembership{Member,~IEEE},
        Yong~Niu,~\IEEEmembership{Member,~IEEE},
        and~Zhangdui~Zhong,~\IEEEmembership{Fellow,~IEEE}
\thanks{B. Ai, Y. Lu, R. He, W. Chen, J. Zhang, G. Ma, Y. Niu and Z. Zhong are with the School of Electronic and Information Engineering, Beijing Jiaotong University, Beijing 100044, China.}
\thanks{Y. Fang is with the Department of Computer Science, City University of Hong Kong, Kowloon, Hong Kong, China.}
\thanks{D. Niyato is with the School of Computer Science and Engineering, Nanyang Technological University, Singapore.}
}
\maketitle
\begin{abstract}
Smart railways integrate advanced information technologies into railway operating systems to improve efficiency and reliability. Although the development of 5G has enhanced railway services, future smart railways require ultra-high speeds, ultra-low latency, ultra-high security, full coverage, and ultra-high positioning accuracy, which 5G cannot fully meet. Therefore, 6G is envisioned to provide green and efficient all-day operations, strong information security, fully automatic driving, and low-cost intelligent maintenance. To achieve these requirements, we propose an integrated network architecture leveraging communications, computing, edge intelligence, and caching in railway systems. We have conducted in-depth investigations on key enabling technologies for reliable transmissions and wireless coverage. For high-speed mobile scenarios, we propose an AI-enabled cross-domain channel modeling and orthogonal time-frequency space-time spread multiple access mechanism to alleviate the conflict between limited spectrum availability and massive user access. The roles of blockchain, edge intelligence, and privacy technologies in endogenously secure rail communications are also evaluated. We further explore the application of emerging paradigms such as integrated sensing and communications, AI-assisted Internet of Things, semantic communications, and digital twin networks for railway maintenance, monitoring, prediction, and accident warning. Finally, possible future research and development directions are discussed.
\end{abstract}

\begin{IEEEkeywords}
The sixth generation (6G) railway communications; edge intelligence; cross-domain channel modeling; integrated sensing and communications; endogenous security.
\end{IEEEkeywords}
\IEEEpeerreviewmaketitle

\section{Introduction}
Rail transportation has offered effective options for the efficient transport of people and goods throughout the past few decades. Compared with other transportation manners, rail transportation has considerable advantages in terms of safety, efficiency, low energy, and environmental friendliness. With the rapid development of industrial technologies, the concept of smart railways is ushering in a new era of transportation intelligentization. This evolution involves the seamless integration of emerging information and computing technologies such as cloud/edge computing and the Internet of Things (IoT) with railway planning and management. Railway communication, one of the fundamental transportation infrastructures, plays an important role in accelerating the intelligence capacity of railway networks. 

In what follows, we will provide a quick overview on the related issues pertinent to smart rail systems. 

\subsection{Development Progress of 5G Smart Rail Communications}

Recent extensive deployment of the fifth generation (5G) mobile communications further provides broadband wireless access, real-time monitoring, remote operation, and intelligent control to smart railways \cite{9103348}. The 5G wireless communication guarantees high-quality interactions between trains and control center, and enables a series of new smart railway applications including onboard and trackside ultra-high-definition video surveillance, multimedia train scheduling, autonomous driving, railway IoT, live streaming, and ultra HD video services. Recent international standardization activities further emphasize the strategic importance of advanced railway communication systems. The 3GPP has initiated dedicated work items for future railway mobile communication system (FRMCS), aiming to define a 5G-based successor to GSM-R that meets evolving railway requirements for safety, performance, and service flexibility \cite{3gpprailway}. In parallel, the International Telecommunication Union (ITU) and the European Telecommunications Standards Institute (ETSI) have also launched collaborative efforts—coordinating closely with the International Union of Railways (UIC)—to develop technical specifications for rail communications \cite{etsirailway}. Industry groups, including Horizon 2020, ETSI TC RT, have been working on the applications of 5G in smart railways for FRMCS. For example, the European Commission has presented a 5G action plan to promote the deployment and uninterrupted coverage of 5G in railways \cite{9454557}. Ericsson has also deployed about 100 live 5G networks that can provide accurate positioning, reliable data transmissions, and end-to-end connectivity for railway operators.

Current communication systems are evolving to follow the 5G-R standards, which are designed based on 5G with specific railway functionalities. 5G-R provides railway communication systems with enhanced mobile broadband, massive machine-type communications, and ultra-reliable low communications. However, the rapid development of communication networks presents new requirements and challenges that will exceed the capabilities of 5G-R systems in future smart railways. First, the growth of connected devices and machine-to-machine (M2M) communication requires extremely high communication speed and scalability for real-time operations. 5G is expected to reach its limits in fewer than 10 years \cite{10054381}. Next-generation rail networks are expected to demand frequencies in excess of 100GHz and increase the capacity to accommodate more than billions of devices. Second, emerging innovative services, such as extended reality (XR), ultra-high-definition video surveillance, and fully autonomous operations, require 10 to 50 times improvements in latency and reliability over 5G \cite{10054381}. Additionally, certain areas of a rail network, including remote mountainous areas and tunnels, require significantly enhanced coverage for reliable smart operations and improved passenger experience. Specifically, in ultra-high-speed traffic scenarios, the mobile speed of some terminals will exceed 1000km/h. Meeting the demands of ultra-high security and ultra-high-precision positioning has become a daunting task. However, the ITU indicators defined by 5G only support the mobile speed of 500km/h, and there is no definition of safety and positioning accuracy \cite{8985528}. Therefore, it is difficult for existing 5G networks and technologies to meet the requirements brought by the future intelligent transportation application scenarios.

\subsection{Design Challenges and Visions for 6G Smart Railways}

The sixth generation (6G) mobile communication system emerges as the future generation communication paradigm, with significantly enhanced communication performance in terms of ultra-low latency, ultra-high security, full coverage, and ultra-high positioning accuracy. 6G-enabled smart railways are envisioned to be characterized by green and efficient all-day operations, strong operational security and information security, fully automatic operations, low-cost intelligent maintenance, and a fully connected real-time service experience for passengers. More and more attention from worldwide organizations and institutions has been paid to the investigation and development of 6G key technologies. A variety of technical solutions have been proposed in the field of intelligent transportation systems (ITSs). The European railway research advisory committee has formulated the ``Rail Route 2050'' plan and proposed a 6G-based blueprint for the development of high-resource-efficiency and intelligent rail transit in 2050 \cite{rail2050}. Oriented to high-speed mobile scenarios and requirements, the EU Hexa-X 6G project defines a new 6G intelligent network architecture and is developing various key enabling technologies for 6G networks \cite{9482430}. The 6G research program issued by Finland takes autonomous driving as one of the main application scenarios, and aims to design 6G networks that can support 99.99999\% reliability, 1ms latency, and 1000km/h ultra-high mobility \cite{6Genesis}. 

For high-speed railway (HSR), 6G gives rise to a wide range of new business requirements in terms of all-day automatic train operations, full lifecycle intelligent operations and maintenance, fully connected self-organizing rail networks, ultra HD 8K/VR videos, ultra-accurate (centimeter-level) train positioning, and real-time digital twin (DT) monitoring and alerting \cite{8766143}. To achieve these safety-critical and sustainable operations, the following unique challenges need to be addressed in applying 6G to future smart railways:
 \begin{itemize}
        \item \textit{High-speed mobility support:} 6G systems must handle speeds exceeding 300 km/h, requiring ultra-reliable, low-latency communication (URLLC) and the ability to mitigate Doppler effects during frequent handovers.
        \item \textit{Wide-area and continuous coverage:} Unlike other applications, railways traverse remote areas, necessitating robust space/air-to-ground communications and integrated sensing to maintain coverage.
        \item \textit{Real-time data exchange:} Railways demand instant data transfer between trains, control centers, and signaling systems to ensure safety and operational efficiency.
    \end{itemize}

To meet the requirements and address the challenges of 6G intelligent transportation applications, new mobile communication architectures and technological breakthroughs are essential. Integrating advanced technologies within the 6G framework will enable near-instantaneous communication, supporting autonomous operations and high-quality services.


\subsection{Comparisons and Key Contributions}
Smart railways have garnered significant attention lately, leading to many survey papers dedicated to this evolving field. For example, Chen et al. \cite{8418701} provided a concise review of the GSM-R system and its limitations, along with evolving user demands and technical challenges for future railway communication systems. However, this paper primarily focused on heterogeneous network architecture without addressing active perception or optimal resource allocation for communication, computing, and caching, and lacked a comprehensive view of 6G technologies. Nold and Corman \cite{10373408} conducted multi-stage interviews with 30 experts to provide a holistic technological view of the railway system. While this study targeted at the future development and feasibility of technology, it did not delve deeper into the unique characteristics of smart railways.

In addition, several specialized overview papers \cite{9582617,9232926,10740531} examine smart railways from different perspectives. For instance, Sheng et al. \cite{9582617} discussed the technical development and applications of HSR based on the space-air-ground integrated network. Noh et al. \cite{9232926} focused on the technical specifications of 5G NR, with an emphasis on its application in high-speed train scenarios. Safitri et al. \cite{10740531} concentrated on quality of service (QoS) considerations in railway communication. Jo and Kim \cite{8026132} proposed a cost-effective IoT solution, including a device platform, IoT network, gateway, and platform server for smart railway infrastructure, further highlighting its potential and feasibility. These studies offer valuable insights into the smart railways. However, they primarily address individual enabling technologies or specific aspects, lacking comprehensive and holistic overviews as well as technical tutorials that integrate multiple perspectives and technologies within the smart railway systems.

In this paper, we will discuss possible 6G key technologies for smart railways, including integrated 6G network architectures, cross-domain channel modeling, wireless coverage technologies, reliable transmission technologies, and computing and caching. We will also present the architecture and key technologies to manage endogenous railway network security. Moreover, the IoT monitoring and transmission technologies, massive machine communications, and integrated sensing and communication (ISAC) will be elaborated. Finally, we will discuss future research and development. Our major contributions are summarized as follows.

\begin{itemize}
	\item We provide a state-of-the-art review of 5G railway communications, including its development progress, benefits, and drawbacks in future smart railway systems.
	\item We analyze the new business requirements and characteristics of 6G railway communications, as well as the design challenges of smart railway systems, including the possible 6G services and enabling technologies.
	\item We give a comprehensive presentation of the key enabling technologies for 6G railway communications, including channel modeling, wireless coverage, ubiquitous edge intelligence, and endogenous security. We further discuss corresponding challenges and promising solutions associated with deploying these technologies in smart railways.
	\item By jointly considering the integrated network architecture, the business requirements, and the integration challenges, we present different deployment methods for a series of key enabling technologies, e.g., cell-free massive multiple-input multiple-output (MIMO), reconfigurable intelligent surface (RIS), DT, to guarantee the ultra-high business requirements and service quality of future smart railway communications.
	\item We comprehensively elaborate the integration techniques of communications, computing, and caching in 6G-enabled railway networks, and design the integrated communication systems by jointly considering the active sensing and optimal allocation of available resources.
	\item We discuss the possible future research and development directions in 6G-enabled smart rail systems.
\end{itemize}

We provide a more detailed comparison between our paper and existing survey papers in Table \ref{table:comparisons}.
\begin{table*}[htbp]
       \centering
       \caption{Comparisons of Relevant Survey Papers}
       \label{table:comparisons}
       \renewcommand{\arraystretch}{1.3}
       \resizebox{\textwidth}{!}{%
              \begin{tabular}{|p{0.8cm}|p{6.5cm}|p{6.5cm}|}
                     \hline
                     \textbf{Survey} & \textbf{Key Contributions} & \textbf{Main Limitations} \\ \hline
                     \multicolumn{3}{|c|}{\textbf{Comprehensive Survey}} \\ \hline
                     \textbf{\cite{8418701}, 2018} 
                     & 
                     \vspace{-0.3cm} 
                     \begin{itemize}
                            \item Review the existing GSM-R and its limitations
                            \item Discuss the future development of user demands and various data services, as well as the key performance indicators for railways
                            \item Survey the recent wireless technologies and network architectures
                            \item Summarize the technical challenges of future mobile communication systems for railways
                     \end{itemize} 
                     & 
                     \vspace{-0.3cm} 
                     \begin{itemize}
                            \item Focus mainly on heterogeneous network architecture, which is a relatively basic part. The active perception and resource allocation of communication, computing and caching are not integrated
                            \item Review the performance indicators and wireless technologies of railway mobile communication systems, lacking a comprehensive consideration of key enabling technologies
                     \end{itemize} \\ \hline
                     \textbf{\cite{10373408}, 2024} 
                     & 
                     \vspace{-0.3cm}
                     \begin{itemize}
                            \item Report on multi-stage interviews of 30 experts concerning a holistic technological view of the railway system
                            \item Provide insights on improving technological aspects into intelligent transport systems
                     \end{itemize} 
                     & 
                     \vspace{-0.3cm}
                     \begin{itemize}
                            \item Focus mainly on the future development and feasibility of technology, without further excavating the unique characteristics of smart railways
                            \item More theoretical considerations and deployment methods of key technologies should be investigated
                     \end{itemize} \\ \hline
              \end{tabular}%
       }
       
       \vspace{1em}
       \resizebox{\textwidth}{!}{%
              \begin{tabular}{|p{0.8cm}|p{13cm}|}
                     \hline
                     \multicolumn{2}{|c|}{\textbf{Special Survey}} \\ \hline
                     \textbf{\cite{9582617}, 2022} & 
                     \vspace{-0.3cm}
                     \begin{itemize}
                            \item Discuss the network architecture of the SAGIN for smart HSRs
                            \item Summarize the research direction for the future development of AI technologies in SAGIN in HSR
                     \end{itemize}  
                     \\ \hline
                     \textbf{\cite{9232926}, 2020} & 
                     \vspace{-0.3cm}
                     \begin{itemize}
                            \item Provide an overview of various 5G NR key design aspects
                            \item Introduces the technical specifications of 5G NR with particular attention on the high speed train scenario
                            \item Discuss the challenges and perspectives for high speed train communications of Beyond 5G
                     \end{itemize} 
                     \\ \hline
                     \textbf{\cite{10740531}, 2024} & 
                     \vspace{-0.3cm}
                     \begin{itemize}
                            \item Highlight the parameters that need to be considered to achieve quality of service in three fields
                            \item Introduce the methods to achieve quality of service
                            \item Propose future research directions in Railway communication, ICN, and Li-Fi scenarios
                     \end{itemize}        \\ \hline
                     \textbf{\cite{8418701}, 2018} & 
                     \vspace{-0.3cm}
                     \begin{itemize}
                            \item Investigates the major challenges and opportunities associated with the smart railway infrastructure
                            \item Propose the network architecture of IoT solution to deduce the potential and feasibility
                     \end{itemize}        \\ \hline
              \end{tabular}%
       }

\end{table*}

\section{Related Works}
In this section, we provide a thorough review of works closely related to this paper. We also introduce key enabling technologies from the perspective of network layers, which enhance wireless transmission and coverage while enabling integrated computing, storage, and intelligent services tailored for smart railways.

\subsection{Network Architecture}
\begin{figure}[htb]
	\centering
	\includegraphics[scale=0.32]{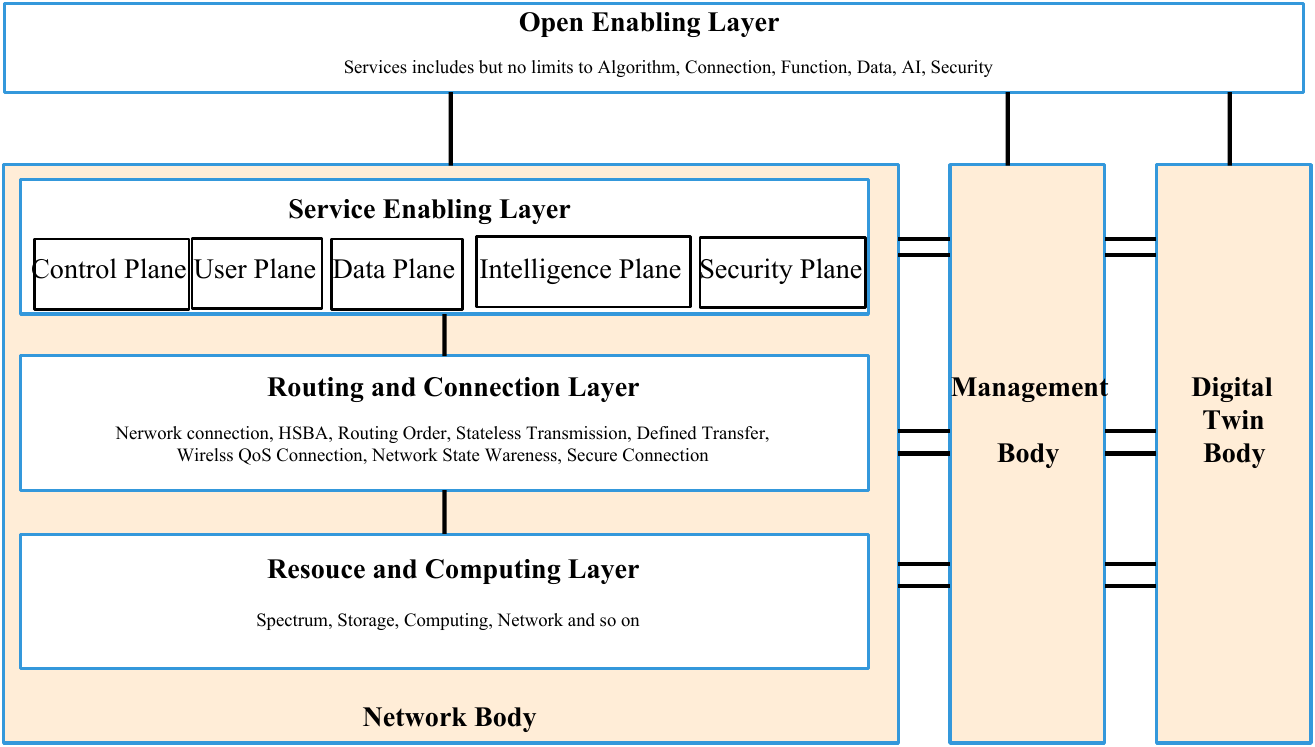}
	\caption{Architecture of 6G communication networks\cite{9651548}}
	\label{fig:li1}
\end{figure}

The design objective of a 6G network is to create an integrated network with lightweight, secure, efficient, and intelligent network entities. 
In the 6G wireless networks White Paper, the logical architecture (Fig.\ref{fig:li1}) is structured into three bodies, four layers, and five planes. The body represents the spatial view, defining key entities such as the DT and management choreography. The layer provides a logical stratification of functions, enhancing control over computing and communication resources while supporting cross-domain integration. The plane represents functional roles, extending the traditional control and user planes with additional data, intelligence, and security planes.

The future railway networks with 6G communication will be integrated with terrestrial networks, aerial networks, and satellite networks. The 6G integrated network is called space-air-ground integrated network (SAGIN) \cite{li1}. Incremental or patch-like enhancements are insufficient to meet the diverse service demands of large-scale, heterogeneous 6G railway networks. Thus, core capabilities such as security, artificial intelligence (AI), and DT must be integrated into the network architecture. Through endogenous design, core technologies can be combined with communication networks to further empower efficient autonomous management, improving user experience and realizing the leap from the Internet of everything (IoE) to the Intelligent Connection of Everything \cite{li2}.

The emerging network architectures that can be adopted for railway communications build a unified control plane and data plane through the integration of radio access network (RAN) and core network (CN) \cite{li7}, including a unified terminal and air interface. Wang et al. \cite{li9} investigated the hybrid network architecture, radio resource management, and transparent handover for emergency settings in detail. A 5G-satellite network architecture based on software-defined network (SDN)/network function virtualization (NFV) was proposed in \cite{li10}. 
Since 2017, 3GPP has defined scenarios, functions, and performance requirements for integrating satellites with 5G \cite{li10,li11}. Rel-16 focused on standardizing 5G non-terrestrial networks (NTNs), while Rel-17 introduced key architectures, including transparent bent-pipe and regenerative models \cite{li11}. These efforts lay a solid foundation for future railway communication networks.

Applying the existing 5G service-based architecture (SBA) to 6G railway networks is challenging due to factors like dynamic topology, long distances, and complex networking. To address this, the holistic service-based architecture (HSBA) has been proposed for 6G railways \cite{li3}, extending service functions across the control plane, user plane, and access network. The network repository function (NRF) enables service registration, discovery, and authentication, ensuring only authorized access to services \cite{li13}. Furthermore, effective network management is essential for handover, offloading, routing, and resource allocation, but integrating terrestrial, aerial, and satellite networks remains challenging. Terrestrial and satellite BSs can jointly serve diverse users. Li et al. \cite{li16} explored SDN-based SAGIN architectures, highlighting their agility in a three-layer satellite constellation. Additionally, SI-STIN \cite{li17} was proposed to coordinate heterogeneous users and systems within SAGIN.


\subsection{Wireless Channels}

To support diversified 6G services for smart railways, wireless channels must be well characterized. Radio wave propagation characteristics and channel modeling provide important bases for the design and evaluation of wireless communication systems. With the expansion of 6G technology to full-scenario, full-band, and full-coverage dimensions \cite{Zhangping141}, 6G railway communication scenarios are more diversified. Thus, an integrated SCCSI (sensing, communication, computing, storage, and intelligence) architecture is needed to unify services \cite{chen2023vehicle}, tailored to the specific needs of rail environments. Beyond traditional low-frequency vehicle-ground communications, 6G rail transit will involve air-ground communications, high-frequency communications, IoT, and vehicle-to-everything (V2X) communications. This diversity makes cross-domain channel modeling a key challenge, requiring consideration of space-ground channels, RIS channels, millimeter wave (mmWave) channels, unmanned aerial vehicle (UAV) channels, and IoT channels. Research on channel measurements and modeling for smart railways mainly focuses on the following aspects.

\begin{itemize}
    \item \textit{Vehicle-to-vehicle (V2V) communication channels:} V2V communication is crucial in smart railways to ensure 6G railway applications, such as unmanned and automatic driving, as well as fully automatic operations. However, in V2V communication scenarios, the rapid dual movement of transmitting and receiving trains results in more complex Doppler effect changes and channel time variability. This makes the generalized stationary assumption of communications more significantly challenging. Moreover, the long communication distance between the transceivers, the distribution characteristics of backscatters at the near end of the train body, and non-line-of-sight (NLoS) propagation in complex environments greatly affect the characteristics of V2V wireless channels during train operation, presenting additional challenges.
    \item \textit{Millimeter wave channels:} Millimeter-wave communication provides higher bandwidth and data transfer rates for railway communications \cite{song2016millimeter}. However, the propagation loss of the millimeter-wave frequency band is much higher than that of the low-frequency band. In addition to the extremely high free-space propagation loss and vehicle penetration loss, it is also affected by factors such as atmospheric attenuation, rain attenuation, and vegetation attenuation. Furthermore, millimeter-wave communication in railway systems needs to overcome more severe Doppler effects than in the low-frequency band. The high mobility characteristics of rail transportation in the high-frequency band will intensify the cluster fading characteristics of the channel, make the non-stationary features more significant, and cause channel variations hard to model. All these present greater challenges for the application of millimeter-wave communication technology in 6G railway scenarios.
    \item \textit{IoT channels:} 
    In railway systems, large-scale IoT deployments enable real-time monitoring of equipment, personnel, and environments, enhancing connectivity, intelligence, and operational efficiency. However, massive device access, diverse deployments, and fluctuating passenger density increase wireless propagation complexity and impact IoT connectivity. The resulting heterogeneous propagation paths—varying in distance, scattering, power, bandwidth, and fading—pose significant challenges for accurate channel characterization and standardized modeling \cite{Youxiaohu3}.
    \item \textit{Air-to-ground channels:} UAVs and other aerial platforms have received widespread attention in future 6G communications \cite{9768113}. Due to their versatility and high mobility, UAVs can act as airborne BSs to assist in improving communication quality in 6G rail transportation. Compared to traditional terrestrial mobile communications, aerial-to-ground communication with UAVs has many new characteristics, such as higher flying heights, three-dimensional high mobility, and high line-of-sight propagation probabilities. However, due to the limitations of UAV payloads, power supplies, and special no-fly zones, it is difficult to measure air-to-ground channels in rail transportation scenarios.  Existing studies mostly focus on static non-rail transportation scenarios, below 6GHz frequency bands, and single-antenna configurations. In addition, fading effects from weather conditions remain underexplored. To fully explore the air-to-ground channel characteristics of 6G railways, it is urgent to carry out channel measurement campaigns in typical scenarios, analyze the multipath distribution and clustering rules of the air-to-ground channel, and create an accurate high-mobility air-to-ground channel model to support the design of air-ground cooperative railway communication systems.
    \item \textit{RIS channels:} RIS technology offers new options to improve communication coverage and performance in 6G railway networks, particularly for ensuring ultra-reliable coverage in complex rail transit scenarios. While most research focuses on deploying RIS on building surfaces, in rail transit, RIS can be more flexibly deployed along railways or on UAVs to address communication challenges. However, RIS channel research is still in its early stages. Key gaps include understanding the electromagnetic radiation and propagation loss of large-scale RIS, analyzing intelligent reflection processes, developing stochastic channel models, and addressing issues like wave control during high-speed movement and the interaction between RIS and wireless channels. These gaps limit the application of RIS in future railway communication scenarios.
\end{itemize}

Moreover, the channel modeling processes of new communication technologies such as terahertz (THz) channel and integrated sensing and communications (ISAC) channel have also been widely explored \cite{9799524}, including THz channel modeling based on IEEE 802.15 and 3GPP models. AI technology has recently been applied to 6G railway wireless channels, enhancing tasks like channel feature extraction, learning, state analysis, and radio wave prediction \cite{9277535}. AI helps mine data features from rail transit channels, establish complex relationships, and build predictive models based on historical data. This enables intelligent radio wave prediction by leveraging the fixed nature of rail transit tracks. The shift towards AI-driven prediction involves creating a radio wave prediction database and model library, improving the accuracy and optimization of rail transit coverage.

\subsection{Wireless Coverage}
Rail transit systems rely on efficient and reliable dedicated communication systems, and the quality of communication network coverage is a critical indicator of their performance. With the increasing number of scenarios and business opportunities for 6G railways, the demand for highly reliable and deeply-covered communication networks has significantly increased. However, due to the high mobility of rail transit scenarios, complex environments, large-scale time-space-frequency networks, heterogeneous services, and limited resources, achieving highly reliable radio wave coverage under capacity and energy efficiency constraints has become one of the main challenges in the future development of rail transit construction. As future 6G networks evolve from 2D to 3D, local to global, and from low- and medium-frequency bands to high-frequency bands, technologies such as drones, smart metasurfaces, millimeter waves, ultra-large-scale antennas, and precise beam tracking will be key to expanding the coverage of 6G railway networks.

Handover is another critical element in a rail transit wireless communication system, and it is essential to ensure high reliability and continuous coverage. In high-speed mobile scenarios, due to the fast time-varying channel characteristics, mobile terminals may encounter frequent handovers between adjacent base stations, resulting in a high handover failure rate and a high probability of communication interruptions. This affects the smoothness of 6G rail transit wireless coverage and fails to meet the user's high-speed and stable transmission rate requirements. Furthermore, actual railways often have to encounter complex and diverse terrains such as viaducts, tunnels, plains, hills, and mountains, which poses higher demands for switching algorithms. In location-based handover algorithms, when the train enters the overlapping zone, two adjacent base stations can dynamically adjust the beamforming gain to achieve beam tracking with the mobile terminal. This method can improve the signal strength received by the terminal equipment, thereby achieving a successful handover. Additionally, handover can be triggered by using the predicted reference switching point to reduce the probability of interruptions during the handover process, providing smoother 6G wireless coverage for the rail transit communication system. The use of advanced technologies such as AI, distributed machine learning, and data analytics can further improve handover performance and enable seamless and uninterrupted communication in smart railway scenarios \cite{yan2019machine}.

\subsection{Cell-free Massive MIMO}
In high-speed mobile scenarios, the Doppler frequency shift caused by mobility can result in significant performance degradation in communication systems. To overcome this issue, cell-free massive multiple-input multiple-output (MIMO) technology can leverage the spatial degree of freedom and macro-diversity gain to achieve high system performance and uniform network coverage, making it a promising technology for future wireless communications with high mobility requirements.

Cell-free massive MIMO is an advanced technology that represents a significant leap from the traditional massive MIMO and cellular network architecture \cite{9585108}. In a cell-free massive MIMO system, multiple distributed access points (APs) are connected to a central processing unit (CPU) via a fronthaul link, and all users in the network utilize the same time-frequency resources simultaneously. On the other hand, cell-free massive MIMO enhances network coverage and provides uniform spectrum efficiency without the conventional cell and cell edge effect in the network architecture. In high-speed rail scenarios, the wireless propagation environment is highly complex and diverse, with trains passing through a variety of terrains such as viaducts, tunnels, stations, mountains, forests, and urban areas. These wireless propagation scenarios contain a rich set of scattering paths, making it necessary to consider the NLoS paths to accurately describe the wireless communication channel of the high-speed rail. Moreover, the application of cell-free MIMO can alleviate the inter-subcarrier interference caused by Doppler frequency offset and mitigate the problem of fast signal processing and frequent base station switching in a high-speed moving environment. The possible deployment of cell-free MIMO in railway communication networks requires effective solutions to two major challenges: scalable deployment and severe Doppler frequency shift. More comprehensive investigations are expected in designing scalable cell-free massive MIMO-OFDM systems and using different transmit power to eliminate interference caused by high-speed movement.

\subsection{RIS in High-Speed Railway}
A RIS is a planar array of numerous passive reflecting elements, capable of individually modifying the amplitude and/or phase of incident electromagnetic waves. By deploying RISs in a wireless network and intelligently coordinating its reflections, the wireless channel between transmitter and receiver can be flexibly adjusted, providing a promising solution to addressing channel fading impairment and interference issues.

As carrier frequency increases, wireless propagation becomes more challenging due to higher penetration loss and lower scattering, reducing direct propagation paths between the transmitter and receiver. To address this, RIS can be deployed to provide additional scattering and enhance beamforming towards the desired receiver. In high-speed rail scenarios with multiple terrains and obstacles, RIS improves propagation conditions. The Doppler effect from high-speed movement and penetration loss from train carriages necessitate RIS in HSR communication systems. RIS-assisted HSR systems enable base stations (BSs) to use fewer antennas, lowering hardware costs and energy consumption. By creating virtual line-of-sight paths, RIS mitigates occlusion, expands coverage, and reduces cell handovers. Its reconfigurable properties, including amplitude and phase shift, help counter the Doppler effect and fading, boosting spectral efficiency of HSR communication systems.
By leveraging its channel enhancement capabilities, RIS can support various applications in high-speed railway scenarios. For example, it can enhance computing and caching performance by expediting the data transmission process \cite{10186000}.

\subsection{Computing and Caching} 
6G cannot be possible without touching upon computing, storage, and intelligence. Hence the integrated design of communications, computing, storage, and intelligence (CCSI) is needed. The integrated design of CCSI for railways needs to be carefully investigated and performed. As we move towards 6G railway communication, computing and caching must go beyond handling high-speed mobility to address domain-specific challenges. Railways exhibit predictable, track-bound mobility, which allows for proactive and location-aware edge caching and resource allocation, unlike UAV or satellite networks where mobility patterns are less deterministic. Furthermore, railway systems are safety-critical, requiring ultra-reliable and low-latency edge computing as well as fault-tolerant caching mechanisms to support functions like autonomous driving, obstacle detection, and emergency communications.
Additionally, there is a growing demand for massive, ubiquitous, and dynamic access, along with enhanced security and privacy measures to ensure the safe and reliable operations of these systems. New technical solutions for scenarios of dynamically changing channels for train-ground communication, and concurrent computing offloading services for massive user terminals also need to be explored. These offloading strategies must account for periodic handovers, tunnel-induced signal shadowing, and high-speed transitions across heterogeneous networks. Moreover, the developed new paradigm can be integrated with railway communication networks to enhance communication performance. For example, the rail DT networks have been jointly designed by integrating wireless communication, edge computing, and mobile caching with DT in rail communication networks \cite{10238401}.

\subsection{Security and Privacy}
The ubiquitous connectivity and improved coverage of 6G raise new challenges to the development of rail communications. Emerging security threats to cloud management platforms, train control systems, edge of rail networks, and massive accessing terminals need to be addressed carefully. The endogenous security architecture and related techniques have been designed and developed for railway networks, to achieve rapid consensus, intelligent active defense, trusted security enhancements, and ubiquitous collaborative security. Meanwhile, blockchain, as a fundamental technique to enable trust establishment among untrusted parties in fully distributed environments, will provide a reliable collaboration platform to manage the trust among untrusted entities in a fully distributed network. The key technologies for endogenous railway security have been discussed and explored in railway communications, including spatiotemporal matching-based trust enhancement, trustworthy authentication, seamless fast switching, intelligent intrusion detection, and big data security and privacy protection.

\subsection{Emerging Technologies}

Emerging technologies play crucial roles in driving the development of the next-generation smart railways. To meet the future demands of more diverse business applications and extreme performance, 6G railway communications will need to make breakthroughs in the integration of various disruptive technologies, such as ubiquitous AI, satellite-ground integration, terahertz communication, and DT.


\subsubsection{Ubiquitous AI} The integration of AI into next-generation mobile communication systems has garnered significant attention in recent years. In railway-specific scenarios, AI-driven solutions play a vital role in optimizing network performance and operational efficiency for HSR communications. Researchers in the communication field have started looking into how to leverage AI to improve the operational efficiency of mobile communication networks \cite{li23,li24,li25,li26}. For example, a deep learning (DL)-based routing strategy was proposed to reduce the routing delay in the integrated satellite-terrestrial network \cite{li23}. To manage the various resources in the integrated satellite-terrestrial network efficiently, Qiu et al. \cite{li24} proposed a deep Q-learning-based method to jointly allocate communication, caching, and computing resources among users. In \cite{li25}, two spectrum-sharing schemes based on the support vector machine and conventional neural network were proposed to achieve less interference and higher spectrum efficiency. 


In V2X and railway-to-everything (R2X) scenarios, AI has demonstrated significant potential. For instance, deep learning models have been applied to HSR channel estimation by training neural networks to handle rapid mobility and frequent handovers \cite{li36}. Additionally, AI-based tools facilitate data-driven decisions to improve the performance of railway networks by adapting to dynamic operational conditions, such as traffic fluctuations \cite{li37}. The scarcity of qualified datasets for vehicular networks motivated the adoption of reinforcement learning (RL) for a vertical handover strategy \cite{li39}. Furthermore, federated learning, as an emerging distributed machine learning model, was proposed to protect user privacy and reduce the communication and computing burden \cite{li41}. AI technologies can also be used to enhance the security of future 6G networks. In \cite{li27}, authors discussed the secure machine learning scenarios and presented two encryption algorithms for the integrated network. 

\subsubsection {Cell-Free Massive MIMO}
To enhance the performance of wireless communication networks, cell-free massive multi-input multi-output (CF m-MIMO) was proposed to support massive ultra-reliable and low-latency communications (mURLLC) \cite{li52}. Such a distributed CF m-MIMO technology significantly increases throughput as well as user coverage, and the co-processing at multiple APs can suppress inter-cell interference. In addition, CF m-MIMO does provide a higher capacity of backhaul connections, but the co-processing at APs increases backhaul overheads. A user-centric approach was proposed for CF m-MIMO to reduce the backhaul overhead \cite{li54}. Nayebi et al. \cite{li55} studied the downlink performance of CF m-MIMO in terms of the minimum data rate. Zhang, Wang, and Poor \cite{li51} integrated mmWave user-centric CF m-MIMO with hybrid automatic repeat request using incremental redundancy to enhance QoS in 6G networks. They also proposed finite block length coding (FBC) based schemes for statistical delay and error-rate bounded QoS provisioning in SWIPT-enabled CF m-MIMO 6G networks \cite{li56}. Table \ref{tab:cf-mimo} summarizes the related studies regarding CF-MIMO systems.

\begin{table*}[t]
    \caption{Comparisons of Related Work on CF-MIMO}
    \label{tab:cf-mimo}
    \centering
    \begin{tabular}{|c|p{6cm}|p{5cm}|p{5cm}|}
        \hline
        \textbf{Ref.} & \textbf{Objectives} & \textbf{Main Factors} &  \textbf{Algorithm} \\
        \hline
        {\cite{li51}} & Minimum the delay violation probability & Finite blocklength coding  & Statistical delay and error-rate bounded QoS provisioning schemes  \\\hline
        {\cite{li52}} & Maximize the smallest of all user rates & Channel estimation errors, power control, and non-orthogonality of pilot sequences & Max-min fairness power control algorithms  \\\hline
        {\cite{li54}} & Less backhaul overhead and  & So-called cell-free (CF) massive MIMO & A user-centric (UC) virtual cell approach  \\\hline
        {\cite{li55}} & Minimum rate among all users and 5\%-outage rate & Max-min power control  & Heuristic algorithms  \\
        \hline
    \end{tabular}
\end{table*}

\subsubsection {Terahertz Communications}
How to increase wireless transmission rate is always an important research direction in wireless communication. Future networks require a terabit-per-second communication link to meet strict delay requirements. Traditional wireless communication systems (below 5GH) and even millimeter wave communication systems (30-300GHz) cannot achieve such high data rates \cite{li45}, and hence the exploration of higher frequency bands was inspired. In this context, Tataria et al. \cite{li48} discussed the use cases and enabling technologies in 6G networks. Chaccour et al. \cite{li49} provided a discussion on the defining features of THz wireless systems and potential use cases related to 6G communications and the IoE. Han and Chen \cite{li50} considered modeling methods in THz along with current issues and future considerations relating to potential 6G applications.

\subsubsection{Semantic Communication}
As an important technology of 6G, semantic communication (SC) provides robust potential for the intelligent transformation of railways by enhancing transmission efficiency and reliability and enabling multi-scenario applications. SC employs a ``transmission after comprehension'' strategy, transmitting only essential semantic information. This significantly reduces bandwidth consumption, boosts data transmission efficiency, and ensures precise data interactions in smart railway systems. It supports a wide range of applications, including autonomous driving, intelligent scheduling, and passenger services. For example, video surveillance is essential in smart railways for real-time monitoring of infrastructure, passenger safety, and operations \cite{10565301}. Traditional systems often face challenges transmitting high-bandwidth video in constrained private railway networks. SC addresses this by sending only contextually relevant information—such as security threats or anomalies—rather than full video streams, thereby improving efficiency and reducing bandwidth usage in spectrum-limited environments. In the future, the deep integration of semantic communication and smart railway systems is poised to become a key driver of smart railway development.

\subsubsection {Cooperative Secure Transmission}
Security is critical for ensuring the safe operation of railway communication networks, protecting infrastructure, devices, data, and assets. Key objectives include confidentiality to prevent unauthorized access, data integrity to ensure information remains accurate, and authentication to verify legitimate users. Due to the inheritance, many of 6G security features are expected to evolve from the 5G framework. Neshenko et al. \cite{li65} presented security-related challenges and sources of threats and highlighted various prospective technologies for 6G. Physical layer security, quantum-safe security, AI-driven security, and trust establishment are supposed to be the top priorities in 6G. Some researchers \cite{li66} focused on the physical layer security, while others \cite{li69} applied AI technologies to protect communication security. Yao et al. \cite{li73} recommended open authentication protocols supporting the integrated network. Blockchain and distributed ledger technologies \cite{li74} were also used to protect the confidentiality and integrity of 6G networks.

\subsubsection {Digital Twin}
With the support of DT technology, 6G networks can establish a parallel physical-virtual system, enabling integrated management across both domains. This allows real-time modeling of networks and applications \cite{9711524}, as well as dynamic control strategies that operate from the virtual space to physical systems \cite{li59}. The DT-driven architecture also enables predictive operations and iterative optimization. Ultimately, it facilitates seamless and intelligent information exchange—between people, devices, and systems—creating a more responsive and adaptive railway communication environment. DT has been widely studied \cite{li61,li63} for improving the performance of mobile edge computing (MEC) systems. Sun et al. \cite{li61} proposed a mobile offloading scheme in DT edge networks to minimize the offloading latency under constraints on service migration and user mobility. Furthermore, Yu et al. \cite{li64} designed a wireless DT network model for 6G networks to exploit DTs to mitigate unreliable and long-distance communication between users and edge servers. Considering the varying DT deviations and network dynamics, Sun et al. \cite{li60} designed a dynamic incentive scheme to adaptively adjust the selection of optimal clients and their participation level for SAGIN. Table \ref{tab:DT} summarizes the related studies regarding DT networks. Through optimization, predictive analysis, fault-tolerant control, automated processes, etc., DT networks will enable real-time perception and close monitoring of the status of rail systems and drivers. The operation and maintenance costs can be reduced and the security and reliability of HSR operations can be improved.

\begin{table*}[t]
    \caption{Comparisons of Related Work on DT}
    \label{tab:DT}
    \centering
    \begin{tabular}{|c|p{3cm}|p{5cm}|p{5cm}|}
        \hline
        \textbf{Ref.} & \textbf{Scenario} & \textbf{Objectives} &  \textbf{Algorithm} \\
        \hline
        {\cite{li61}} & MEC & Minimize the offloading latency  & DRL  \\\hline
        {\cite{li63}} & IoT, MEC & Minimize the weighted cost function & Blockchain-empowered FL + DRL  \\\hline
        {\cite{li64}} & 6G, MEC &Minimize the average system latency & DRL + TL  \\\hline
        {\cite{li60}} & Air-ground network & Maximize the utility  & Stackelberg game + FL  \\
        \hline
    \end{tabular}
\end{table*}
\subsection {Lessons Learned}
The review provides key insights into the development of 6G-enabled smart railway systems. An integrated network architecture is essential for seamless connectivity, while advanced channel modeling and emerging technologies like RIS, cell-free MIMO, and terahertz communications offer significant potential but require tailored solutions to address high mobility and complex railway environments. Additionally, emerging technologies including AI-driven techniques, digital twin networks, and semantic communications play vital roles in improving resource management, security, and predictive capabilities. Future efforts should prioritize improving and innovating these technologies and tackling railway-specific challenges to ensure effective deployment.

\section{Network Architecture and Key Enabling Technologies for 6G Smart Railways}

To leverage emerging 6G technologies, a robust new network architecture for smart railways is essential. In this section, we present a flexible network architecture that facilitates the integration of enabling technologies.
\subsection{Integrated 6G Network Architecture}
\begin{figure*}[htb]
    \centering
    \includegraphics[scale=0.37]{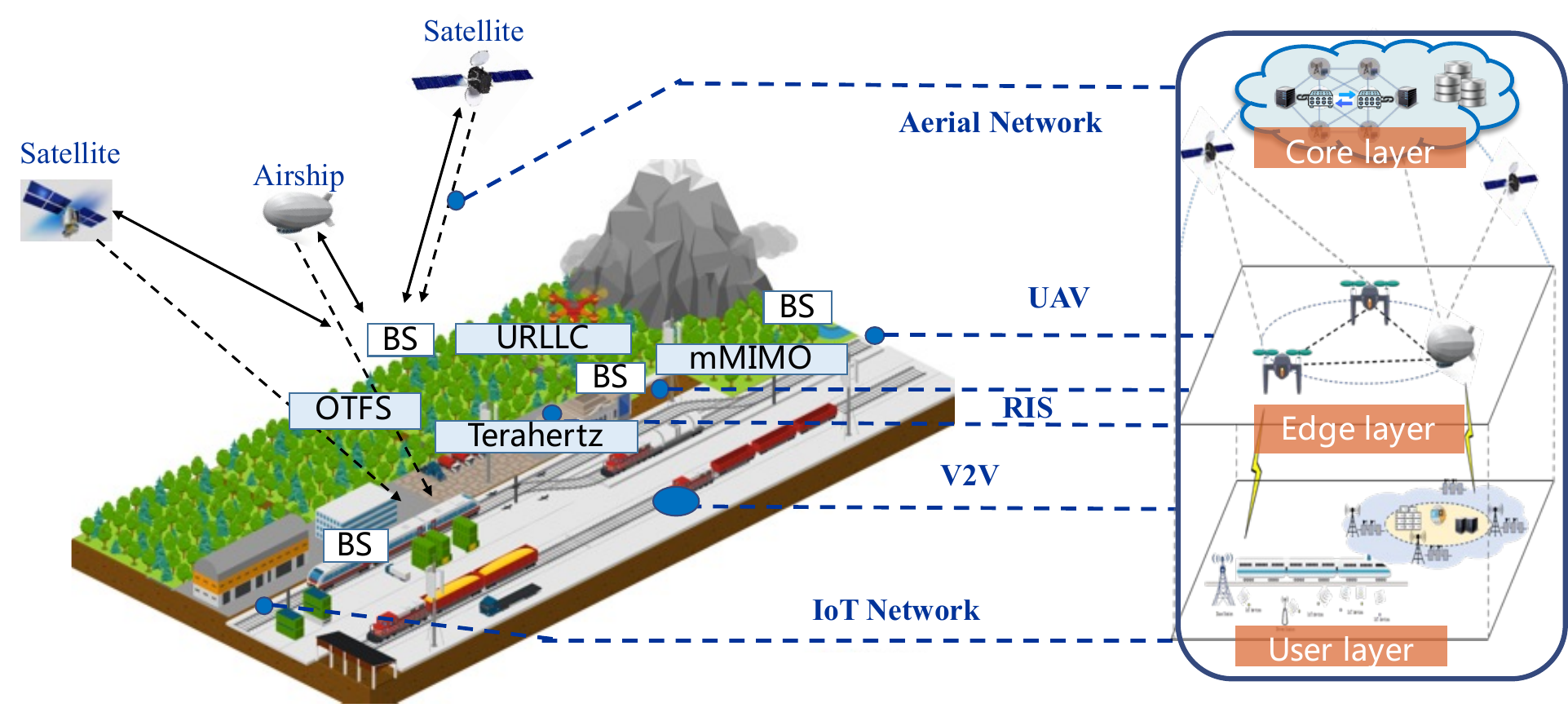}
    \caption{Network architecture of 6G smart railways}
    \label{fig:fig1-l}
\end{figure*}
To achieve the design goal for an ultra-reliable, resource-efficient, and ubiquitous intelligent communication network for smart railways, several challenges must be addressed. Firstly, the network must be fully connected and compatible with various terrain scenarios, including tunnels, mountainous areas, and remote regions. This can be achieved by utilizing various communication technologies such as satellite communication, RIS, and THz communication to enhance coverage and transmission quality. Secondly, to establish intelligent connections between humans, machines, and objects, emerging distributed and cooperative AI techniques should be integrated into communication networks. This integration will optimize the operation of communication systems and the allocation of network resources. Thirdly, the network architecture must guarantee the security and reliability of automated railway operations. Endogenous security measures should be integrated into the network at the architecture level.

To achieve the vision for intelligent railway systems, it is essential to create an integrated network that optimizes limited communications, computing, and storage. This network should address the challenges of full connectivity, massive intelligent connections, and strong endogenous security. The network design should leverage diverse emerging communication techniques to provide high-speed data transmissions and wide-area coverage across various frequencies, including THz and mmWave. Additionally, it should support fast and concurrent access to a vast number of terminal devices and enable them to realize distributed, interactive edge intelligence. Furthermore, the network should ensure the secure and efficient automatic operations of railway systems under various weather conditions and over various terrains.

The architecture of 6G railway systems consists of three layers: the user layer, access layer, and core layer, as depicted in Fig. \ref{fig:fig1-l}. The user layer facilitates the connectivity of trains, mobile devices, trackside equipment, and IoT sensors to the edge layer through various communication methods, including satellite communications, UAV or RIS-assisted communications, and THz communications. The user layer also includes distributed resources such as storage and computing that can be collaboratively utilized to perform multiple railway operations and applications. The edge layer comprises BSs and MEC servers with resources, enabling reliable on-demand connections and real-time computation services for end users. Furthermore, the edge layer provides ubiquitous intelligence to users, allowing users with limited capabilities to access high-quality personalized services. The core layer consists of servers and cloud nodes that provide significant computational resources for executing computing-intensive tasks, supporting large-scale services and applications across the entire railway system. Tasks that cannot be processed by the edge layer are offloaded to the core layer for efficient execution. The network nodes in the core layer perform not only conventional transmission flow routing but also the routing of network resources to improve the utility of resources.

\begin{figure}[htb]
    \centering
    \includegraphics[scale=0.28]{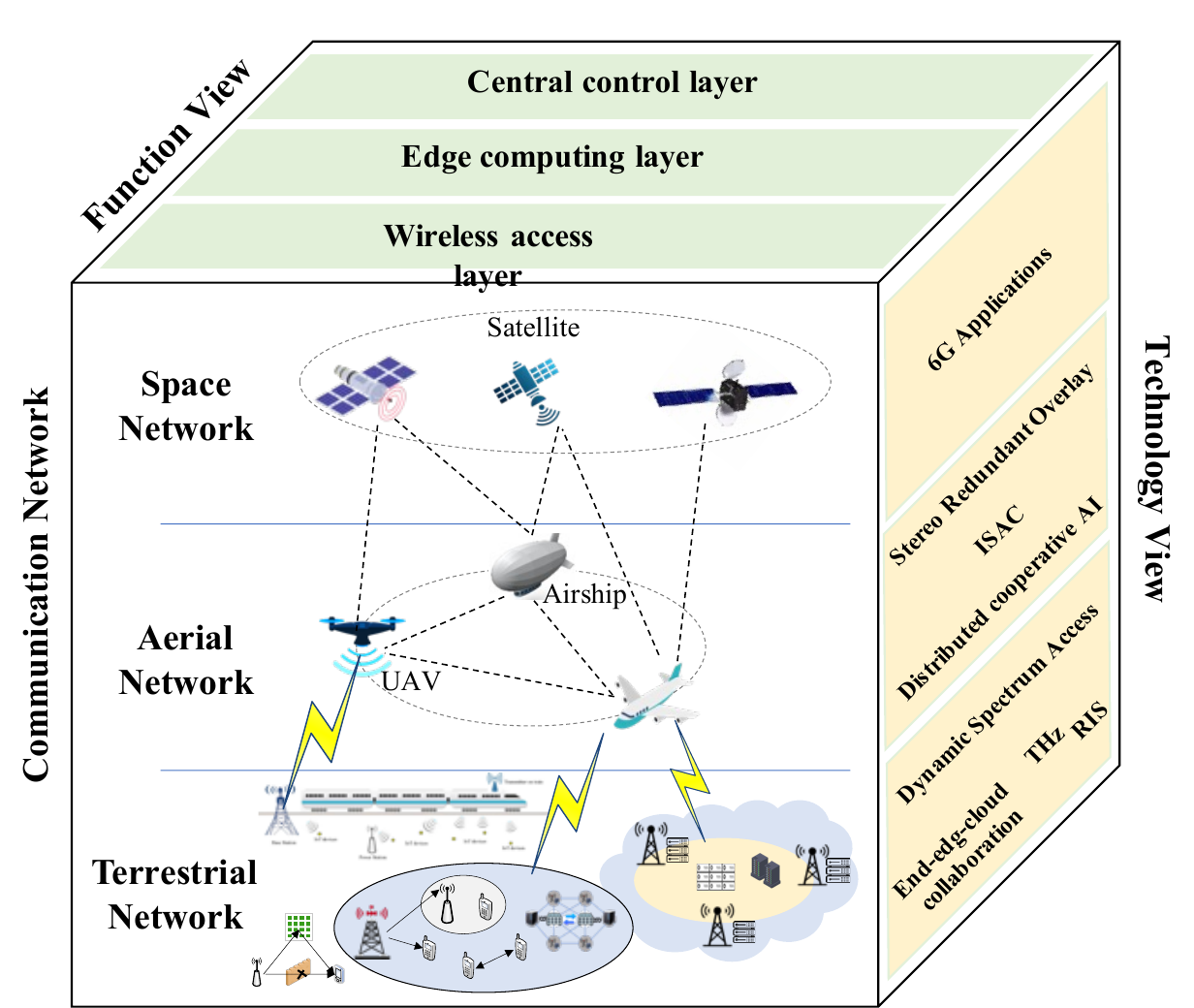}
    \caption{Different views of 6G railway architecture}
    \label{fig:fig2-l}
\end{figure}

Fig. \ref{fig:fig2-l} illustrates the detailed communication network architecture from different perspectives. The network can be divided into three segments, namely the terrestrial, aerial, and space network, based on their connection modes. In the terrestrial network, trains, mobile devices, and IoT sensors connect to trackside BSs through wireless links such as V2V and V2X links. Emerging technologies such as terahertz communication, visible light communication (VLC), and RIS can be used to improve communication efficiency, transmission rate, and connection reliability.

The aerial network employs airships and UAVs to enlarge the wireless coverage area and enhance the communication quality between the user layer and edge layer. In this scenario, distributed AI techniques can be used to cooperatively learn from vast data and optimize the network performance. Moreover, using distributed AI for event prediction and operation scheduling can significantly enhance the operations of railway systems.

The space network employs satellite communication to provide users with constant access to the railway network anytime and anywhere, particularly in plateaus, tunnels, and remote areas. The integrated communication between space and space will play a crucial role in establishing connections through inter-satellite routing and satellite-ground communications.

\subsection{Cross-Domain Channel Modeling}
Dedicated mobile communication is the necessary infrastructure of a smart railway, which undertakes key safety-related services such as railway dispatching command, train operation control, and early fault warning. Characterization and modeling of wireless channels play an important role in the standardization and deployment of wireless communication systems. Understanding radio wave propagation mechanisms and establishing an accurate channel model are the cornerstones of wireless communication system design, evaluation, and standardization of 6G railway systems.

\subsubsection {Train-to-Train Channel}
For railway train-to-train (T2T) communications, the unique propagation environment and requirements bring challenges to channel modeling. In T2T links, the relative speeds of the transmitter and receiver are high, and the Doppler effect causes time-varying channels. With the increase in train speed, channels show fast fading in time domain and frequency dispersion in frequency domain. Carrier frequency offset will cause rapid degradation of signal demodulation performance and make communication links non-stationary. In addition, fast fading will seriously degrade the performance of symbol detection and increase bit error rate, thus greatly reducing the performance of railway communication systems. On the other hand, when trains move at high speed, multipath structure changes rapidly, which results in the intensification of non-stationary characteristics. T2T link needs to update channel state information quickly, and it is sensitive to the change of multipath characteristics. To sum up, it is important to accurately characterize the fast time-varying channel to ensure the performance of T2T communications. There are three main challenges in the T2T channel:

\begin{enumerate}[a)]
	\item High mobility of trains: the high relative speeds in HSR systems intensify Doppler effects and fast fading, increasing channel non-stationarity.
	\item NLOS propagation: long distances and obstructed environments (e.g., mountainous areas) lead to frequent NLOS links, causing higher attenuation and unpredictable LOS/NLOS switching, which degrades reliability. This calls for dedicated NLOS models, coverage extension, and link prediction techniques.
	\item Complexities of environment: railway scenarios (e.g., viaducts, stations, crossings) feature unique scatterers and dynamic elements like overtaking trains, resulting in complex and unpredictable multipath propagation.
\end{enumerate}

Facing the above challenges, there have already existed several related works. First, channel measurement is the main tool of channel characterization and measurements of railway scenarios have received special attention. For example, Mozo et al. \cite{9984956} carried out T2T measurement with the relative speed of 50 km/h, and proposed six-tapped delay line (TDL) models suitable for T2T communication in hilly areas and railway stations. Unterhuber et al. \cite{9474457} derived power delay profiles (PDP) and Doppler spectral densities based on T2T channel measurements in various scenarios. Furthermore, distance-variant statistics about the K-factor, delay spread, and Doppler frequency spread are presented. In general, T2T channel measurement is still not sufficient since most railway channel measurements are for train-to-ground (T2G) links \cite{9127790, 6378492}. There are significant differences between T2T and T2G links. Although other methods, such as ray tracing \cite{8438326} and stochastic geometric models \cite{9786750}, can largely alleviate the difficulties caused by the lack of measurement data. Targeted T2T channel measurements, especially high-speed ($>$300 km/h) and multi-frequency measurements, are still necessary and urgently needed.

In terms of frequency, the existing railway communications mainly aim at sub-6 GHz, such as 2.6GHz and 5.2GHz \cite{9135487}, and only a small amount of work involves high-frequency bands \cite{7794737}. In terms of scenarios, typical railway scenarios have received some attention, such as viaduct \cite{8392735}, cutting \cite{6987277}, tunnel \cite{9933809}, and mountainous areas \cite{9378519}. However, the gap is the lack of investigation on T2T links, especially in high-speed cases. T2T links can lead to extremely high relative speeds, which require additional research. In terms of channel characteristics, most existing works focus on large-scale parameters. Small-scale characteristics, especially multipath characteristics, have not received sufficient attention, mainly due to the limited measurement ability. Fortunately, ray tracing or stochastic geometric methods complement the study of small-scale characteristics to some extent. For example, in \cite{9397335}, railway channel simulation in a THz band was carried out based on ray tracing. The multipath distribution, parameter correlation, and channel capacity are analyzed in detail. In \cite{8114238}, time-frequency correlation functions, PDP, and Doppler power spectrums of mobile-to-mobile channels were analyzed according to the scatterer cluster based on geometric simplification. Nevertheless, simulation calibration and model verification based on measured data are still necessary.

To sum up, despite all the efforts, the research challenges of T2T channels can be summarized as follows: multi-frequency (especially millimeter wave) T2T channel measurements at high speed, modeling multipath distribution and multipath birth-death process in fast time-varying channel, NLOS channel modeling for long-distance railway scenario, classification and modeling of complex propagation scenarios, feature extraction, and hybrid modeling of time-space-frequency multi-dimensional channel. Therefore, there is no doubt that further research on T2T channels is still one of the key topics. It also plays an important role in developing key technologies for the next-generation railway communication systems.

\subsubsection {Space/Air-to-Ground Channels for Railways}
With the integration of new generation information technology and high-speed railway, smart HSR has gained special attention lately and requires low-delay and high-data-rate communication systems to realize full interconnection of infrastructure, trains, passengers, and goods \cite{9316444}. However, traditional ground communication systems cannot provide real-time and reliable communications to meet the growing communication demands of HSR. Therefore, 6G railway communication systems will combine ground wireless communications and space/air communications to adjust coverage area, improve communication quality, and achieve seamless connection. With this background, the concept of SAGIN is conceived and has attracted intensive attention. There are mainly three segments in SAGIN multidimensional network: space segment with satellite communication network, air communication segment with aerial network, and ground communication network \cite{9130836}. Compared with traditional ground communication, space/air-ground communication is an effective solution to high mobility and all-area challenges of future HSR communication and has broad application prospects in future smart high-speed railway communication systems. Specifically, space/air-ground communication can offer wide coverage and seamless access capability for sparsely populated areas, especially HSR lines in remote areas without additional infrastructure deployment \cite{9631953}. It can also assist services such as public emergency rescue, earth observation and navigation through terminal communication to enable the high availability and safety objectives required by railway transportation.

Satellite systems include geosynchronous (GEO), medium earth orbit (MEO), and low earth orbit (LEO) satellites, each with distinct channel and handover characteristics \cite{9120643}. GEO satellites, orbiting at about 35,800 km, offer wide coverage and continuous visibility but suffer from high latency and signal attenuation, particularly in urban and high-latitude areas. LEO and MEO satellites, orbiting at 200–2000 km and 2000–35,000km respectively, offer stronger signals and lower latency than GEO systems, but require more frequent handovers \cite{9275613}.

Combining UAVs with satellite communications offers a promising solution for global coverage and has drawn growing interest. Once limited to military use, UAVs are now widely adopted in civilian sectors such as weather monitoring, traffic control, cargo delivery, emergency response, and communication relaying \cite{7470933}. UAVs are generally categorized as fixed-wing or rotary-wing. Fixed-wing UAVs support higher speed and payloads but require continuous motion, making them unsuitable for stationary tasks. Rotary-wing UAVs, though limited in range and load, can hover and maneuver flexibly. In railway communication, UAVs equipped with transceivers can serve as aerial platforms or nodes to extend coverage and enhance connectivity in high-speed, high-traffic scenarios.

It is worth noting that accurate channel characterization is essential for reliable space/air-ground communications, particularly given their unique propagation features. In space-ground links, long distances lead to both LOS and NLOS components reflected by surrounding objects, with multipath fading and shadowing further intensified in HSR scenarios due to high speeds and obstacles. Several fading models have been proposed, including the C. Loo model \cite{1623307} for ideal channel state information, the Corazza model \cite{corazza1994statistical} for imperfect conditions, and the Lutz model \cite{289418}. Finite-state Markov models have also been applied to capture dynamic channel states in satellite links \cite{8840846}. For railway satellite-to-ground channels, it is crucial to consider the effects of fading effect caused by high speeds of both satellite and train. Some literatures have established satellite-to-ground channel for railway scenarios. In \cite{arapoglou2012railway}, a stochastic dynamic model for Ku-band and above incorporates fast fading, rain attenuation, and diffraction from metallic structures. In \cite{8739589}, a two-state MIMO model for Ka-band mobile satellite channels is developed based on data across 1–20 GHz for railway and highway scenarios.

It is widely known that UAVs possess the characteristics of good mobility and flexibility, and can dynamically optimize their position, which is suitable for improving channel conditions in ground communication, especially in high-speed railway communication scenarios. Another salient attribute of air-ground communication is a strong line-of-sight link, which reduces small-scale fading between a UAV and a ground and facilitates highly reliable transmissions over long distance. In current studies, some channel models for UAV communication have been proposed. To investigate the influence of UAV altitudes on channel characterization, Huang et al. \cite{9909265} carried out a series of channel measurements for four typical UAV altitudes: level, dive, climb, and right bank, which provide valuable insights into the system design. Furthermore, \cite{7501562,7835273} conducted air-ground channel measurements in hilly/mountainous, suburban, and near-urban scenarios, which are common communication scenarios in high-speed railway systems. On the other hand, the geometry-based stochastic channel model (GBSM) is used to mimic the non-stationary channel characteristics, including time-space-frequency correlation function and Doppler power spectral density \cite{8725553}. For air-to-ground channel models in HSR scenarios, fasting fading and the impact of ground feature (i.e., signal obstructions and reflections from buildings and terrain) need to be further considered. At the same time, some literatures describe UAV-to-ground channel for railway scenarios. In \cite{8569308}, the UAV-aided amplify-and-forward relay system for high-speed train communications over generalized $k-\mu$ fading channels is presented. In \cite{8885447}, UAV-Based FSO Communications for HSR backhauling is presented, detailing a composite fading model for UAV-to-HST channel.

However, space/air-ground propagation channels can be hampered by obstacles (such as mountains, buildings, and trees) and affected by factors such as weather, flashing and scintillation. For example, the rain attenuation from Ka-band may reach above 10dB, resulting in poor quality. Moreover, severe weather (tsunamis and hurricanes) might cause deviations in the predetermined paths. Another significant problem is serious delay, propagation loss, and Doppler effect, due to the mobility and long-distance communication in space/air-ground communications. In addition, the UAV deployment and path planning should be noted. Based on channel characteristics and communication requirements, how to realize the optimal 3D deployment and flight trajectory of UAV to reduce physical collisions and maintain sufficient coverage of the entire area is challenging.
\subsubsection{Railway Channels at High Frequency Bands}
With high carrier frequencies, HSR communication systems can provide high-rate, safe and reliable communication services. With the popularity of entertainment facilities in trains, especially for long-distance HSR travel, the demand for higher-rate communication services has also significantly increased. Therefore, HSR communication systems with mmWave frequency bands are expected to become the key technology of HSR communications in the future. However, the propagation characteristics of mmWave are quite different from sub-6 GHz bands. Meanwhile, the communication system in HSR scenario needs to work with high mobility, frequent handovers, and changing complex surrounding environments. Therefore, mmWave channel measurement and modeling in HSR scenarios is still one important direction to be studied urgently.

To handle critical communications scenarios over HSR channel, it is necessary to deal with two different communication scenarios: train-to-ground communications and railway hotspot communications. In the following subsections, we describe the main requirements of these two scenarios.

\subsubsection{Train-to-Ground Communications}
Train-to-ground communications are the most important and complex service scenarios. On one hand, train control information needs to be transmitted with high reliability by train-to-ground communication systems. On the other hand, activities such as entertainment for passengers inside trains require large-capacity and high-rate communications. Therefore, how to meet the above requirements simultaneously will become the key factor to the next generation of train-to-ground communication systems.

With the advancement of 6G standardization, the era of full spectrum is coming \cite{9193934}. Hence, the application of the full spectrum in cellular networks can also be extended to train-to-ground communication networks. Sub-6 GHz and other low-frequency bands generally have lower propagation loss than mmWave frequency bands and they also have excellent performance against Doppler shift. Therefore, low frequency bands such as sub-6 GHz can provide reliable transmissions for train control information. The mmWave communication can provide high-rate data services for passengers inside train due to its large bandwidth. Based on the full spectrum of train-to-ground communication networks, exploring the multi-frequency channel characteristics in HSR scenarios will be the key to the deployment of train-to-ground networks.

Most existing investigations still focus on sub-6 GHz measurement and modeling in HSR scenarios. For example, a stochastic channel model based on 930MHz channel measurement in different HSR scenarios \cite{6858019} was proposed. In addition, Zhang et al. \cite{8392735} established a Markov-based multi-link tapped-delay-line channel model for railway communications based on 460MHz measurement. In the area of mmWave HSR channel modeling, most research used theoretical and simulation methods due to the limitations of mmWave HSR measurements. In mmWave HSR channel measurement campaign, Soliman et al. \cite{8568640} performed channel measurements at 63GHz and analyzed channel data to model wagon-to-wagon channels with train vibrations. Meanwhile, the impact of Doppler frequency shifts caused by vibrations is also analyzed. However, the measurements were not performed for trains running at high speed. Park et al. \cite{9411172} conducted mmWave HSR measurements at 28GHz with a speed up to 170 km/h in two different HSR scenarios: viaduct and tunnel scenarios. Among them, the transmitter is fixed on the ground, and the receiver is installed on the train roof with a monopole-type omnidirectional antenna. Based on 28GHz mmWave HSR measurements, large-scale and small-scale fading characteristics were analyzed. 

Due to the limitations in conducting HSR channel measurements, GBSM is a commonly used method. For example, a 3D non-stationary wideband tunnel channel model for HSR communications was proposed based on GBSM \cite{8641489}. It models the tunnel as a cuboid and assumes that scatterers are randomly distributed on the surface of the cuboid. Based on the established model, various correlation functions are simulated and analyzed. In addition, researchers used ray tracing (RT) or measurement-calibrated channel generators to generate channel data in HSR scenarios. The HSR channel is then analyzed and modeled based on the generated channel data. For example, Dutty et al. \cite{9303579} used the calibrated NYUSIM channel generator to generate channel data in mmWave HST backhaul networks. Among them, 28GHz and 60GHz LOS and NLOS channels were analyzed, and a statistical mmWave channel model was proposed. However, to improve the performance of mmWave channel model, more channel measurement campaigns in HSR scenario are still required.

\subsubsection{Railway Hotspot Communications}
In the 6G railway networks, railway hotspot scenarios usually require large-capacity communication coverage, such as railway waiting halls and hub areas. In these scenarios, the number of communication users is large and data traffic is heavy, while mobility is slow. Thus, high-frequency and large bandwidth communication systems can be used to provide large-capacity data services in railway hotspot scenarios. MmWave and sub-terahertz (sub-THz) communications are expected to be the potential key technologies due to their large bandwidth and feasible coverage. Meanwhile, in hotspot scenarios such as waiting halls, the environment is more open. This can provide more LOS propagation for mmWave and sub-THz. For example, Zhao et al. \cite{zhao2019channel} measured 28 GHz channels in a large HSR station hall and validated channel parameters using the QuaDRiGa simulator. Similarly, \cite{8438925} conducted HSR channel measurements in subway and rural settings and applied a calibrated QuaDRiGa model to simulate high-mobility mmWave mobile hotspot networks (MHNs).

In addition, in the railway hotspot scenario, it will be possible to use THz access points to enable ultra-fast data transmission for passengers and smart devices in 6G era. THz access points deployed in railway hotspot can work at THz frequencies, and provide at least 20 Gbps data rates through beamforming and short-range transmission technology \cite{8597759}. However, limited by the performance of ultra-high frequency THz devices, commercialization and deployment of THz short-range communication still face great challenges. Therefore, THz materials, antennas, and related channels still need further investigation in the future.

\subsubsection{RIS Channels for Railway}
As a new paradigm of 6G, RIS can realize reconstruction of the communication environment by adjusting RIS elements phase in real time \cite{10333669}, providing more options for the design of high-speed train communication systems. Meanwhile, RIS can effectively combat high penetration loss and mitigate fast fading for mmWave communications by introducing a virtual LoS component. In \cite{9613722}, a RIS-assisted high-speed train system was proposed, in which RIS was embedded into the train window, to overcome high path loss and mitigate channel fading. To address fast time-varying penetration loss, Zhang et al. \cite{9690475} explored various RIS deployments, including on tracks, carriage windows, and interior walls. Therefore, to exploit the advantages of RIS for mmWave HST communications, channel modeling, and characterization analysis are crucial for the design and evolution of RIS-aided HSR communications.

Currently, few studies focus on RIS-assisted channel modeling for high-speed trains. Existing investigations mainly cover measurement-based statistical models \cite{9769365, 9206044, 9880837} and geometry-based stochastic models (GBSMs) \cite{9541182, 9775205}. Measurement is an efficient and straightforward method to obtain channel characteristics. For instance, Zhou et al. \cite{9769365} proposed a two-path model considering LoS and RIS-assisted links, showing RIS can mitigate fast fading caused by multi-path effect. Moreover, RIS can achieve different performances due to the configured degrees of freedom. Tang et al. \cite{9206044} conducted path loss measurements in an anechoic chamber for both far-field and near-field RIS configurations. In \cite{9880837}, path loss was modeled using floating-intercept and close-in models. However, these studies are limited to static scenarios and lack scalability for high-mobility environments. GBSMs, known for low complexity and flexibility, are widely adopted in RIS-assisted channel modeling and performance analysis. Basar et al. \cite{9541182} introduced a cluster-based model for indoor and outdoor mmWave scenarios, while Sun et al. \cite{9775205} proposed a three-dimensional cylinder channel model analyzing the impacts of RIS reflection, phase shifts, and location on channel characteristics. Yet, these models rarely combine RIS-assisted channel modeling with high-speed train communication scenarios.

\begin{figure}[ht]
	\centering
	\includegraphics[scale=0.7]{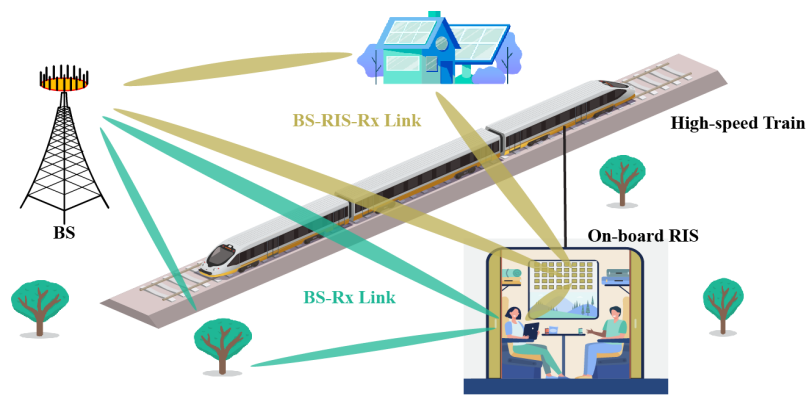}
	\caption{On-board RIS-assisted high-speed train channels.}
	\label{fig:fig2}
\end{figure}

RIS can be flexibly deployed in high-speed train communication scenarios. A straightforward deployment solution is to install RIS on railway tracks to aid high-speed train communication systems, which effectively increases receiver power by introducing the virtual LoS path. Accordingly, since RIS link is a cascaded channel, the path loss of RIS link is related to the product of path losses of BS-RIS and RIS-Rx links \cite{9690144}. As a result, the received power of passengers becomes lower for deploying RIS on railway tracks when the distance between RIS and the transceiver is farther. This means that more RIS elements are needed to achieve the desired effect. Different from placing RIS on railway tracks, RIS is attached to the surface of train windows can increase received power with equal number of RIS elements. In addition, on-board RIS can guarantee the channel between RIS and receiver is quasi-static, which can mitigate the channel fading of high-speed train communications. Because of this, we propose a 3D cluster-based channel model for onboard RIS-assisted high-speed train communications, as illustrated in Fig. \ref{fig:fig2}. Herein, the RIS-assisted high-speed train channel model is divided into BS-RIS-Rx and BS-Rx links, and both high-speed train and onboard RIS are moving with speed $v_R$ in horizontal direction $\gamma_R$. Five propagation cases are abstracted to describe real channels: (i) single bounce at RIS side (SBR) case, (ii) multi bounce at RIS side (MBR) case, (iii) LoS case, (iv) single bounce (SB) case, (v) multi bounce (MB) case. Channel impulse responses (CIRs) from the $p_{t h}$ BS antenna  $A_T^{(p)}$ to the $q_{t h}$ Rx antenna $A_R^{(q)}$ is a superposition of above five propagation cases, which can be written as (in dB) \cite{submitted-1}

\begin{small}
	\begin{equation}
		h_{p q}(t)=h_{p q}^{\mathrm{SBR}}(t)+h_{p q}^{\mathrm{MBR}}(t)+h_{p q}^{\mathrm{LoS}}(t)+h_{p q}^{\mathrm{SB}}(t)+h_{p q}^{\mathrm{MB}}(t) .
	\end{equation}
\end{small}

Since the power of SB case accounts for a larger proportion of the total CIR, the total received power is enhanced by maximizing the power sum of LoS, SBR and SB cases. The objective of the optimization can be written as
\begin{equation}
\max _{\phi_n(t)} \quad \mathrm{E}\left[\left|h_{p q}^{\mathrm{SBR}}(t)+h_{p q}^{\mathrm{LoS}}(t)+h_{p q}^{\mathrm{SB}}(t)\right|^2\right].
\end{equation}

\begin{figure}[H]
	\centering
	\includegraphics[scale=0.05]{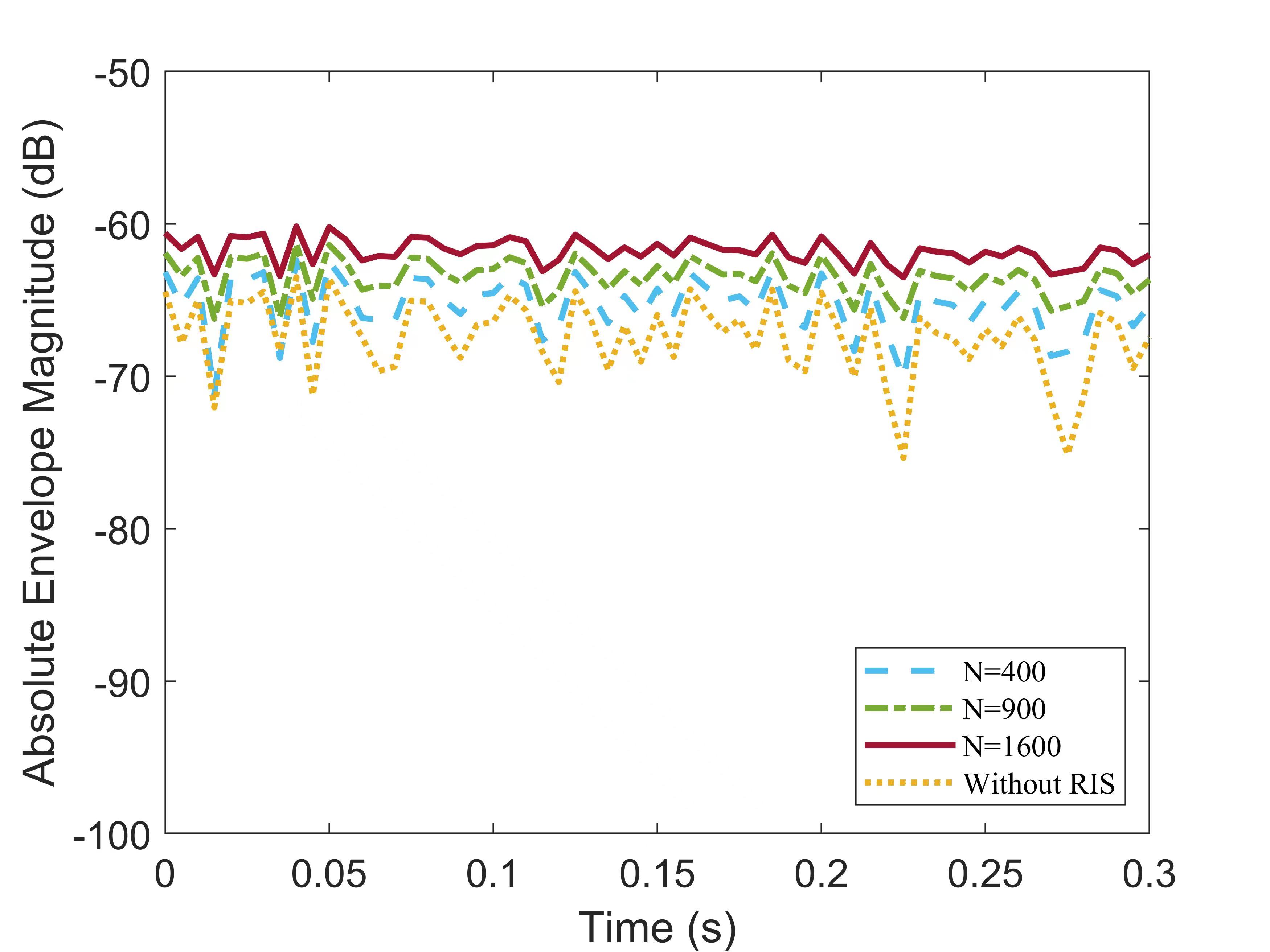}
	\caption{Absolute envelope magnitude of CIR using RIS optimization phase with different numbers of RIS elements.}
	\label{fig:fig3}
\end{figure}

Fig. \ref{fig:fig3} compares the impacts of the number of the RIS elements on absolute envelope magnitude of CIR using RIS optimization phase. It can be found that a larger number of RIS elements can increase the magnitude of CIR and decrease the degree of variations in CIR. The reason is that a larger number of RIS elements increases the proportion of SBR components in CIR, weakening the influence of other multipath components on high-speed train channels.

However, RIS-assisted high-speed train channel modeling also has the following challenges: First, real-world measurements are essential for validating models, but are difficult to conduct due to limited access to HSR experiments and the size limitations of RIS at mmWave frequencies. Additionally, high-speed train channels exhibit distinct characteristics such as high Doppler shifts and severe penetration losses, necessitating specialized RIS phase optimization strategies to address the dynamic and demanding communication environment.

\subsubsection{Railway IoT Channels}
In recent years, IoT for railways (IoT-R) has gained significant attention, aiming to enable environmental perception and decision-making for smart railways. Applications of IoT-R include natural disaster detection, perimeter intrusion alerting, and railway infrastructure monitoring \cite{9496190}. By deploying numerous sensors, IoT-R enables real-time tracking of trains and cargo, and monitors rail infrastructure, playing a key role in smart railways. 
Accurate channel models help improve energy efficiency, node localization, reduce interference, and enhance service quality \cite{9714331}, making it essential to study wireless channel modeling in IoT-R.

IoT-R encompasses ground IoT and intra-train IoT scenarios. In ground IoT, cellular-based IoT technologies are applicable, while device-to-device (D2D) communications \cite{8340813} are ideal for intra-train IoT. Studies have examined path loss models for cellular IoT, such as NB-IoT and D2D communication. For instance, Caso et al. \cite{9383771} proposed an empirical path loss model for NB-IoT in urban areas, improving path loss estimation accuracy. Additionally, Bedeer et al. \cite{8287901} introduced log-distance path loss models for D2D communication in VHF/UHF bands. These models showed path loss exponents between 4.13 and 4.80 and shadow fading standard deviations between 8.87 and 10.96 dB. 
Due to the complex and unique environment, the path loss model for IoT-R needs to be further explored.

In intra-train IoT scenarios, the movement of passengers in train carriages causes stochastic shadowing events from the human body, which impacts D2D communication links. To study this effect, the shadowed $\kappa-\mu$ fading model was introduced in \cite{6953146}. This model represents the received signal statistics by clustering multipath components, with a dominant signal component in each cluster following a Nakagami-m distribution.This model, designed for D2D communication, provides insights into shadow fading in IoT-R systems.

With limited spectrum resources in railways, mmWave and THz communications offer promising solutions to alleviate the strain of spectral resources. The two-dimensional geometrical model for short-range sub-THz D2D communications was proposed in \cite{7500097}, while \cite{7400962} analyzed channel propagation at 28GHz using 3D ray-tracing simulations. Numerous mmWave radio propagation parameters were presented and the corresponding channel models were described. 
Channel sparsity in different scenarios was also evaluated and compared. Due to the importance of mmWave and THz band communications in railways, the propagation characteristics and models of mmWave and THz channels should be investigated in IoT-R scenarios.

In summary, it is necessary to further study propagation characteristics and channel models in IoT-R, including path loss, shadow fading, small-scale fading, and mmWave and THz channel characteristics.

\subsubsection {AI Enabled Railway Channel Modeling}
Traditional channel models fall into two categories: statistical and deterministic. Statistical models fit large-scale measurement data without accounting for geographic details, which limits accuracy in rapidly changing HSR environments. Deterministic models, based on Maxwell equations to obtain channel propagation characteristics, require detailed environmental data and are computationally intensive, with reduced accuracy when the environment changes. Thus, for fast-changing HSR scenarios, the channel model needs to have strong generalization capability, low computational complexity, and guaranteed accuracy.

With the development of AI and computer vision technology, visual data (such as images, RGB-D camera data, and point cloud data) are used to assist wireless communication systems in predicting wireless channels. Using depth camera, lidar, and other sensors, the communication system can effectively obtain physical information of scatterers in the environment, which can help the communication system to predict channel state information (CSI) (such as receiving power and link blockage \cite{8792137}) or select the best beamforming strategies to improve communication quality. Due to the limitation of coverage of millimeter wave and terahertz communication systems, it has become an important research content of channel modeling to accurately model radio transmission by reconstructing a wireless communication environment. Environmental reconstruction typically relies on LiDAR sensor data, from which channel parameters—such as path loss, time delay, and multipath effects—are derived based on radio wave propagation principles \cite{8608834}. Besides, simulation tools, such as ray tracing, are also used to estimate electromagnetic characteristics of environmental materials \cite{8345613} and obtain channel parameters \cite{9325887}. However, channel modeling based on environment reconstruction still faces great challenges. Sensors such as lidar are expensive and the method of channel modeling using point cloud and ray tracing is complex and computational, which is not suitable for real-time processing. Moreover, railway communication channel has high dynamic modeling and real-time requirements, and it is thus difficult to reconstruct the environment.

The AI-based channel modeling approach provides a new solution. Deep neural networks are capable of learning complex relationships from input to output. Therefore, in channel modeling, AI models can directly use environmental information to map them to channel parameters. For supervised learning, a ``data-label pair'' training model using ``environmental information-channel feature parameters" enables the model to learn how to map complex environmental information to channel characteristics.

The model accuracy largely depends on how effectively environmental information is utilized. This information can range from point cloud data and grid maps to building maps and satellite maps. While high-precision data can improve modeling accuracy, it often involves redundancy, complex preprocessing, and limited availability. In contrast, satellite images offer coarser granularity but are easy to access and suitable for training models with good generalization. Given these advantages, we use satellite images as input to a deep neural network to predict path loss at specific transceiver locations, which can be further applied to railway scenarios. The model is trained using open-source 5.9GHz channel measurement data with GPS coordinates. We use static map API from MAPBOX to download satellite images from the location information corresponding to the channel measurement data \cite{6194400}.


We design a multi-input model that can simultaneously consider system state information such as distance, transmit power, transmit antenna height, and environmental information. The model can be described as
\begin{equation}
\mathrm{PL}=\mathrm{y}(\mathrm{x}, \omega, \theta)+\epsilon
\end{equation}
where $y$ is the mapping relation learned by the model, $\theta$ is the hyperparameter in the model, $\omega$ is the weight learned by the model, and $\epsilon$ is the shadow fading. $x = [p, h, d, M]$ is the input feature of the model, where $ p $ is the transmit power, $ h $ is the transmitter height, $ d $ is the transceiver distance, and $ M $ is the input image.

The model uses a convolutional neural network to extract satellite image features, a multilayer perceptron (MLP) to extract system state information, and finally combines the higher-order features of the above two components and then uses the MLP in order to calculate PL.

A key challenge arises from the variable size of satellite image inputs, which traditional neural networks struggle with due to fully connected layers requiring fixed-size inputs. The principle is that the convolutional network operates by sliding the convolutional kernel over the image, and when the size of the satellite image becomes larger, no new type of environmental information is introduced, so the convolutional kernel can recognize the same patterns. After the convolutional layer, we design a spatial pyramid pooling (SPP) layer to handle its output, which can output feature maps of different sizes as feature vectors of the same size while guaranteeing a certain amount of spatial information. This allows feature extraction to continue through the fully-connected layer.

\begin{figure}[htb]
	\centering
	\includegraphics[scale=0.6]{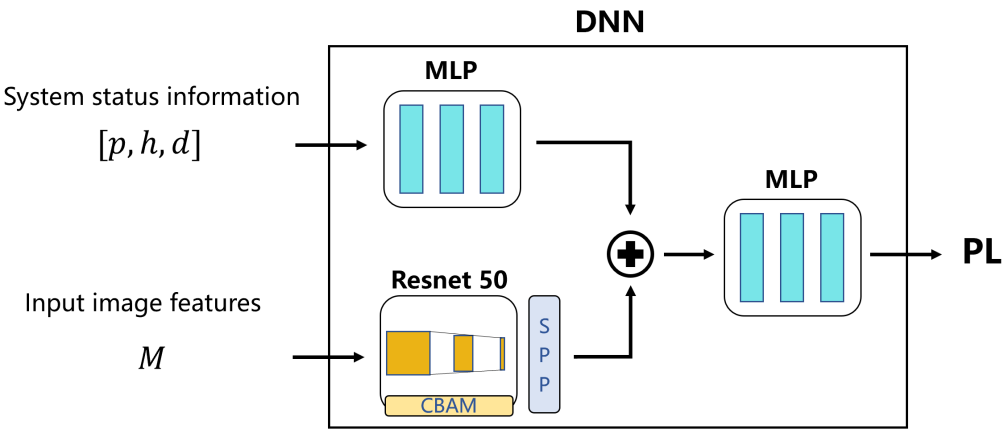}
	\caption{The proposed model architecture}
	\label{fig:fig5}
\end{figure}

In order for the model to effectively and accurately extract features from satellite images, a 50-layer deep residual convolutional network based on CBAM's spatial and channel dimensional attention mechanism is used on the satellite image part \cite{7780459}. Since the satellite images still have high redundancy, the attention mechanism in channel dimension is used to filter the feature patterns and retain effective features; the attention mechanism in spatial dimension enables the model to focus on the effective part of satellite images. The architecture diagram of the model is shown in Fig. \ref{fig:fig5}.

\begin{figure}[htb]
	\centering
	\includegraphics[scale=1.2]{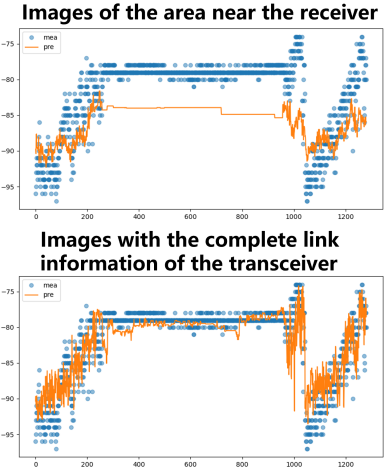}
	\caption{Comparison of model results under NLOS conditions}
	\label{fig:fig6}
\end{figure}

The model achieves a root mean square error (RMSE) of 6.36dB on the test set, with a 2.5dB relative improvement in RMSE when compared to a model trained using only satellite images of the area near the receiver. The prediction results of the two models for a test data under NLOS conditions are shown in  Fig. \ref{fig:fig6} with a 2.2dB relative improvement in RMSE for the model using the satellite image with the complete link information of the transceiver as input.

\begin{figure}[htb]
	\centering
	\includegraphics[scale=0.5]{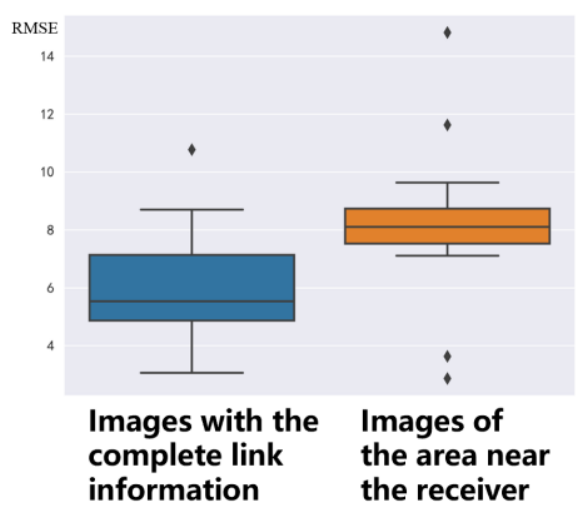}
	\caption{RMSE box line graph comparison}
	\label{fig:fig7}
\end{figure}

The RMSE of both models for each street in the test set are plotted as box plots, as shown in  Fig. \ref{fig:fig7}. According to this figure, the RMSE in the scenario with the maximum error does not exceed 11dB using the model containing the complete link information of the transceiver, and the RMSE in the scenario with the maximum error is 14.2dB using the model with the information of the area near the receiver, so the proposed model has better robustness. 

\subsection{RIS for High-Mobility Scenarios}
\begin{figure}[htb]
	\centering
	\includegraphics[scale=0.3]{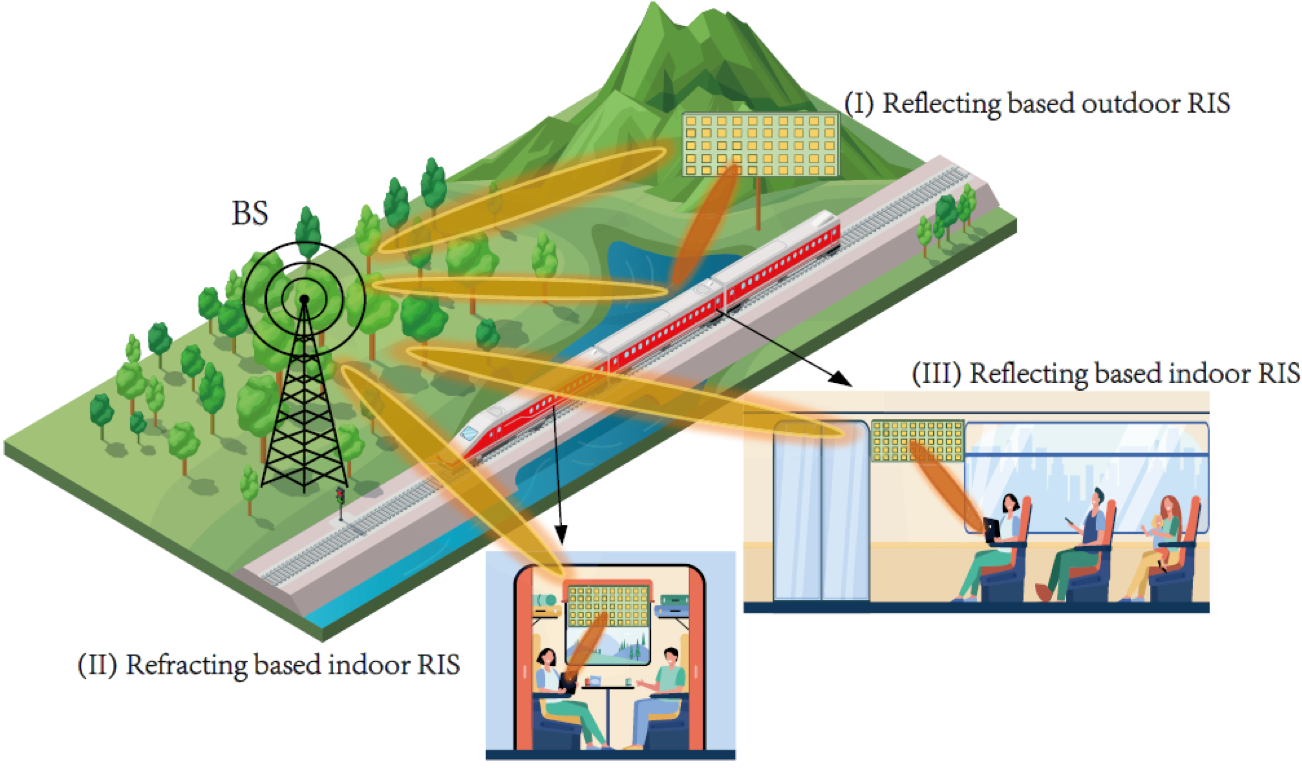}
	\caption{RIS application scenarios}
	\label{fig:fig9}
\end{figure}

With the global rollout of 5G, research on 6G-enabled smart rail scenarios is gaining momentum. However, high train mobility leads to frequent handovers, significant Doppler shifts, and short channel coherence times. These factors cause non-stationary and rapidly time-varying wireless channels, which degrade the performance of train-ground communications.

To address these challenges, RIS is emerging as a key technology for future 6G communications. RIS can be widely applied in smart rail scenarios, as they enable real-time control of signal reflections through a large array of passive elements, creating a smart and reconfigurable radio environment for enhanced wireless performance \cite{10556753}. The typical RIS deployment scenarios in railway networks are depicted in Fig. \ref{fig:fig9}. Despite its great potential, RIS faces new and unique challenges in its efficient integration into wireless communication systems, such as performance analysis, channel estimation, and beamforming.

\begin{enumerate}[a)]
	\item \textit{Performance Analysis:} 
       To evaluate the performance of RIS-assisted smart rail scenarios, including the effects of Doppler shift, imperfect CSI, and RIS phase shift errors, theoretical performance analysis is essential but challenging. The impact of channel variations caused by users' relative movements at RIS-assisted multiple-input single-output wireless communication system was examined in \cite{9765773}. Chapala and Zafaruddin in \cite{chapala2022reconfigurable} developed exact analysis on the performance of RIS-assisted vehicular communication system considering phase noise with mobility over asymmetric fading channels by coherently combining received signals reflected by RIS elements. 
	\item \textit {Channel Estimation:} Due to the harsh nature of smart rail scenarios environment and the unpredictable hardware impairment of RIS, it is challenging to acquire accurate and complete CSI, accurate channel estimation methods must be investigated. Kalman filter and minimum mean square error (MMSE) criterion were employed to design a channel estimation scheme for the RIS-assisted communication system under mobility scenarios in \cite{9384497}. Sun and Yan \cite{9292080} proposed two wideband channel estimation schemes with Doppler shift adjustment for multipath and single-path propagation environments in RIS-assisted systems.
        An MMSE-based interpolation channel estimation method considering the time-varying nature of fading even within the coherence time was developed in \cite{9875062}.
	\item \textit{Beamforming:} To compensate for the wireless channel fading of smart rail scenarios and adapt to dynamic channel conditions, active beamforming at BS and passive beamforming at RIS needs to be appropriately designed. A deep reinforcement learning framework for designing a BS beamforming and RIS phase shifts for RIS-aided millimeter-wave high-speed railway networks was proposed in \cite{9651545}. Di et al. \cite{9110889} proposed a hybrid beamforming scheme for RIS-based multi-user communications with limited discrete phase shifts. 
\end{enumerate}

\begin{figure}[htbp] 
	\centering	
	\includegraphics[width=0.7\linewidth]{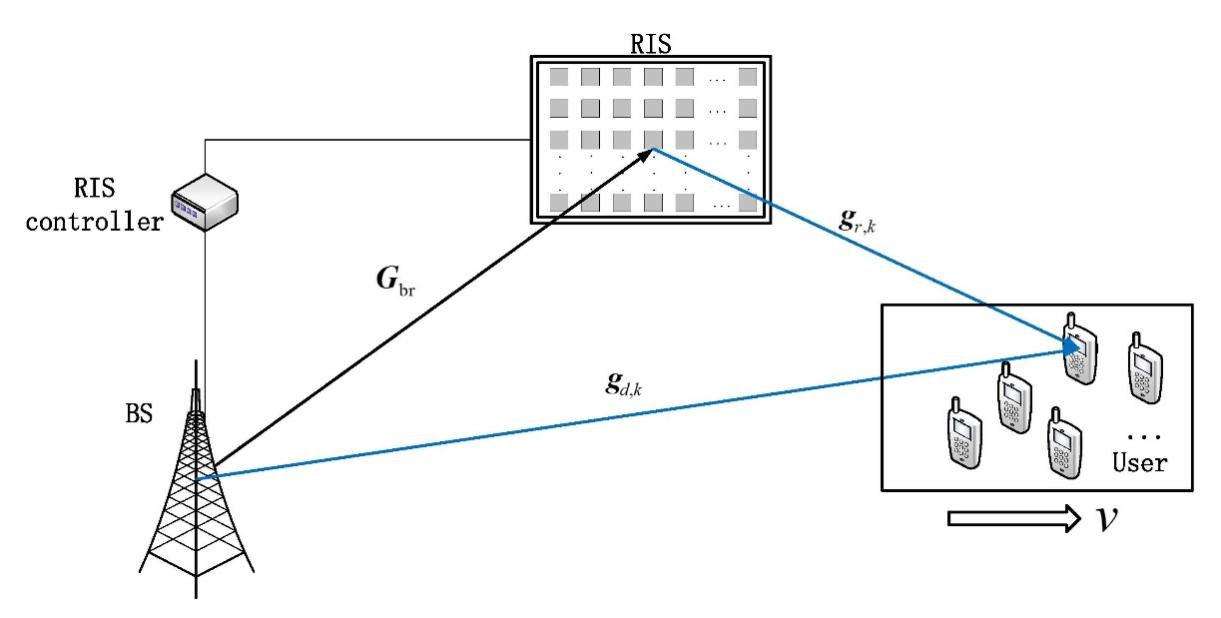} 	 
	\caption{An RIS-assisted MISO system.} 
	\label{fig:fig10}	 
\end{figure}

\begin{figure}[htbp]
	\centering
	\includegraphics[width=0.6\linewidth]{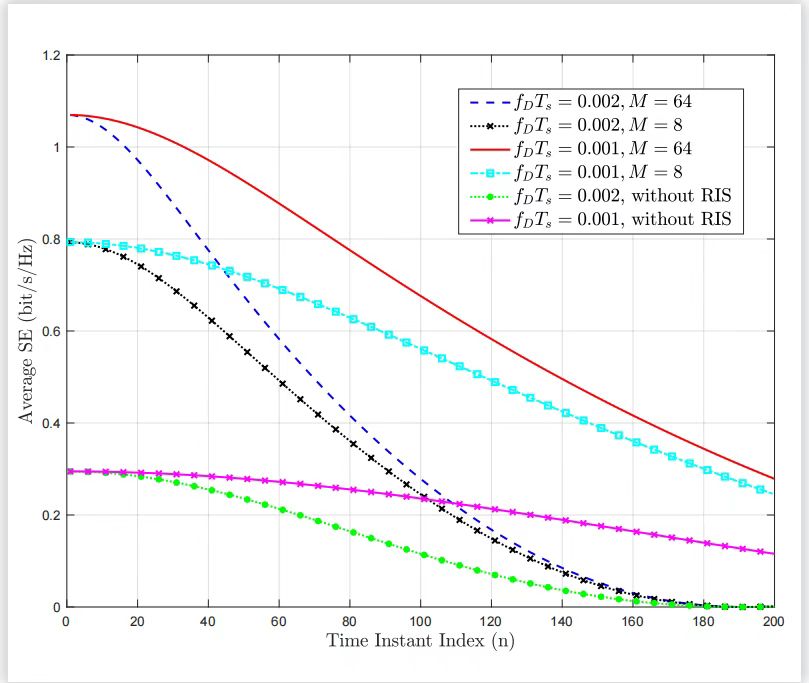}
	\caption{Average SE versus time instant}
	\label{fig:fig11}
\end{figure}

\begin{figure}[htbp]
	\centering
	\includegraphics[width=0.6\linewidth]{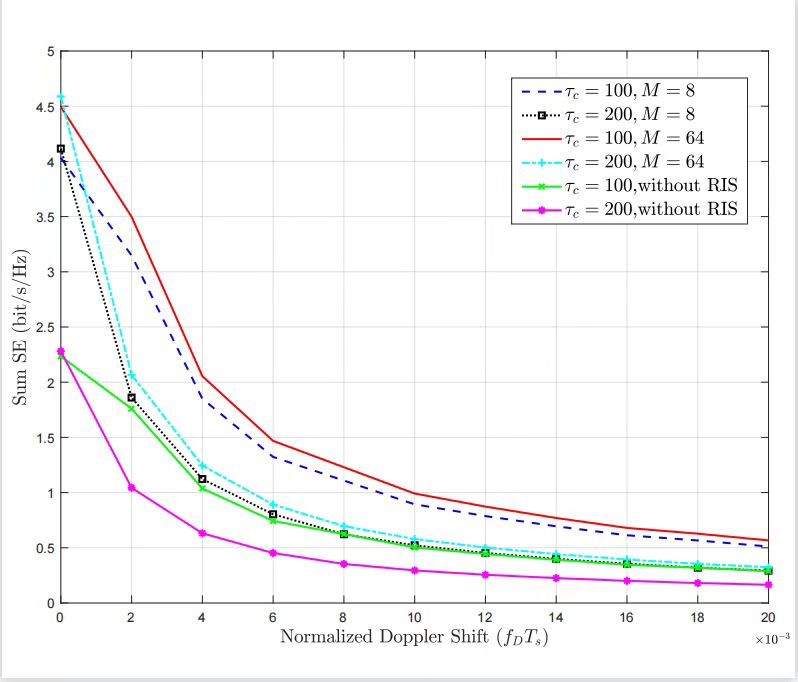}
	\caption{Sum SE versus normalized Doppler shift}
	\label{fig:fig12}
\end{figure}

To overcome the above challenges, a multiple-input single-output(MISO) system model shown in Fig. \ref{fig:fig10} is considered. 
The channel characteristics between BS/RIS and users change over time due to the relative movement of users. 
Fig. \ref{fig:fig11} and Fig. \ref{fig:fig12} show the effect of the channel aging on the spectral efficiency (SE). We use the normalized Doppler shift $f_DT_s$ to represent the impact of channel aging on system performance. The larger the value of $f_DT_s$, i.e., the faster the users move, the more seriously the impact of the channel aging on the system. Fig. \ref{fig:fig11} shows the relationship between the users’ average SE and the transmission time under different $f_DT_s$ and the number of RIS elements. We observe that the average SE decreases as time increases because of the channel aging effect, and increasing $f_DT_s$ will accelerate the decline of the average SE. However, the increase in the number of RIS elements will increase the average SE. This shows that the use of an RIS can alleviate the channel aging effect. 
Then we analyze the impact of transmission time on the system performance. Fig. \ref{fig:fig12} shows the relationship between the sum SE and the normalized Doppler shift $f_DT_s$ under different transmission times and the number of RIS elements. With the increase of the Doppler shift, the spectral efficiency of the system with longer signal transmission time decreases more sharply. Moreover, although the sum SE degrades with the increase of the transmission time and the normalized Doppler shift, the usage of a RIS can effectively alleviate the negative effect brought by channel aging.

\subsection {Lessons Learned}
The advancement of 6G-enabled smart railways highlights the necessity for a unified network architecture that incorporates advanced technologies to overcome challenges such as high mobility, complex environments, and dynamic conditions. This architecture ensures seamless connectivity through satellite, UAV, and THz technologies, while cross-domain channel modeling addresses propagation issues like Doppler effects and non-line-of-sight conditions. Technologies such as RIS, AI-driven tools, and IoT-R significantly enhance channel performance, predictive analysis, and real-time monitoring capabilities. Emerging solutions such as THz communications and VLC offer potential for high-capacity demands, although practical deployment challenges remain unaddressed.

\section{Reliable Transmission Technologies for 6G Smart Railways}
After characterizing the channels, we then discuss how to effectively conduct reliable transmissions over such complicated transmission environments for 6G smart railways.

\subsection{OTFS in Smart High-Speed Railway}

IoT is one of the typical services in smart railways to enable environmental sensing. Developing railway-oriented IoT communication systems has been scheduled by China, the European Union, France, and Switzerland \cite{fraga2017towards,li2019railway}, etc. In 6G, ultra-massive machine-type communication (umMTC) for IoT is a promising candidate, which is with $10^7$ devices per kilometer \cite{8766143}. Developing umMTC in smart railways faces two main challenges: high mobility of HSR and massive connections constrained by the limited spectrum.

On the one hand, orthogonal frequency division multiplexing (OFDM) is the basic multicarrier modulation scheme in 5G New Radio (NR). However, OFDM is fragile to Doppler shift brought by high mobility, which could be more notable in the high-frequency band. Recently, delay Doppler (DD) domain multicarrier (DDMC) modulation schemes are proposed, including the OTFS modulation \cite{hadani2017orthogonal} and orthogonal delay Doppler multiplexing (ODDM) modulation \cite{9829188}, which have been recognized as a promising waveform for communications with significant delay and Doppler spread \cite{10806729}. 
By exploring the multipath diversity, OTFS modulation is with lower probability of experiencing deep fading and the reliability of which under doubly-selective channels is higher than OFDM \cite{8671740}. For high-speed railways, the typical scenario that requires high transmission reliability under high mobility, DDMC, and related modulation schemes are required.

On the other hand, the scarce spectrum limits the number of umMTC devices in smart railways. Researchers have proposed to increase the system capacity to enable massive access. For example, it has been a hot topic to design non-orthogonal multiple access (NOMA) schemes to support a large number of devices \cite{7842433}. Also, grant-free random access is well researched \cite{wang2019massive}, in which devices transmit immediately once they have collected data without the permission of BS. 
Currently, there have various works on designing multiple access schemes based on OTFS modulation for 6G \cite{9411900,9928043}.
However, there still exist some key challenges to be considered.

\begin{enumerate}[a)]
	\item \textit{Complexity of grant-free random access schemes based on OTFS:} In the updated NOMA and grant-free access schemes, iterative user identification and data detection are usually considered, which could bring in significant computation and latency to the 6G smart railway system.
	\item \textit{Channel characterization of the channel spreading function in smart railways:} The performance of OTFS-based multiple access schemes depends on the characteristics of the channel spreading function. In the literature, the channel spreading function is usually generated by the ideal tapped-delay-line (TDL) model, while its characteristics are rarely investigated in the realistic channel environment. To apply OTFS modulation into smart railways, it remains to be an essential topic to characterize the channel spreading function practically, and verify the feasibility of OTFS modulation accordingly.
\end{enumerate}

To overcome the above challenges, the following scheme is proposed.
\subsubsection{Orthogonal time frequency space modulation enabled tandem spreading multiple access (OTFS-TSMA)}

\begin{figure*}[htbp]
	\begin{center}	
		\includegraphics[scale=0.45]{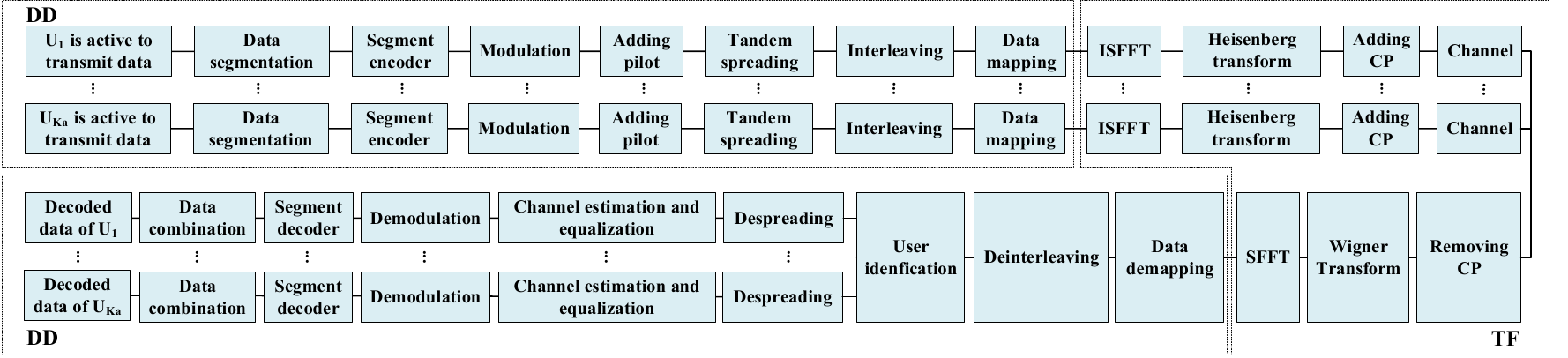}	
	\end{center}	
	\caption{OTFS-TSMA transmission diagram.}	
	\label{fig:fig13}	
\end{figure*}

To realize massive connections with high reliability and low complexity for umMTC in smart railways, OTFS-TSMA was proposed in \cite{9496190}, whose transmission diagram is given in Fig. \ref{fig:fig13}.


OTFS-TSMA has indeed addressed the aforementioned design challenges as elaborated below. 
\begin{enumerate}[a)]
     \item \textit{Massive connections under limited spectrum:}  OTFS-TSMA proposes tandem spreading combinations for massive users, which are unique for users and can be regarded as the user identity. 
     Meanwhile, the number of colliding segments among users is limited considering the nature of two-dimensional (2D) convolution due to OTFS modulation, so the collision can be recovered by employing the erasure coding \cite{8352626}.

    \item \textit{Reliable communication under high mobility:} In addition to the TSMA designs in OTFS-TSMA introduced above, we note that there are OTFS operations and the chip-level interleaving/deinterleaving additionally, as shown in Fig. \ref{fig:fig13}. In particular, we propose the interleaving strategy to maintain the orthogonality of spreading sequences under the influence of 2D cyclic shift of spreading chips in OTFS modulation. 
    In addition, the multiple diversity along the Doppler domain is utilized to mitigate the probability of false alarm (FA) based on the proposed scheme for deleting FA users.

\end{enumerate}

\subsubsection{Channel spreading function characterization in smart railways}

\begin{figure*}[!t]
	\centering
	\subfloat[]{\includegraphics[scale=0.35]{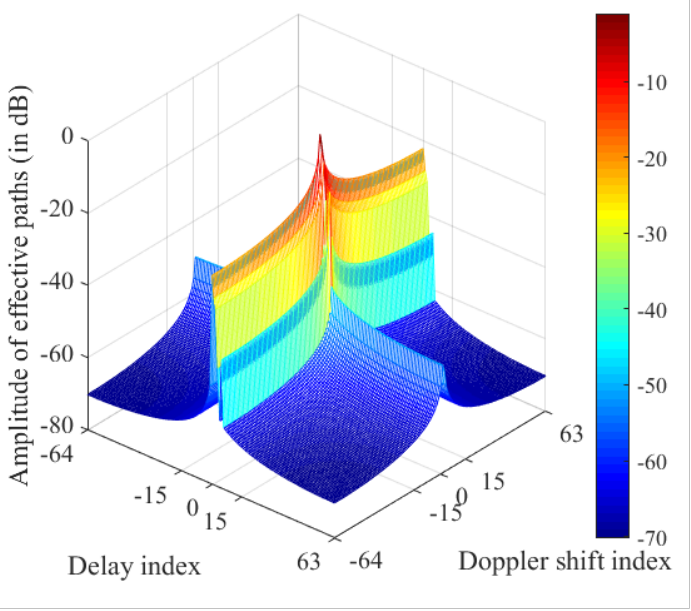}%
		\label{fig_first_case-14}}
	\hfil
	\subfloat[]{\includegraphics[scale=0.35]{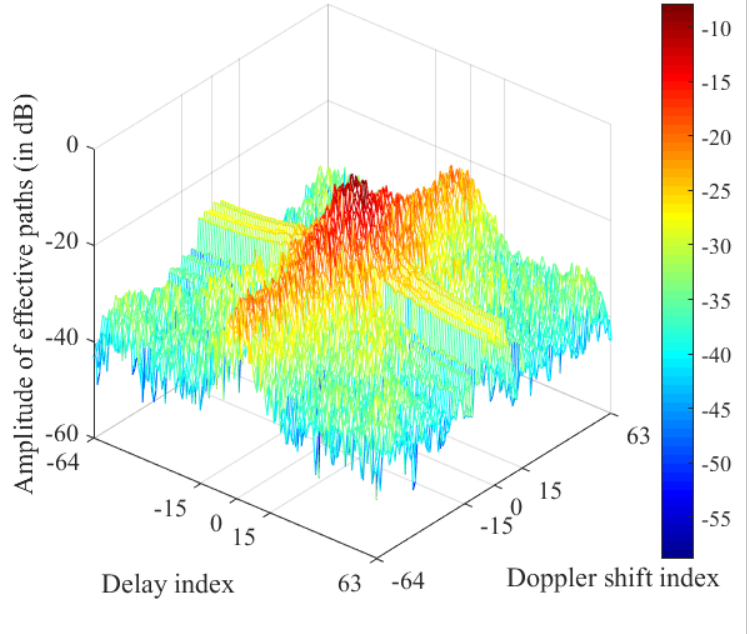}%
		\label{fig_second_case-14}}
	\caption{Channel spreading function of (a) TDL channel model and (b) practical HSR scenario measured in \cite{8888372}.}
	\label{fig:fig14}
\end{figure*}

The performance of OTFS is related to the characteristics of the channel spreading function, including the number of multipath components (MPC), and degrees of freedom (DoF) \cite{1386525}. For the application of OTFS in smart railways, the HSR channel spreading function is characterized with channel transfer function measurement in the TF domain along the Beijing-Shenyang railway in China \cite{10056866}. Fig. \ref{fig:fig14} illustrates the comparison between the measured discrete channel spreading function in the HSR scenario and the ideal discrete channel spreading function generated by the tapped-delay-line (TDL) model. It is revealed that factors affecting the observed discrete channel spreading function include sampling error due to the limited frame size of OTFS, symplectic fast Fourier transform (SFFT), and small-scale fading in time domain.

\begin{enumerate}[a)]
	\item \textit{Influence of sampling error}: The system resolutions regarding delay and Doppler shift are the reciprocal of bandwidth and time duration, respectively. When the delay and Doppler shift paths are not divisible by the corresponding resolutions, sampling error occurs and effective paths are generated.
	\item \textit{Influence of SFFT}: The discrete channel spreading function is not a band-limited periodic function, which could be discontinuous at the boundaries when interpreted as a periodic function by SFFT. Hence, paths without explicit physical meaning occur.
	\item \textit{Influence of time domain small-scale fading}: Since CIR and discrete channel spreading functions are Fourier transform pairs \cite{1386525}, so discrete channel spreading function varies with CIR. Based on the observation that CIR can be regarded as invariable during channel coherence time, the small-scale fading makes the discrete channel spreading function vary if the time duration of the OTFS data block is longer than the coherence time.
\end{enumerate}

As a result of the aforementioned factors, the sparsity and compactness of the discrete channel spreading function degrade greatly compared to the continuous channel spreading function generated by the TDL channel model, while the latter one is the base of the mainstream OTFS-related literature \cite{8786203,9181410,10274133}. Therefore, it is pointed out that properties of the discrete channel spreading function in the spectrum-limited system should guide the design of OTFS modulation correspondingly. 

\begin{figure*}[!t]
	\centering
	\subfloat[]{\includegraphics[width=2.4in]{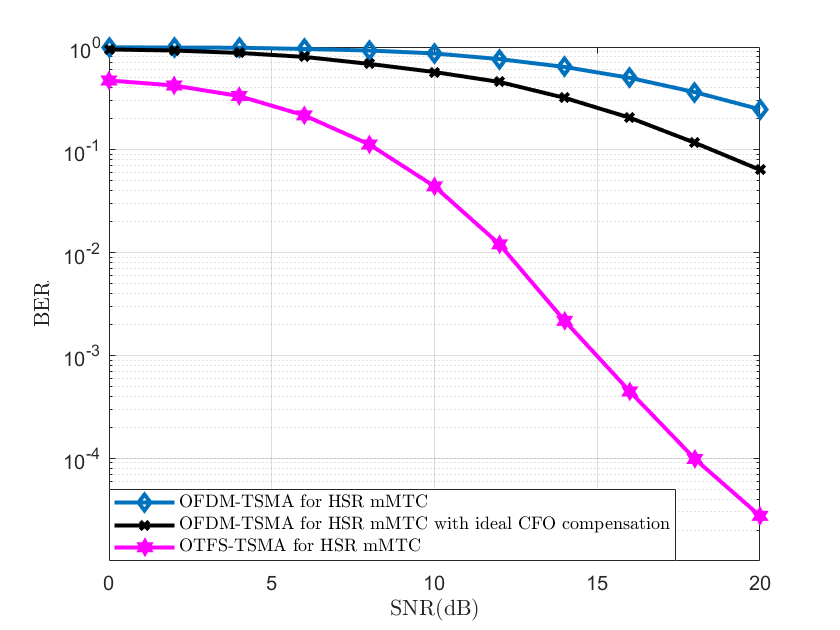}%
		\label{fig_first_case-15}}
	\hfil
	\subfloat[]{\includegraphics[width=2.4in]{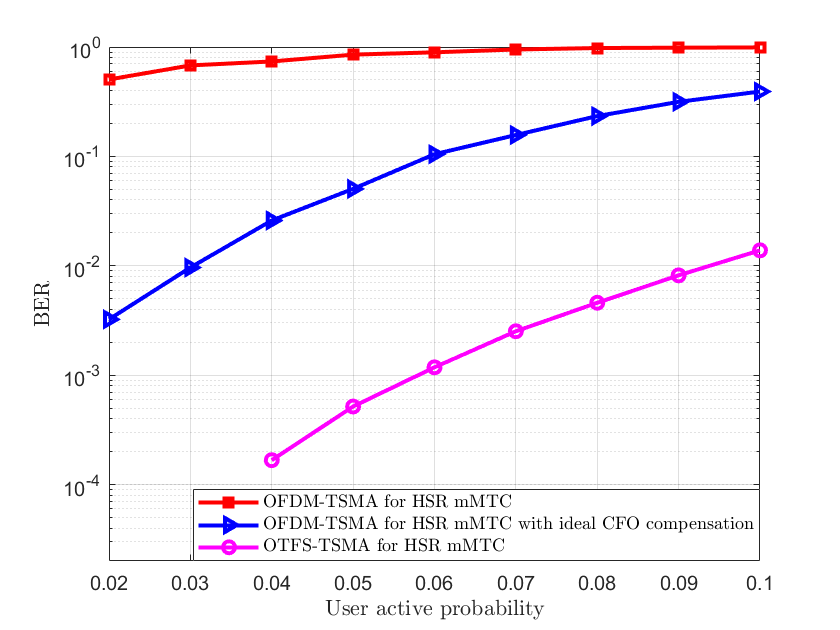}%
		\label{fig_second_case-15}}
	\caption{Performance of OTFS-TSMA versus TSMA and OFDM-TSMA in terrestrial mMTC for smart railways: (a) BER performance with SNR varies and (b) BER performance with user activation probability varies.}
	\label{fig:fig15}
\end{figure*}

The performance of OTFS-TSMA in terms of bit error rate (BER) is demonstrated in Fig. \ref{fig:fig15}. Based on the proposed tandem spreading combinations in OTFS-TSMA, the system can access 560 users \cite{9496190}. In addition, OFDM-TSMA with/without ideal carrier frequency offset (CFO) compensation are selected as benchmarks.

As given in Fig. \ref{fig:fig15}(a), the performance of OTFS-TSMA overtakes OFDM-TSMA in HSR scenarios. On the one hand, OFDM-TSMA is vulnerable to CFO, since the de-spreading-based user identification scheme is affected by the energy leakage due to the loss of orthogonality of subcarriers. On the other hand, the channel transfer function in TF domain is fast-varying in HSR scenario, so certain segments experiencing deep fading cannot be detected and miss detection (MD) occurs. On the contrary, OTFS-TSMA experiences stable channel gain over DD domain, and the influence of deep fading is mitigated. Fig. 16(b) compares the performance of OTFS-TSMA and OFDM-TSMA regarding BER under different user activation probabilities. Based on the user grouping strategy in OTFS-TSMA, it can be observed that the multipath diversity in Doppler domain not only enhances the system capacity but improves the transmission reliability by deleting FA users. However, the discrete channel spreading function depicted in aforementioned section is not considered in the primary research regarding OTFS-TSMA. The transmission diagram of OTFS-TSMA is still evolving regarding receiver design for the sparsity-reduced and compactness-reduced channel spreading function.


\subsection{THz Communications for High-Speed Mobile Scenarios}


In recent years, the demand for railway services around the world is growing continuously. In the future, HSR will develop along the direction of passenger transport service networking, intelligent transport organization, and automatic safety monitoring. In train operation control, new services are emerging, such as the railway IoT, railway multimedia scheduling, and high-definition video security monitoring. Facing the existing vacuum tube HSR technology, the T2G wireless communication system is the key to ensure its safe operation. It makes the maglev train run in the low pressure pipe close to the vacuum, with the mode of ultra-high speed, all-weather, low air resistance \cite{8634036}. In the face of the explosive growth of new mobile applications, the demand for broadband mobile communications among passengers is surging and the proportion of streaming media services is rising, which brings unimaginable challenges to the operation and maintenance of wireless networks. In view of such growing requirements, it is urgent to investigate a railway mobile communication system with high bandwidth and high data transmission rate.

The EFH frequency band is used for vacuum pipeline transmission, with wavelengths ranging from 1 cm to 1 mm. In future wireless communications, THz wave is an electromagnetic wave whose wavelength lies between mmWave and infrared light wave. By integrating THz communication into the future HSR wireless communication system, the system can extend into the higher THz frequency band (0.1-10THz), effectively addressing the issue of spectrum scarcity. Dozens to hundreds of GHz bandwidth resources will greatly improve the current wireless system capacity, and achieve terabit peak data rate and low transmission delay \cite{latva2020key}.

Because of the high frequency of THz waves, its communication will face high propagation loss. In the LOS path, the attenuation of radio signals in atmospheric propagation depends on the free space path loss, molecular absorption and the attenuation caused by rainfall. Friis' transmission formula shows that doubling the frequency will increase the free space path loss by four times \cite{9269930}. It requires both the transmitter and receiver to use extremely thin highly directional beams to compensate for the link loss. In addition, THz has low beam characteristics, so the link is sensitive to obstacles. Railway transportation is widely deployed, and the environment along the track is complex and time-varying. In many typical scenarios such as cities, villages, and mountains, buildings or terrain structures will cause random short-term blockage of wireless links, seriously reducing the quality of T2G links. Considering the high-definition video security monitoring and other related services, as well as the large number of users gathered in high-speed trains, there is a huge demand for data services in T2G and T2T scenarios. Therefore, how to achieve efficient and robust transmission scheduling in T2G and T2T scenarios when the train is moving at a high speed has become one of the key challenges faced by HSR communication systems. The position information of train can be obtained through the train control system, or predicted through CSI information of the low frequency communication that is guaranteed to be covered. With the extremely high transmission rate of THz communications, T2G and T2T communications can make use of THz frequency bands to carry out opportunistic communications when the channel conditions are good, and complete large capacity data transmission in a short time to achieve robust transmission scheduling.

Jiang et al. \cite{9267779} reviewed channel modeling and characteristic analysis for the future 6G communication system, including geometric random model, high dynamic channel characteristics, optimization of channel characteristics, and expansion of bandwidth and spatial dimensions. Lv and You \cite{9254121} studied HSR and T2T communication, showing that 6G technology can well meet the time and space requirements on vehicle interconnection. Wei et al. \cite{8634036} applied the solution of a leaky wave system to achieve uniform phase distribution along the track direction to effectively suppress the Doppler effect in 5G communication vacuum tubes. 
Given the high propagation loss of mmWave and its vulnerability to link blockage, Ma et al. \cite{9833304} proposed a relay-assisted algorithm using UAV and mobile relays (MRs) in a T2G communication system to effectively overcome link blockage. The directional antenna and UAV are used to enhance the robustness of the mmWave T2G communication system and maximize the number of transmission flows on the premise of meeting the QoS requirements and channel quality. 
To overcome the interference, eavesdropping and other problems in T2G communications, Wang et al. \cite{9779990} designed a UAV-assisted scheduling algorithm, which uses two mmbands to achieve high system throughput. 

\begin{figure}[htb]
	\centering
	\includegraphics[scale=0.25]{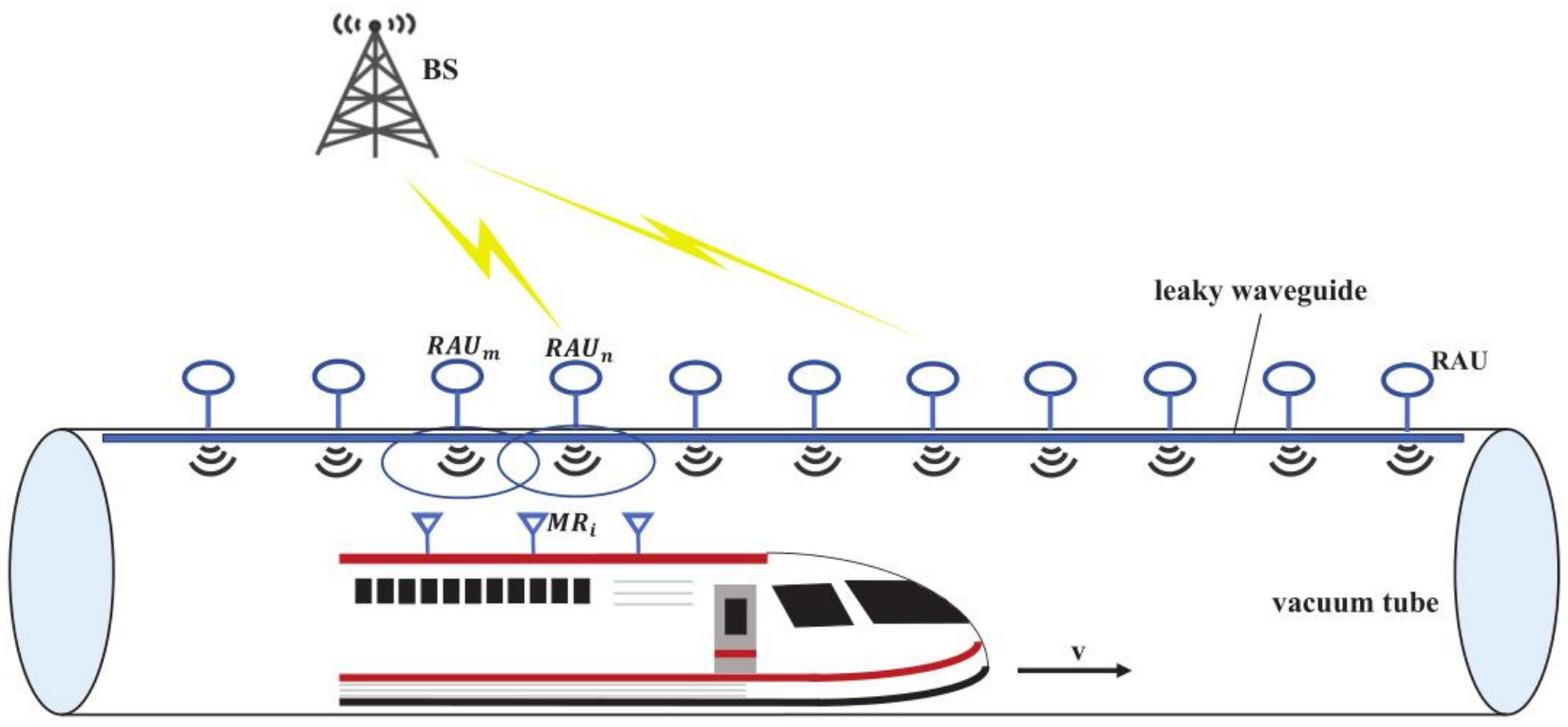}
	\vspace{-23pt} 
	\caption{HSR communication system in the vacuum tube.}
	\label{fig:fig17}
\end{figure}

In order to achieve a faster, cleaner, more environmentally friendly mode of transportation, vacuum tube transportation has also attracted significant attention. The tightness of vacuum tube train can reduce the aerodynamic noise and resistance to a large extent \cite{zhang2011key}. Thus, we focus on THz communication for vacuum tube high-speed trains. In the vacuum tube HSR system, THz wireless communication is detailed in Fig. \ref{fig:fig17}. Multiple remote access units (RAUs) are placed at equal distances outside the metal tube, interconnecting and allowing communication between them. The leaky waveguide minimizes the Doppler effect, even when the train is traveling at extremely high speed \cite{8634036}. Each MR at the outer top of the train is equipped with one steerable directional antenna for receiving. And different content is transmitted to the passengers through the intra-vehicle LAN. As passengers request various services, each flow has distinct QoS requirements, defined as the minimum throughput required for each flow. 
\vspace{-25pt}

The unit of frequency is GHz and the unit of distance is km.
According to Shannon’s channel capacity, the transmission rate of the link can be calculated as
\begin{equation}
	R_i^{T H {\mathrm{z}}}=\eta W^{T H z} \log _2\left(1+\frac{P_r^{T H z}(i)}{N_0 W^{T H z}+\sum_j I_{j i}^{T H z}}\right), \\
\end{equation}
where $W^{THz}$ is the bandwidth of the THz band. $\eta$ is in the range of (0,1) and describes the efficiency of the transceiver design. $P_r^{THz}(i)$ is the received power at the receiver of link $i$. $\sum_jI_{ji}^{THz}$ is the interference of co-band concurrent flows. There is a tradeoff in the choice of the carrier-frequency of the THz transmission, which affects both the bandwidth and the path loss of the THz transmission. The higher the bandwidth utilized, the higher the molecular absorption \cite{4455844}. We choose $f^{THz}$ = 340GHz. Experiments show that the group velocity dispersion of the channel in this frequency band is very small, and the signal is not easy to be broadened. The maximum transmission rate of this frequency band can achieve higher than 10Gbps.

The effective scheduling aims to minimize the transmission time. Since different flows carry different kinds of service for passengers and each flow has its own QoS requirement also known as the minimum throughput. The set of the flows requested by different MRs is defined as $\boldsymbol F$. The set of the MRs is defined as $\boldsymbol I$, so ${\lvert{\boldsymbol F}\rvert}\;{\boldsymbol \leq}\;{\lvert{\boldsymbol I}\rvert}$. For the flow $i\;{\boldsymbol \in}\;{\boldsymbol F}$, its QoS requirement is $q_i$. We denote the number of TSs that complete all the transmission traffic requirements as $\delta_i$. Hence, the objective function of the scheduler is
\begin{equation}
	\min\;\delta_i.
\end{equation}

If the flows want to be scheduled successfully, they must rely on the links between the BS and RAUs, as well as the links between the RAUs and MRs. For link $l$, we define a binary variable $\alpha_l^t$ to indicate whether it is scheduled for transmission in slot $t$. If that is the case, $\alpha_l^t=1$; otherwise, $\alpha_l^t=0$. For the RAUs that can both transmit or receive, the links to and from the same RAU can not be scheduled concurrently. The half-duplex transmission condition of each RAU is enforced by the following constraint:
\begin{equation}
	\alpha_l^t+\alpha_{l^{\prime}}^t \leq 1, \quad \text { if } t_l=r_{l^{\prime}} \text { or } r_l=t_{l^{\prime}} \text {. }
\end{equation}

The RAU and MR are both with only one receiving antenna and can only receive a single link at a time. This means that, any two concurrent links $l$ and $l^{\prime}$ must have different receivers. 

Each link $l$ also has its QoS requirement $q_l$, and it is the same as the QoS requirement of the flow which $l$ transmits for. For each flow, its actual achieved throughput must satisfy its QoS requirement, and its corresponding links are also the same. For link $l$, we have the following QoS constraint.
\begin{equation}
	q_l^a=\frac{\sum_{t=1}^M R_l^t \Delta T}{T_s+M \Delta T} \geq q_l,
\end{equation}
where $R_l^t$ is the actual transmission rate of link $l$ in time slot $t$. Regardless whether it is BS, RAU or MR, the number of its transmit antennas limits the number of simultaneous transmitted links. However, the number of transmit antennas deployed at BS is $N_{B_t}$, which is different from the deployment situations of RAU and MR. We define the set of links that are transmitted from BS in time slot $t$ as $S_B^t$. Thus, we have the following constraint.
\begin{equation}
	\sum\nolimits_{l \in S_B^t} \alpha_l^t \leq N_{B_t}.
\end{equation}

Each RAU can only transmit on a single link at a time. 
Some flows are relayed by RAUs and then forwarded to the target MRs. The sum throughput of the links transmitted from BS is larger than the total throughput of the links transmitted from all RAUs. The set of links which are transmitted from RAU $m$ in time slot $t$ as $S_{R_m}^t$. Hence, the throughput constraint is as follows.
\begin{equation}
	\sum\nolimits_{l \in \sum_{t=1}^\tau S_B^t} q_l^a \geq \sum\nolimits_{l^{\prime} \in \sum_{t=1}^\tau{ }\sum_m S_{R_m}^t}  q_{l^\prime}^a .
\end{equation}

It can be seen that the problem is a nonlinear mixed-integer programming problem. The complex nonlinear terms are also embodied in the constraints. The problem is actually NP-hard \cite{8598887}. Thus, the final goal is to develop competitive solutions with low computational complexity, leading to the following heuristic algorithm.

We propose the location-aware scheduling scheme with QoS consideration. We determine the proper transmission order to complete the link scheduling relayed by RAUs to minimize the scheduling time for completing a certain number of flows with their QoS requirements satisfied.

The links in $S_B^t$ are transmitted from BS in time slot $t$, and the links in $S_R^t$ are transmitted from RAUs in time slot $t$. A parameter is defined to be the inverse of the required TS number of links in a frame to satisfy its QoS requirement, and this parameter is to preliminarily prioritize the scheduling order of the links. For link $l$, the priority value can be expressed as
\begin{equation}
	\text {Pr}(l)=\frac{R_l^{T H z} \cdot \Delta T}{q_l \cdot\left(T_s+M \Delta T\right)} .
\end{equation}

For each link, we calculate the priority value and schedule the corresponding flows in the descending order in the subsequent process. We indicate the links from BS to RAUs and from RAUs to MRs as two independent sets, i.e., $S_{P_B}$, $S_{P_R}$. The links in $S_{P_R}$ are mainly generated from the scheduled links in the set $S_{P_B}$. We also propose a parameter $N_c$ to record the TS number of flow scheduling. 

\begin{figure}[ht]
	\centering
	\setlength{\parskip}{0.1cm plus4mm minus3mm}
	\includegraphics[scale=0.45]{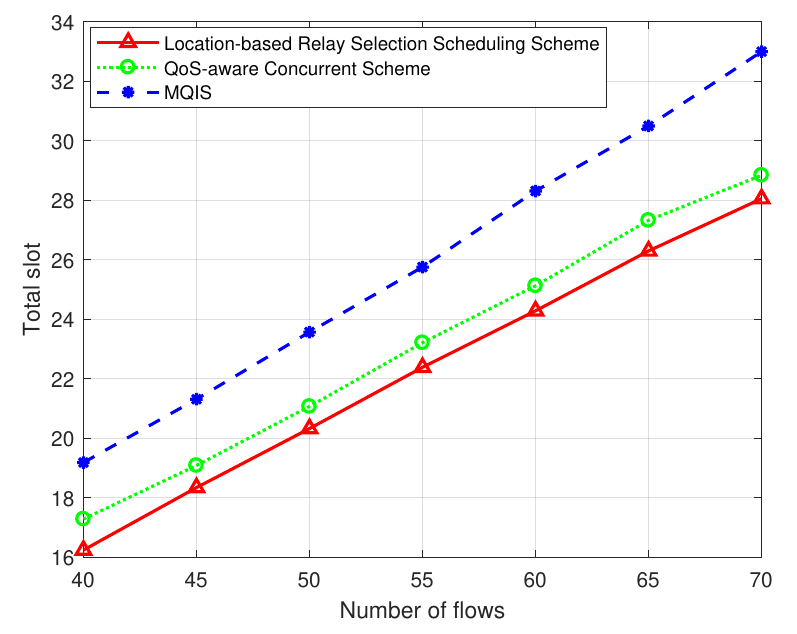}
	\caption{Number of time slots versus different numbers of requested flows}
	\label{fig:fig18}
\end{figure}

We consider an HSR system realized by the vacuum tube. There are 24 MRs at the top of the train. Each carriage can accommodate 50-80 seats, then each MR needs to deal with 10-20 passengers’ data traffic at most. Moreover, the QoS requirement of each flow is uniformly distributed between 10Mbps and 500Mbps.
For comparison purposes, we implement the state-of-the-art QoS-aware concurrent scheme \cite{8344113} and maximum QoS-aware independent set (MQIS) based scheduling algorithm \cite{6364219} as baseline schemes. From Fig. \ref{fig:fig18}, we can see that the total number of slots of the three schemes is increasing as more flows are requested. And the total number of time slots needed by the location-based relay selection scheduling scheme is the least. This shows that the proposed scheme has a superior performance on transmission efficiency compared with the QoS-aware concurrent scheme and MQIS. 

\begin{figure}[ht]
	\centering
	\includegraphics[width=0.7\linewidth]{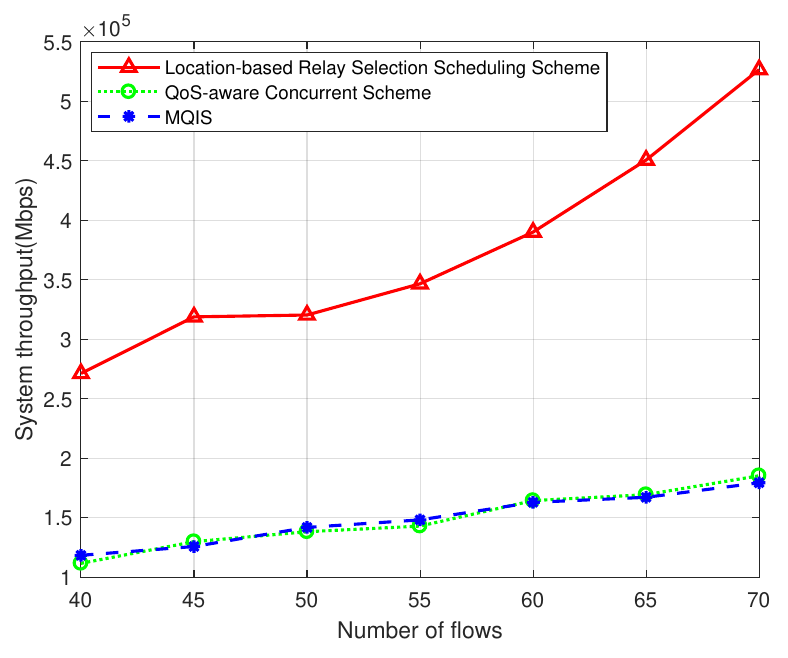}
	\caption{System throughput versus numbers of flows.}
	\label{fig:fig19}
\end{figure}

In Fig. \ref{fig:fig19}, we plot the system throughput under different numbers of requested flows. From Fig. \ref{fig:fig19}, we can see that the system throughput of the three schemes is all increasing with the increased number of requested flows. For each flow, our proposed scheme can continuously schedule the flows that need to be transmitted according to the principle of transmission time minimization. The proposed scheme achieves the highest system throughput among the three schemes. The QoS-aware scheme and MQIS have similar performance on the system throughput.

\subsection {Railway Semantic Communication}
\subsubsection{Semantic Communication in Railway Systems}

In railway systems, applications like intrusion detection require processing large volumes of video data, which demand extensive neural networks. This poses a significant challenge for resource-constrained devices, such as surveillance cameras, in smart railway environments. A practical solution is to offload computation to edge servers, where raw video data is transmitted from cameras for processing and intrusion detection \cite{9261169}. Traditional systems transmit video pixel-by-pixel, resulting in high redundancy and inefficiency. In contrast, semantic communication focuses on transmitting only the task-relevant features. SC extracts semantics-task-specific features—using neural networks, reducing unnecessary data transmission and significantly improving bandwidth efficiency in resource-limited environments.

To enhance SC performance, several advanced techniques have been explored. Deep Joint Source-Channel Coding (DJSCC) combines source and channel coding through neural networks, achieving end-to-end optimization for multi-modal data transmission,  including images, videos, and CSI \cite{10772628}. DJSCC technology can achieve the end-to-end optimization of semantic communication systems and improve their overall performance \cite{xu2023deep}. Additionally, channel-adaptive methods adjust encoding and decoding based on SNR feedback, ensuring SC systems remain effective under varying channel conditions.

Recent advancements have also expanded SC’s application in multi-modal fusion tasks. For example, cross-modal coding and dynamic feature adjustment based on channel conditions optimize transmission efficiency \cite{10431795}. Systems like MU-DeepSC integrate image and text for visual question answering, while layer-wise Transformer architectures support multi-user environments, improving data fusion across modalities \cite{xie2022task}.

\begin{figure}[htb]
       \centering
       \includegraphics[scale=0.35]{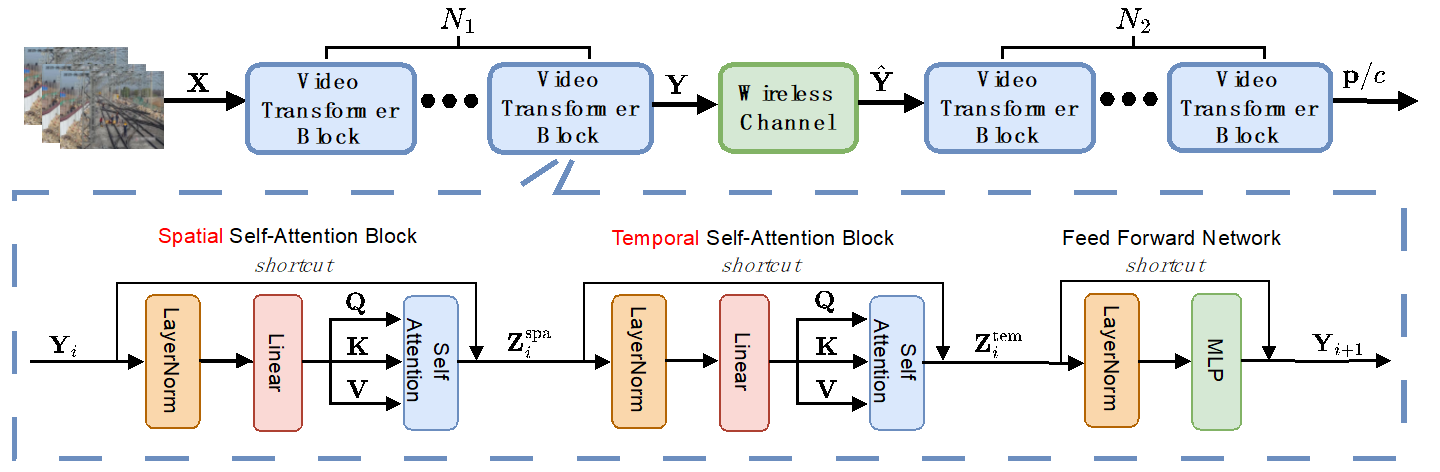}
       \caption{The detailed architecture of Rail-SC}
       \label{fig:2}
\end{figure}

The proposed SC architecture, Rail-SC, consists of a railway device and an edge server that collaboratively perform intrusion detection, as shown in Fig.\ref{fig:2}. The computational-constrained railway device deploys a semantic encoder responsible for a small portion of the computation, while the edge server hosts the semantic decoder, which handles the majority of the computation. The railway device uses the semantic encoder to process a portion of the video and extract task-relevant semantic information. This information is then transmitted over wireless channels to the receiver. The receiver, equipped with the semantic decoder, completes the video processing and generates the intrusion detection results. Since both spatial and temporal correlations are present in videos, Rail-SC leverages a video Transformer, incorporating factorized spatial and temporal attention modules to efficiently manage these spatial-temporal correlations. Additionally, SNR-based layer normalization (SNR-LN) is employed to adjust the intermediate features based on the signal-to-noise ratio (SNR), enabling better channel adaptation.


\subsubsection{Rail-SC System for MIMO with the CSI feedback}

\begin{figure}[htb]
       \centering
       \includegraphics[scale=0.5]{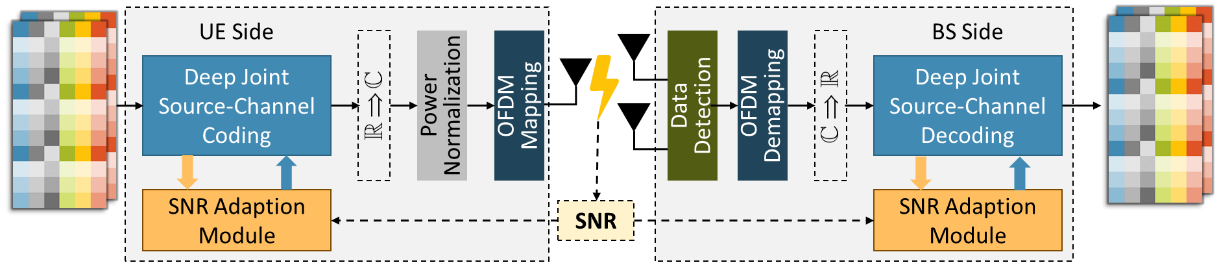}
       \caption{The framework of ADJSCC-CSI}
       \label{fig:3}
\end{figure}

In high-speed scenarios with rapidly varying channels, frequent CSI feedback is necessary to mitigate channel aging, but it introduces significant overhead. Furthermore, continuously monitoring time-varying channels for model switching is challenging, which can reduce the accuracy of CSI reconstruction and, consequently, degrade the system's communication capacity. To address these issues, a deep joint source-channel coding based framework with SNR adaptation for CSI feedback (ADJSCC-CSI) was proposed in \cite{9954153} to adapt to changes in channel SNR while maintaining CSI reconstruction performance with low feedback overhead. The ADJSCC-CSI network architecture, shown in Fig.\ref{fig:3}, jointly considers source coding and channel coding during training to extract more efficient, low-overhead semantic features of CSI for mitigating channel fading. At the receiver, a corresponding joint channel-source decoder reconstructs the CSI. This integrated approach mitigates the ``cliff effect’’ observed in traditional communication systems, where poor channel quality leads to significant degradation in CSI reconstruction performance. The network is trained end-to-end, with an SNR-adaptive module incorporated at both the transmitter and receiver to enhance the network's SNR generalizability. This module integrates channel SNR information into the training process, improving the network's generalization capability and enabling it to adapt to varying channel conditions using a single trained model.

In traditional modular communication systems, DL-based networks are often designed independently to perform specific functions and replace corresponding modules in classical communication systems. However, the optimization spaces of different modules may interact, and directly concatenating individually trained modules can lead to performance degradation. 

\begin{figure}[htb]
       \centering
       \includegraphics[scale=0.45]{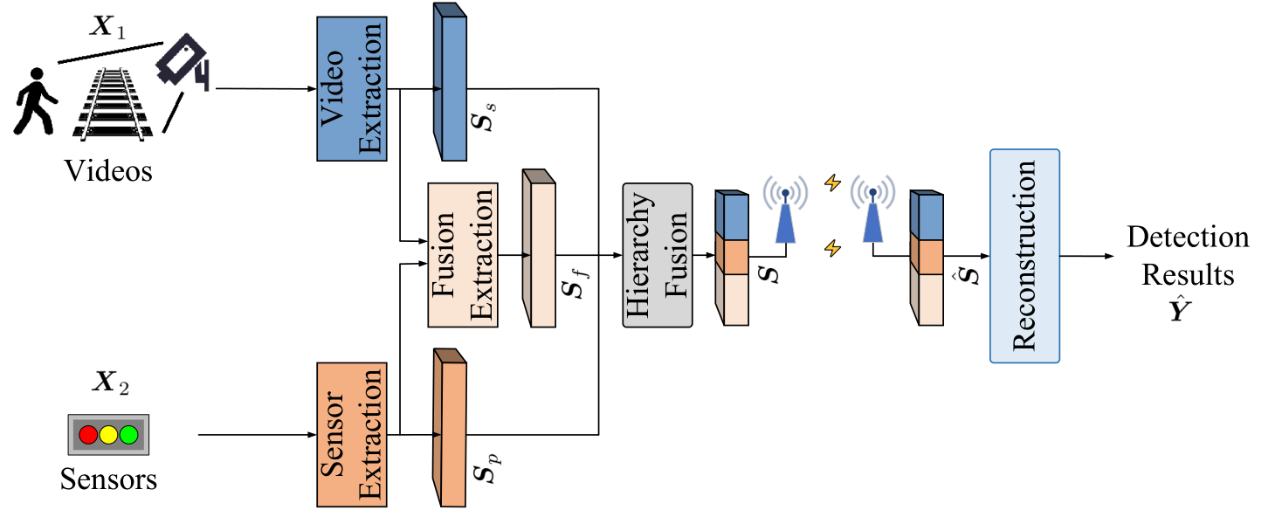}
       \caption{ The framework of hierarchy-aware and channel-adaptive data fusion}
       \label{fig:4}
\end{figure}

In Fig.\ref{fig:4}, a hierarchy-aware and channel-adaptive semantic communication framework is proposed \cite{10660530}. This framework integrates semantic communication techniques with multi-modal data fusion, enabling the system to adapt to varying channel conditions while ensuring efficient transmission. The primary contribution of this approach lies in its ability to dynamically combine fused modal features or deep and shallow features, improving communication efficiency and reducing data redundancy. Specifically, the framework addresses the bandwidth challenges in tasks requiring the fusion of different modalities, such as sensor readings or video signals, which are common in railway systems. By leveraging a semantic communication-based data fusion mechanism that is both hierarchy-aware and channel-adaptive, the system enhances the transmission of fused information without the need for additional bandwidth, enabling high-quality reconstruction of task-relevant features even in bandwidth-constrained railway environments.

\subsection {Lessons Learned}
The development of reliable transmission technologies for 6G smart railways highlights the necessity of addressing high mobility and complex environments with innovative solutions such as OTFS modulation, which effectively manages Doppler shifts and enhances spectrum efficiency. Advanced techniques including OTFS-TSMA, grant-free access, THz communications, and semantic communication enable ultra-massive connectivity and meet high-capacity demands. However, challenges such as path loss, link blockages, and dynamic conditions require sophisticated relay mechanisms and scheduling algorithms. Accurate channel modeling and location-aware scheduling further optimize resource allocation, reduce inefficiencies, and enhance overall system performance. Together, these advancements form the foundation for efficient, reliable, and scalable 6G smart railway networks.

\section{Intelligence and Security Technologies for 6G Railways}
Security and intelligent technologies are closely integrated and mutually reinforcing in smart railway systems. Intelligent technologies, such as AI analytics, and machine learning, enable advanced monitoring, prediction, and optimization, improving both operational efficiency and passenger experience. At the same time, security techniques, including encryption and intrusion detection measures, protect the integrity of these systems and ensure secure, reliable operations. Together, they drive innovation, address current challenges, and lay the groundwork for future smart railway advancements.

\subsection{Edge Intelligence Technologies}
With the swift advancement of MEC, AI, and IoT technologies, the network connectivity for intelligent railways will experience comprehensive acceleration in the era of 6G, promising a bright future for the development of intelligent transportation systems on railways. MEC brings computation and storage closer to end-users, reducing network latency and enabling real-time processing of large data volumes. It also enhances security and privacy by minimizing the need to transmit user data to remote cloud servers. The full utilization of edge resources and the execution of AI applications through edge computing contribute to the development of edge intelligence \cite{9596610}. 
In the realm of intelligent HSR, collaborative and interactive communication between on-board edge intelligence servers and edge data centers underpins the deployment and execution of business scenarios. The foundational support for achieving intelligence originates from the integration of various networks and computing resources facilitated by MEC. To ensure optimal service quality for intelligent rail operations and allocate resources judiciously, the edge intelligence in the 6G era must align with the characteristics of cloudification, intelligence, and service.

Edge intelligence, with its architecture and technological advantages, has emerged as a driving force for the future of intelligent industries across various sectors, especially in the rail transit industry. From the perspective of communication scenarios, Asad et al. \cite{2021Edge} summarized intelligent private network with user, network, and application-centric edge intelligence in next-generation railway systems and discussed some challenges. Chen et al.  \cite{9922178} proposed intelligent train control models deployed in MEC systems to provide reliable and real-time computing services for autonomous train control systems. 
An energy harvesting-driven cloud-edge-device collaboration IoT platform was proposed in \cite{9354351} to overcome the challenges of weak energy harvesting capability and unstable data rate of vehicle-ground communications in online fault detection. Xu et al. \cite{9899356} considered the channel dynamics that change over time in RIoT scenarios and studied the dynamic resource management of joint sub-carrier allocation, offloading ratio, power allocation, and computing resource allocation assisted by multi-access MEC. From the perspective of business requirements, considering the high-speed characteristics of the train, a task offloading scheme based on the task preprocessing deep deterministic policy gradient (TP-DDPG) algorithm was designed in \cite{10061573}, where the computational resources of MEC servers can be easily accessed by the UEs. 
Xu et al. \cite{9506824} investigated how frequent handover in uplink and downlink affects offloading and proposed an algorithm to maximize network throughput in a trackside MEC network.  An edge computing-aided FD framework for traction systems in HSR was investigated in \cite{8926409}, which has the potential value of further integration with important issues such as HSR dispatching and command, transportation organization, and operation management.

However, there is limited research on the combination of edge intelligence and smart HSR in the above works, and there are still many challenges remaining. Firstly, the channel scenario is complex and diverse, involving various scenes such as elevated tracks, tunnels, stations, and mountainous areas. Each type of scene has its distinct physical characteristics, resulting in varying channel conditions. Secondly, the base stations along the train tracks are typically deployed in linear strips, making it difficult to adopt dense mesh coverage and multi-point coordination like in urban settings, while capacity upgrades entail high costs.  On the other hand, HSR and urban rail transit services primarily consist of train operation control and passenger communication services. The train operation control requires high real-time performance, and the data exhibits characteristics such as type heterogeneity, varying business arrival patterns, and diverse service quality requirements. Moreover, For passenger communications, deploying train-mounted base stations and MEC enables improved network coverage, content caching, and computation offloading. However, this also introduces competition for limited wireless and MEC
resources, especially under high user density and heavy traffic loads. In particular, when numerous terminals 
offload tasks simultaneously, this can cause severe interference and MEC queue congestion, thereby increasing transmission errors and processing delays.

\begin{figure}[htb]
	\centering
	\includegraphics[scale=0.29]{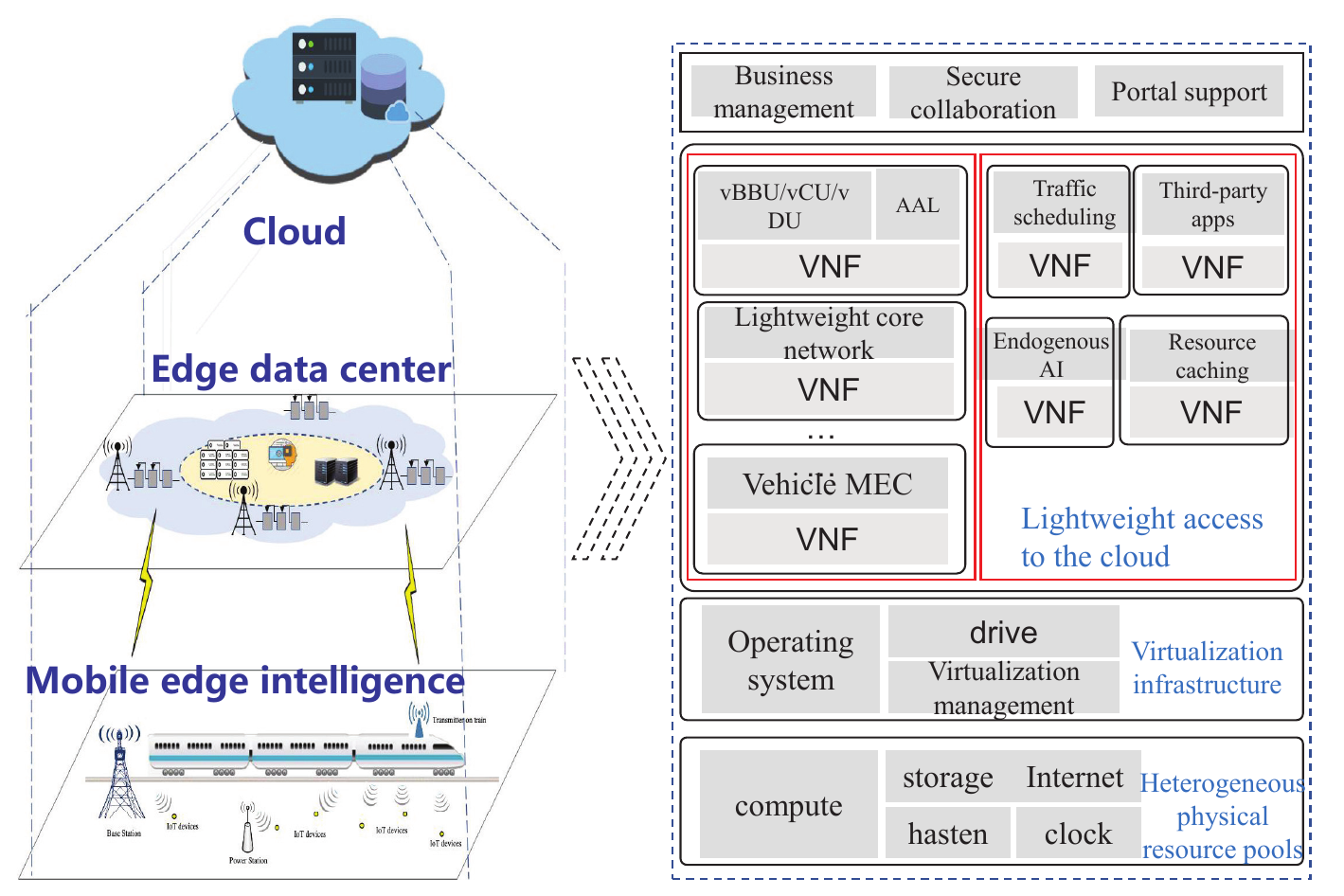}
	\caption{Edge intelligence network architecture}
	\label{fig:fig20}
\end{figure}

To overcome the above challenges, an edge intelligence network architecture for 6G intelligent HSR is proposed. Fig. \ref{fig:fig20} shows a two-layer edge intelligence architecture, including the mobile edge intelligent entity layer, and the edge data center layer. The mobile edge intelligence functions as the entity for onboard edge computing, primarily delivering low-latency services to users within the train and making intelligent predictions regarding real-time channel access for onboard users. The edge data center complements user requirements that cannot be fulfilled by mobile edge intelligent entities due to factors such as constraints imposed by the business scenario. Based on the edge intelligence architecture, the following solutions can be taken to solve the above challenges.
\begin{itemize}
     \item \textit{Dynamic channel scenarios for T2G communication:}  To address the dynamic and time-varying channel scenarios in T2G communication, a machine learning-based channel prediction model is trained in real-time onboard MEC. In addition, the local model parameters of the channel recognition and prediction models on all trains are trained offline based on federated learning, channel identification and prediction models on all trains are converged to a globally optimal unified model. Based on real-time channel prediction during train operation and combined with models of incoming traffic such as sensors, actuators,  and monitoring videos, the onboard MEC can use dynamic stochastic optimization strategies for uplink wireless transmissions to meet the quality of service requirements of different services.
    \item \textit{Concurrent computing offloading services for a large number of user terminals:} In the dynamic offloading scenario of ultra-dense networks, to overcome the uncertainty of interference and MEC computing load, the mean-field game method is adopted to coordinate the uplink computing and transmission strategy of a large number of densely distributed terminals for computation offloading, to minimize the total energy consumption and MEC processing delay.
    \item \textit{Secure and intelligent spectrum sharing:} Given the characteristics of complex heterogeneity and high mobility, distributed federated learning and reliable blockchain can be integrated for spectrum management. The multi-agent collaboration can also be used to optimize the spectrum-sharing strategy and improve efficiency, thereby reducing communication overhead, and realizing secure, reliable, and intelligent spectrum sharing.
\end{itemize}

With the emergence of 6G communication technologies, massive infrastructures will be densely deployed and the number of data will be generated exponentially. Massive data collected by users can be used to train a machine learning model, which is an effective way of implementing intelligent services in railway scenarios. Centralized training leads to communication congestion and user privacy leakage. To address these problems and improve efficiency, federated learning has been proposed. Most current studies on federated learning consider all users performing only a single task at a certain time, which results in low network utility and long task execution time. To address these challenges, we present solutions for scenarios involving a single aggregation server, multiple edge servers, and parameter authentication.
\subsubsection  {Single server application scenario} 
We propose a task-driven user coalition algorithm by jointly optimizing the total network utility and efficiency for the multi-task federated learning model \cite{10258056}. This algorithm can be further deployed in 6G railway networks, especially when trains stop at the station and the server has a wide coverage area.

\begin{figure}[htb]
	\centering
	\includegraphics[scale=0.36]{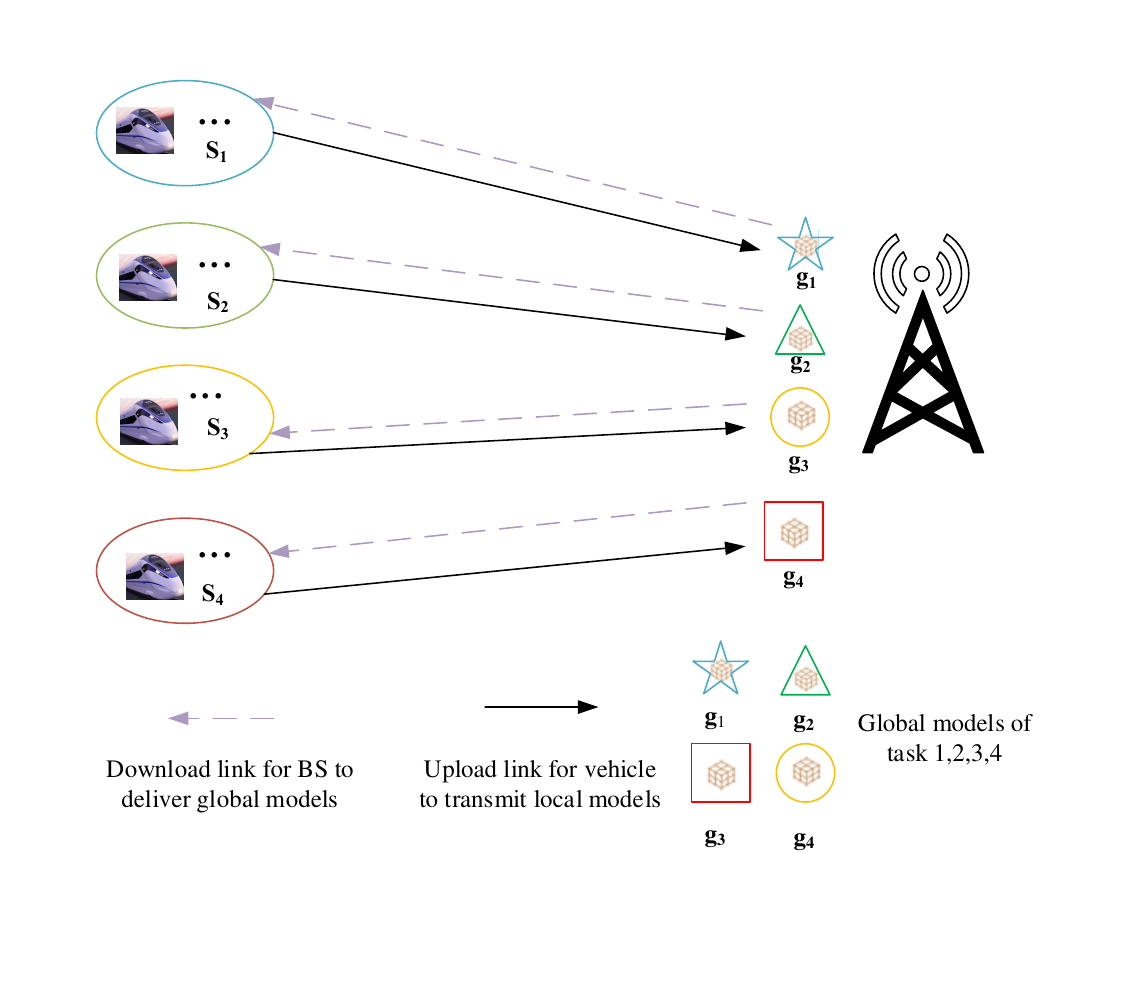}
	\caption{A multi-task FL model for railway networks}
	\label{fig:Flmodel}
\end{figure}

For task $ m $, set $ {\boldsymbol{S}_{m}} $ consists of $ S_m $ users that cooperatively execute data analysis and train a learning model. In the federated learning algorithm, BS releases the global models of different tasks to users. Once a task is selected to perform, users will use their collected data to train the local federated learning model and transmit the parameters of the local federated learning model to BS. These local federated learning models are integrated by BS to update the shared global federated learning model. The final task model is transmitted to the users requesting the service. The communication procedure of a multi-task federated learning algorithm is illustrated in Fig. \ref{fig:Flmodel}. 
The optimization objective in our multi-task FL model is:
\begin{align} 
	&\mathop {\rm{min}} \limits_{\boldsymbol{w_{n{S_m}}}} \sum\limits_{m = 1}^M {\frac{1}{{{K_m}}}\sum\limits_{{n \in {\boldsymbol{S}_m}}} {L({{\boldsymbol{w}}_{{\boldsymbol{nm}}}},{{\boldsymbol{D}}_{\boldsymbol{n}}},{{\boldsymbol{v}}_{{\boldsymbol{nm}}}})} }.
\end{align}


The goal is to minimize the loss function while maximizing the network utility, given as:
\begin{align} 
	& \mathop {\max \  }\limits_{{\boldsymbol{C}},{\boldsymbol{B}}} {{\mathop{\rm G}\nolimits} _N}.
\end{align}
Based on the Lipschitz continuous theory, the two-object optimization problem is transformed into a convex objective function from the network utility function and loss function with constraints through a series of mathematical analyses.
\begin{align} 
	& \mathop {\min }\limits_{{{\boldsymbol{P}}}{\boldsymbol{,a}}} \sum\limits_{m = 1}^M {\sum\limits_{{n \in {\boldsymbol{S}_m}}} {(1 - {a_n} + {a_n}(1-{P_{nm}}))} } \label{eq:twp}.
\end{align}
 From the definition of upload delay, it is inferred that only when $ \sum\limits_{m = 1}^M {l_m^U} $ reach the minimum, the value of ($\ref{eq:twp}$) reaches the minimum. Hence, the maximization problem of accuracy is equivalent to minimizing the total execution delay of tasks.
\begin{align} 
	& \mathop {\min }\limits_{{{\boldsymbol{P}}}{\boldsymbol{,a}}} \sum\limits_{m = 1}^M {l_m^U}.
\end{align}

\begin{figure}[h]
	\begin{center}
		\includegraphics[scale=0.26]{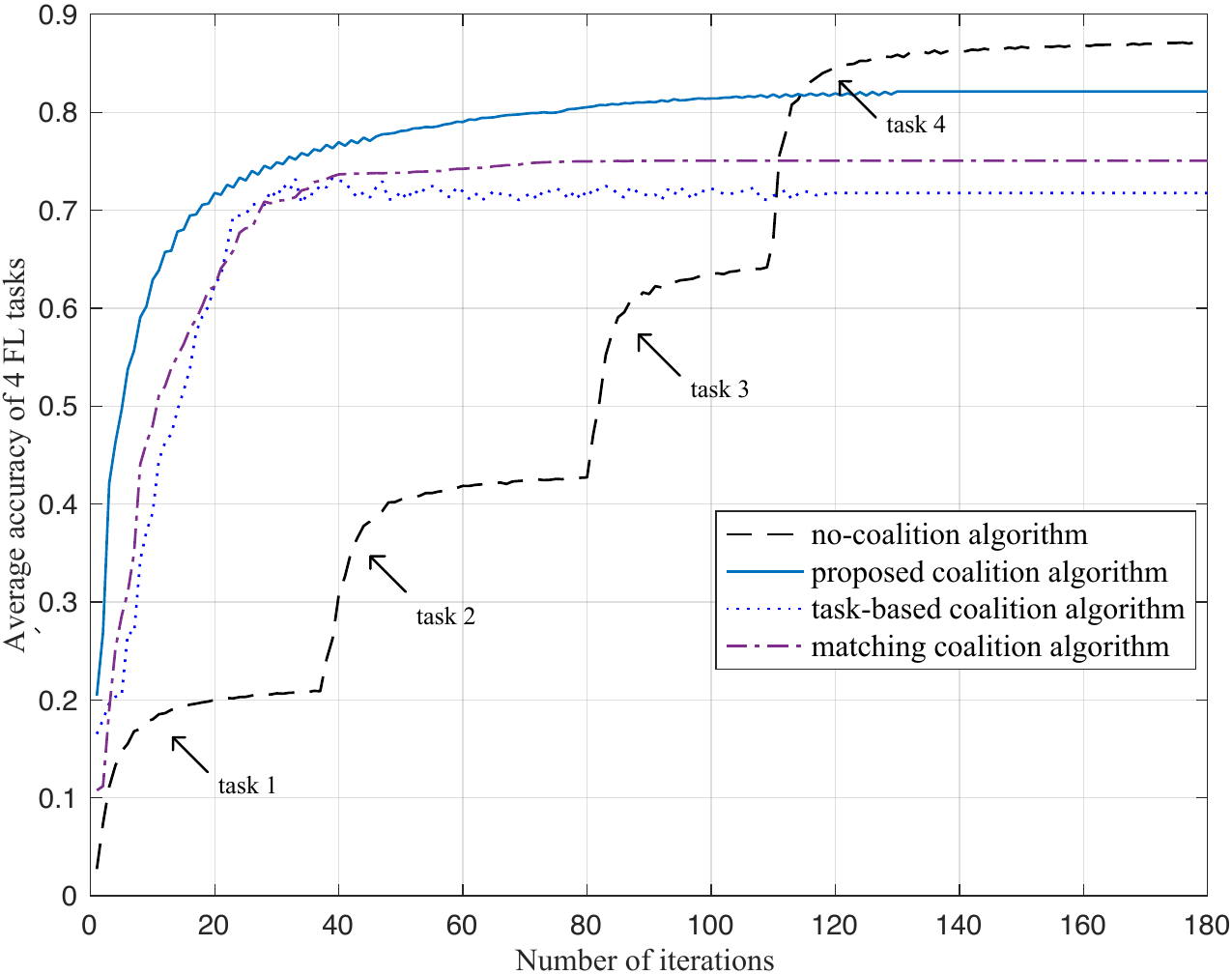}
		\captionsetup{font={small}}
		\caption{Convergence of the multi-task FL in different algorithms} \label{fig:li2}
	\end{center}
\end{figure}

Then, we proposed a task-driven coalition formation that utilizes a not-entirely selfish preference order, jointly considering the user's revenue and the total network utility. Fig. \ref{fig:li2} demonstrates that our proposed coalition algorithm significantly reduced the total execution time of tasks and the average accuracy could exceed 80$\%$. Ultimately, there was a substantial improvement in total network utility by using the proposed coalition algorithm.

\subsubsection{Multiple servers application scenario}
The integration of federated learning in a 6G-enabled railway system can provide reliable and accurate services while reducing the communication resources and central computation burden, as well as protecting the privacy of entities. Due to the high-speed mobility of trains, roadside infrastructure can be equipped with servers for federated learning to reduce communication latency. In \cite{10437116}, we addressed how federated learning tasks and users are matched and how model parameters are exchanged between servers in a multi-servers scenario.


Users upload their local federated learning models to the BS via frequency domain multiple access (FDMA).
The total delay of user $v_i$ to complete a training session in the $t$-th iteration is given by
\begin{align} 
	&{\tau _{{m_i}{v_i}t}} = \tau _{{v_i}t}^{Comp} + \tau _{{m_i}{v_i}t}^{Comm},
\end{align} 
where $\tau _{{m_i}{v_i}t}^{Comm}$ is the transmission delay and  $\tau _{{m_i}{v_i}t}^{Comm}$ is the computing delay.
Our ultimate goal is to reduce execution time while maximizing the network utility. Because the two optimization problems are under the same constraints of communication, computation, and task allocation, they can be expressed as a joint optimization problem  
\begin{align} 
	&\mathop {\max \ }\limits_{{\boldsymbol{A}},{\boldsymbol{B}}}  G \ \ \& \ \ \mathop {\min \ }\limits_{{\boldsymbol{A}},{\boldsymbol{B}}} \tau \label{eq:multwop}.
\end{align} 

Evidently, this optimization problem involves computing and communication settings, which can be realized on the block coordinate descent (BCD) technique. The optimized bandwidth allocation is performed while fixing the computing settings, followed by the optimization of computing resources while fixing the communication setting. The gradient projection is adopted to solve the bandwidth allocation problem.  The proposed stable matching game-based coalition formation algorithm is based on the following to realize many-to-one matching between the sets $\mathcal{V}$ and $\mathcal{M}$. 
\begin{itemize}
\item {\bf{Preference list of tasks:}} The benefit function of a generic task $m_i$, $K_{m_i}(\cdot)$, is evaluated by considering both the training delay and the time of uploading local parameters for a given user $v_i$ to complete task $m_i$. Hence, we have a preference list of tasks:
\begin{align} 
	&{K_{{m_i}}}({v_i}) = \frac{1}{{{\tau _{{m_i}{v_i}t}}}} = \frac{1}{{\tau _{{m_i}{v_i}t}^{Comp} + \tau _{{m_i}{v_i}t}^{Comm}}}.
\end{align}
\item {\bf{Preference list of users:}} Similarly, the benefit function of a generic user $v_i$, $O_{v_i}(\cdot)$ is defined as
\begin{align} 
	&{O_{{v_i}}}({m_i}) = Va{l_{{m_i}{v_i}}} - {\zeta _{{m_i}}}(T \cdot E_{{m_i}{v_i}t}^{Comp} + \sum\limits_{t = 1}^T {E_{{m_i}{v_i}t}^{Comm}} ).
\end{align}  
\end{itemize}

\begin{figure}[h]
	\begin{center}
		\includegraphics[scale=0.35]{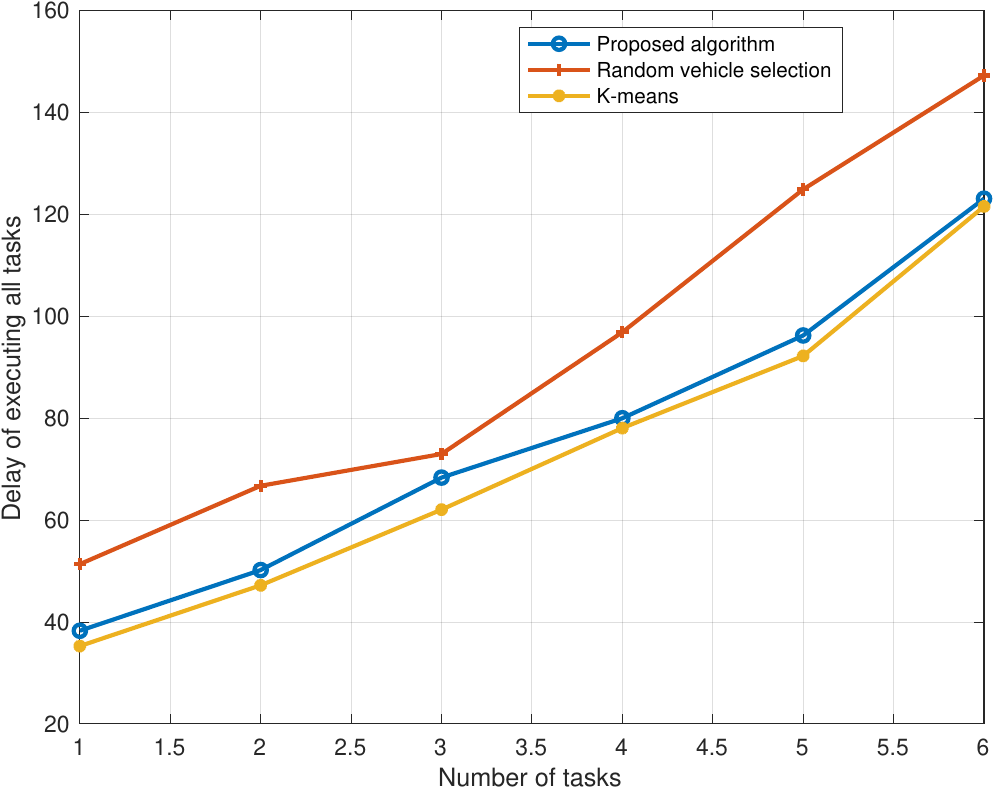}
		\captionsetup{font={small}}
		\caption{Delay of multi-task FL as the user coalition changes}\label{fig:lifig3} 
	\end{center}
\end{figure}

The simulation results show that our proposed algorithm can reach a stable vehicular coalition and yield performance improvements compared to the random algorithm, K-means algorithm, and one-by-one execution algorithm which do not or only take into account resource allocation, wireless factors, and vehicular mobility as shown in Fig.\ref{fig:lifig3}.
\subsubsection{Parameter-authentication federated learning}
There is a dilemma between secure aggregation and privacy protection when federated learning encounters malicious attacks. To solve the dilemma, we are motivated to propose a parameter-authentication federated learning scheme (PAFL) to resist model attacks and protect client privacy simultaneously while maintaining communication and computing efficiency.

In our system model, attackers launch fake global model attacks, risking user privacy and delaying genuine task as selected users unknowingly train on fake tasks. To preserve privacy, users upload local models anonymously via blockchain-generated identities, preventing BSs from verifying model authenticity. Abnormal models may arise from adversarial tampering-through data poisoning or direct parameter manipulation-or unintentionally, due to limited computation, high mobility, or unstable links. Both cases degrade global model accuracy, convergence, and overall reliability.
With the abnormal local models, the global model is updated as:

To solve the conflict between model authentication and privacy protection in the multi-task federated learning system, we design a PAFL scheme defending against malicious attacks. Specifically, we use zkSNARK, a state-of-the-art zero-knowledge proof technique, to authenticate federated learning models. zkSNARK allows the prover to prove its knowledge in a single message, which accords with the high mobility and strict delay requirement.

Various requirements tend to trigger multiple tasks simultaneously. RSUs publish federated learning tasks $m$ and generate global models ${\boldsymbol{\omega}}_m$ for users to download for local training. The Pedersen commitment \cite{5634099} parameters $(p_m,q_m,g_m,h_m)$ are also broadcast to the registered users. $p_m$ and $q_m$ are large primes where $p_m-1$ is divided by $q_m$. ${\mathcal{G}}_{q_m} $ is the unique subgroup of $p_m$'s inter group of order $q$, which is generated by $g_m$. $h_m $ is chosen from ${\mathcal{G}}_{q_m}$ and nobody knows $\log_{g_m}{h_m}$. If user $v$ is interested in task $m$, it will give a response $pc_{mv}^0$ to the RSU. 
\begin{align} 
	&pc_{mv}^0 = g_m^{a_v}h_m^{r_m} \mod \ q_m,
\end{align}
where $r_m \in {\mathcal{Z}}_{q_m}$ and $a_v$ is the private attribute of user $v$. The response $pc_{mv}^0$ is also the basis for user $v$ to decrypt the global aggregation of federated learning. As the heterogeneity of vehicular data, computing resources and reputation value, RSUs attempt to select high-reliability users for each task.

Along with federated global model ${\boldsymbol{\omega}}_{m}$, the RSU also generates a secret function $s_m({\boldsymbol{x}})$ and proving key to be shared with user $v_m$. The variables that satisfy the conditions of the secret function $s_m({\boldsymbol{x}})$ are called the witness. users can generate proof $\pi$ to authenticate themselves based on the secret function. Meanwhile, selected users also generate Pedersen commitment ${pc}_{mv}^t$ to schedule to execute a federated learning task later. 

If malicious or curious third parties want to get users private information from local parameters, they may impersonate the honest RSU to send fake global models to users. This is often the case that fake global models generally consist of simple weights to infer privacy with little resource consumption. When user $v$ receives the parameters of the global federated learning model, it proves the genuineness of the global model based on the zero-knowledge proof. The global models that are not generated by the RSUs registering on the blockchain will be dropped out. Otherwise, users train the learning model based on their local dataset ${\mathcal{D}}_v$. 
User $v$ needs to test the knowledge of the Pedersen commitment and then create its commitment ${pc}_{mv}^t$ in each iteration:
\begin{align} 
	& {pc}_{mv}^t = g_m^{a_v}h_m^{r_m} \mod \ q_m.
\end{align}

Upon completing local training, user $v$ uses the secret function $s_m({{\boldsymbol{\omega}}_{mv}^t})$ to prove its knowledge of the task $m$. 
Then, user $v$ uploads local model ${\boldsymbol{\omega}}_{mv}^t$, Pedersen commitment ${pc}_{vm}$, and proof ${\pi}_v$ to the smart contract. The smart contract uses the verification key to prove whether ${\pi}_v$ is valid. If the proof is invalid, the local model ${\boldsymbol{\omega}}_{mv}^t$ will be dropped out and the vehicular reputation will be affected. Otherwise, the smart contract works as the data authenticator to authenticate the reliability of local model ${\boldsymbol{\omega}}_{mv}^t$ based on the dataset ${\boldsymbol{D_v^{\ast}}}$ uploaded in the registration. The blockchain gives a signature ${\sigma}_v^t$ to each ${\boldsymbol{\omega}}_{mv}^t$ that is authenticated to be trained:
\begin{align}
	&{{\sigma}_v^t} = Hash({{\boldsymbol{\omega}}_{mv}^t}, {\boldsymbol{D_v^{\ast}}}).
\end{align}

The blockchain transmits authenticated parameters ${\boldsymbol{\omega}}_{mv}^t$, Pedersen commitment ${pc}_{vm}$, and signature ${\sigma}_v^t$ to the corresponding RSUs. Aimed to protect vehicular privacy, the identity of each user that accesses the blockchain is anonymous. An honest user can always convince the honest RSU for the given proving key, verifying key, and secret function. For each valid proof ${\pi}_{mv}$, the following is valid: 
\begin{align} 
	&{\mbox{Pr}}[{\mbox{Verify}}(crs, {\pi}_{mv}, s_m(\boldsymbol{x})]=1.
\end{align} 

\begin{figure}[h]
	\begin{center}
		\includegraphics[scale=0.36]{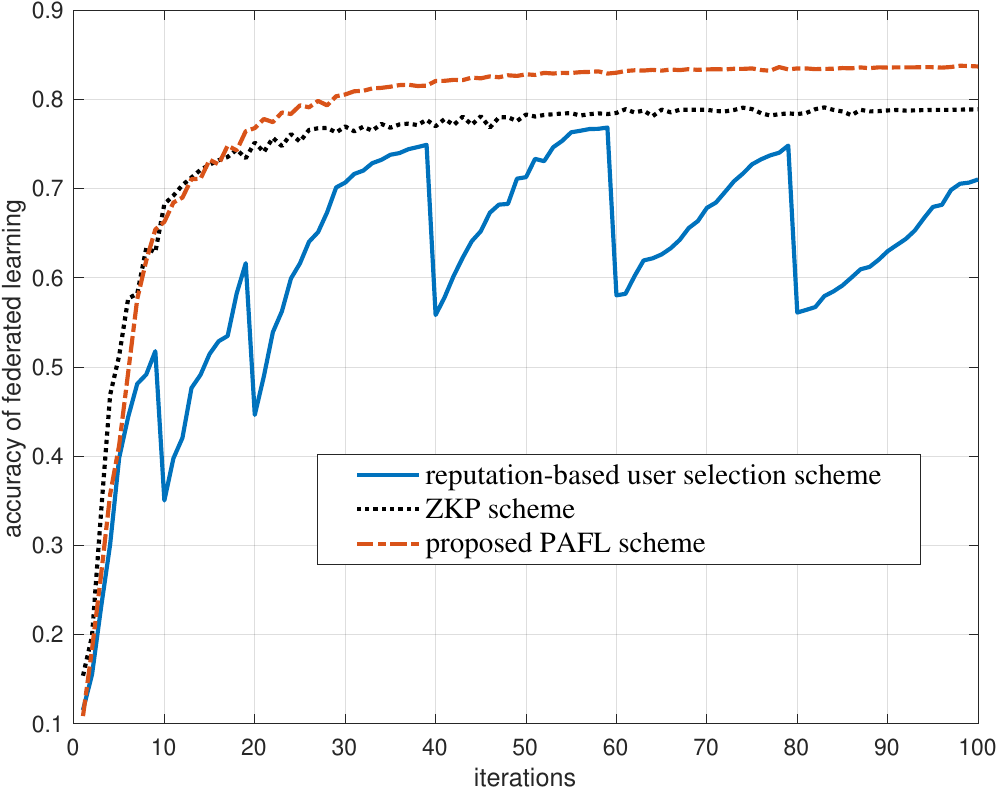}
		\captionsetup{font={small}}
		\caption{Accuracy of FL under defensive schemes} \label{fig:fig10-2}
	\end{center}
\end{figure}

The simulation results shown in Fig.\ref{fig:fig10-2} demonstrate that our proposed scheme achieves better performance than existing privacy-preserving and attack-defense federated learning schemes in terms of accuracy, attack defense, overhead, as well as privacy protection.

\subsection{Endogenous Security Technologies for Smart Railway}
Future smart rail systems will depend on a powerful ``rail transit brain'' to ensure safe operation within a trustworthy traffic environment. As the backbone of information exchange, 6G—an integrated space-air-ground-sea network—will provide on-demand services and support diverse rail applications. Its convergence with big data mining enhances performance but also introduces new security challenges due to increased device diversity and blurred network boundaries. To ensure secure and intelligent rail transit, 6G networks must move beyond traditional planning and adopt robust endogenous security mechanisms. In essence, 6G-enabled smart rails should prioritize both built-in security and intelligence, while integrating technologies such as blockchain and digital twins \cite{dang2020should}.

\subsubsection {Endogenous secure railway network architecture}
The current 5G network adheres to the traditional IP-based architecture, where security modules are retroactively updated through additional patches following attacks. This mechanism results in a slower growth rate of network self-defense capabilities compared to the escalating pace of attacks, rendering the architecture outdated and unsuitable for 6G networks. Furthermore, the conventional design of the rail transit communication network lacks initial considerations for architecture-level security, relying on traditional ``external-mounted" and ``patch-type" security protection mechanisms. Consequently, these traditional methods cannot adequately ensure the security of future smart rails and are ill-equipped to counter potential ubiquitous attacks and uncertain security risks posed by evolving 6G networks.

Endogenous security emphasizes establishing a comprehensive information security system, reinforcing internal network security measures, and thwarting potential attacks from the source. In the context of high-quality services and data privacy enhancement, diverse innovative strategies have been investigated across a broad spectrum of vertical domains within 6G. 
For endogenous secure resource scheduling management, Zhang et al. \cite{zhang2023endogenous} proposed the integration of DT and 6G edge intelligence. This approach involves resource scheduling management through DT-assisted state information evaluation, utilizing the attack detection resource management algorithm to proactively adjust policies based on estimated attack probabilities. Chang et al. \cite{chang20226g} presented a secure endogenous wireless access network architecture based on blockchain for 6G. They designed a unified identity authentication framework using blockchain technology to manage network entity identity certificates and oversee authorization. 
Addressing security challenges related to frequency band scarcity, Ramezanpour et al. \cite{ramezanpour2023security} proposed an unsupervised learning method for the authentication process.

The extensive research on these initiatives lays the foundation for the application of endogenous security technologies in 6G railway communication networks to mitigate cybersecurity risks \cite{10075050}. A series of studies have been undertaken to improve the security of smart rails, focusing on access and authentication aspects. Wang et al. \cite{10234591} introduced a secure and efficient authentication key agreement scheme aimed at bolstering authentication security during both the initial registration and handover processes within space-ground integrated railway communication networks. A lightweight mutual authentication mechanism and a rapid key generation and distribution mechanism were devised to prevent unauthorized access to the integrated networks. 
Zhu et al. \cite{9484083} proposed to use blockchain in communication-based train control (CBTC) systems for distributed key management and communication node authentication. The problem of block producer selection and onboard blockchain client handoff are studied jointly by leveraging a deep reinforcement learning approach. Furthermore, the application of federated learning in HSR systems has been employed to decrease computation costs and enhance fault diagnosis performance, as discussed in \cite{10036973}.

\begin{figure}[htbp]  
	\centering  
	\includegraphics[scale=0.2]{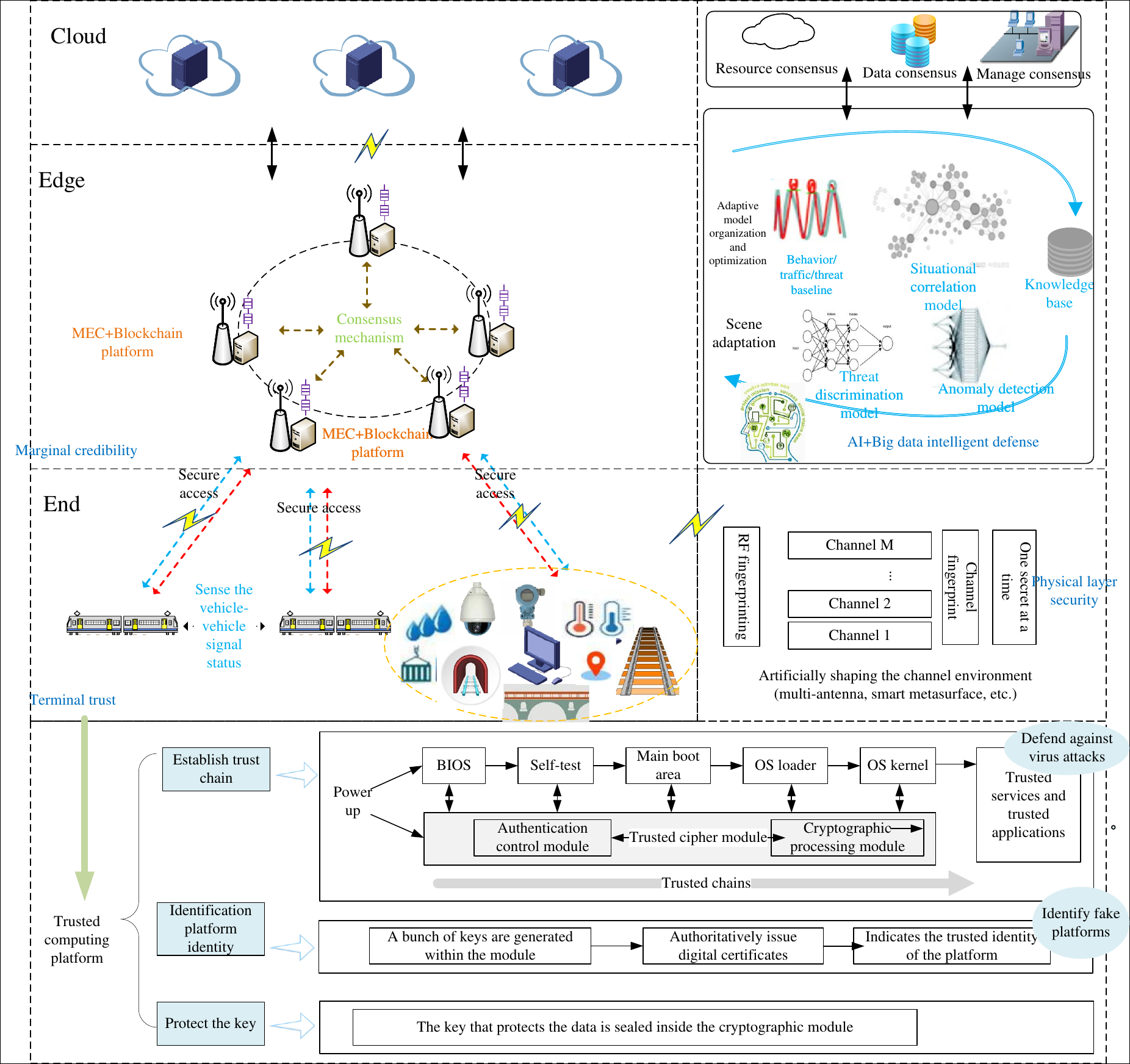}
	\caption{Railway Endogenous Security Architecture}  
	\label{fig:ES-6G}  
\end{figure}

However, current security protection systems face challenges in meeting the evolving 6G security requirements. There is a need to continuously cultivate self-adaptive, autonomous, and self-growing security capabilities within the system. To build an endogenous security network, the envisioned endogenous security architecture for intelligent rail communication networks has been proposed, as illustrated in Fig. \ref{fig:ES-6G}. The network integrates defense mechanisms from the design phase to enhance its inherent ``immunity''. Serving as the engine, AI monitors the system status in real time and identifies potential security risks. By combining proactive threat detection with intrusion prevention, the architecture enables intelligent endogenous security—defined by ``risk prediction and active immunity''. The proposed cloud-edge-end coordinated architecture encompasses the hardware support of a trusted computing platform, data consensus mechanism powered by blockchain, IoT-enabled security situation awareness, and AI-based attack defense models. The proposed framework is designed to fulfill the demands of intelligent consensus, intelligent defense, trust enhancement, and ubiquitous cooperation for the endogenous security of 6G smart rails.

Built upon the endogenous security framework, the smart rails will advance a new generation of identity security, overhaul in-depth network security, and address security concerns across diverse terminals and IoT devices. 

\subsubsection {Key technologies for endogenous security in a smart railway}
In railway network scenarios, the strong complementarity between space and ground networks—particularly in coverage and mobility—highlights the value of space–earth integration in 6G development. As ground Internet, mobile communications, and space-based services converge, a unified information network becomes central to smart rail systems. Within 6G networks tailored for railway applications, addressing security threats in such networks requires more than border defense; it demands end-to-end protection with capabilities in auditing, detection, response, and recovery. Techniques such as vulnerability scanning, threat traceability, situational awareness, intrusion detection, and intelligent analysis are essential \cite{10075050}. In conclusion, future smart railway networks can benefit from a broad range of technologies, including intelligent analysis, trusted authentication, and distributed collaboration, to achieve endogenous security.

\begin{enumerate}[a)]
	\item{\emph{Intelligent analysis and detection:}} Intelligent analysis and detection technology can realize automatic security processing, rapid detection and response for railway network scenarios. Recent developments in AI technology enable it to play a pivotal role in enhancing the security of railway networks through intelligent analysis and detection capabilities. It excels at the accurate detection of network traffic and abnormal behaviors, offering advanced functionalities such as backtracking and analysis. Furthermore, through the thorough analysis of network data traffic characteristics, AI extracts essential data features to identify and understand intrusion behaviors. This intelligent approach contributes to a comprehensive and proactive security strategy for railway networks. For instance, Li et al. \cite{dai2021review} leveraged AI algorithms to enable more accurate detection, backtracking, and root cause analysis of abnormal traffic and behavior in railway networks. Pei et al. \cite{10078088}, on the premise of big data security and privacy protection technology, emphasized that federated learning can improve the ability of the overall endogenous structure of the system to realize privacy protection. 
 	\item{\emph{Trusted authentication:}} Within the realm of 6G rail intelligent transportation systems, access is exclusively granted to verified equipment or devices. This rigorous verification process serves as a protective measure, ensuring the overall integrity and security of the network. Employing a trusted authentication process, any abnormal temporal and spatial information is promptly identified, reported, and addressed. Chen et al. \cite{chen2022zero} proposed the establishment of a decentralized network access control mechanism through adaptive cooperation among the involved control domains. Meanwhile, Loven et al. \cite{loven2019edgeai} advocated for integrating AI into edge computing to enable fine-grained control and management of personal data, thereby implementing effective access control. These proactive strategies not only enhance the intelligence of railway networks but also significantly diminish the probability of railway accidents. 
	\item { \emph{Distributed collaboration:}} Distributed collaboration technology stands as a core key technology within the realm of 6G, introducing endogenous security for distributed and decentralized wireless networks. Blockchain and distributed AI, two pivotal components of this technology, play a significant role in enhancing the safety and reliability of 6G systems in rail transit. Its applications within railway networks span various scenarios, including network management, D2D communication, resource sharing, and spectrum management. In a proposal by Dai et al. \cite{8998330}, a decentralized and secure content caching scheme is presented, leveraging blockchain to facilitate trust among untrusted devices. Additionally, AI is employed to craft an intelligent caching strategy, enabling end-to-end caching. Another innovative approach by Liang et al. \cite{10177944} integrates distributed reinforcement learning with blockchain to establish a trust management and services optimization model. This model effectively ensures the provision of trusted and high-quality services within the distributed collaboration framework.
\end{enumerate}

\subsubsection  {Blockchain empowered trustworthy railway networks}
In conventional railway communication systems, security standards are often overlooked during the initial network design, leaving the fundamental network vulnerable to threats in identity authentication, access control, network communication, and data transmissions. This lack of initial consideration results in serious security issues. Hence, 6G networks demand robust and highly available security protection capabilities, with a particular emphasis on adapting to the distributed and decentralized intelligence of wireless network architecture. To prevent potential attacks on terminals, it is essential to establish a comprehensive protection mechanism for access devices, ensuring the overall security of these entry points.

Blockchain, integrating distributed storage, peer-to-peer transmission, consensus, and encryption, aligns with the decentralized architecture of railway networks. It ensures data integrity, secure operations, and trust among untrusted devices—without relying on costly centralized processing. In future 6G-enabled smart railways, blockchain will play a key role in secure data sharing, distributed security frameworks, and lightweight IoT authentication algorithms.

Presently, blockchain technology has evolved into a certified collaboration application, boasting a system with multi-layer security authentication and a distributed privacy protection function. This has played a positive role in realizing endogenous security. Blockchain-based mechanisms effectively address authentication challenges for distributed ledger technologies, offering a robust solution to the security of 6G endogenous networks through multi-factor authentication technology. In works such as \cite{zhu2019applications}, authors highlighted the threats posed by authentication and security challenges in communication, proposing access control technology as a solution. 
Additionally, Dang et al. \cite{dang2020should} observed that network decentralization based on blockchain technology can simplify network management and enhance network performance.

With the introduction of emerging technologies, railway communication networks have the characteristics of complex structure and ambiguous boundaries, resulting in a series of security threats including data leakage and malicious access. Building upon the mentioned solutions, Feng et al. \cite{feng2023blockchain} proposed an innovative blockchain-enabled solution for building an endogenously secure railway communication system. By integrating the zero-trust security model \cite{10091151}, the approach leverages blockchain and Merkle trees to implement distributed identity storage. The proposed architecture supports a public-private railway network with components such as identity registration, two-way authentication, and reputation evaluation. A bidirectional authentication proxy is introduced between cloud servers to guard against both internal and external threats. Furthermore, zero-knowledge proofs are employed to enable efficient and secure node authentication, reinforcing the system’s trust and resilience.

\begin{figure}[htbp]  
	\centering  
	\includegraphics[scale=0.47]{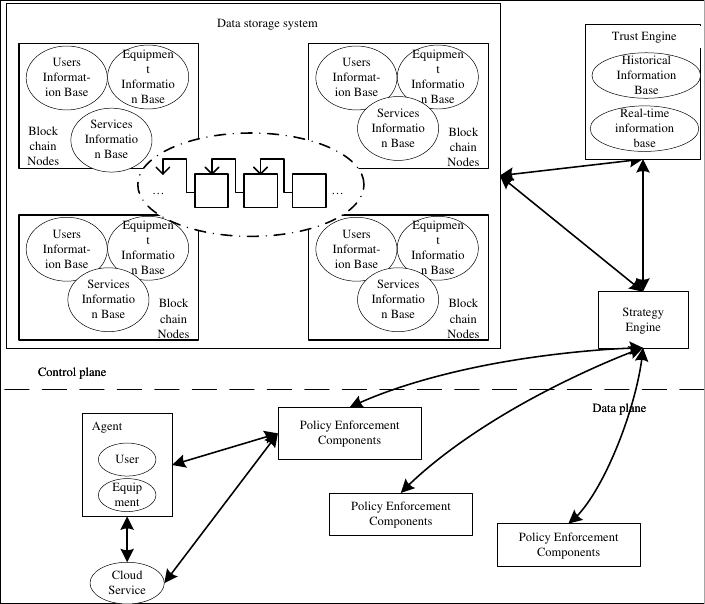}\\  
	\caption{Blockchain-based zero trust architecture.}  
	\label{fig:Blockchain-based zero trust system architecture for railway.}  
\end{figure}

Fig. \ref{fig:Blockchain-based zero trust system architecture for railway.} shows the system architecture of the blockchain-based zero-trust railway network model. The system architecture is divided into a control plane and a data plane. The control plane includes a data storage system, policy engine, and trust engine, which collectively handle information registration, authentication, access control, and reputation evaluation for railway employee accounts and terminal devices. This ensures effective identity and access management. The data plane manages actual data transmission and processing through proxies, public cloud services, and policy enforcement mechanisms tied to user and device identities. Its primary goal is to ensure secure, efficient data flow and support seamless access to application services.

\subsection{Railway Digital Twin Networks}
The design objective of railway wireless networks is to achieve ultra-high capacity and ultra-low-latency communication by integrating various communication and computing technologies. The emergence of 6G technology is propelling rail transportation into a new era characterized by digitization, intelligence, informatization, and automation. However, future railway wireless networks may still encounter challenges in terms of security, intelligence, efficiency, and capacity. The advent of DT networks presents opportunities to address the aforementioned challenges.

As an emerging technology for next-generation networks, DT enables seamless mapping of physical railway systems to their virtual counterparts \cite{9745481}, allowing for comprehensive, real-time analysis. DT facilitates predictive optimization, fault-tolerant control, and process automation, enabling real-time monitoring of railway components—including driver status—to enhance safety, reliability, and operational efficiency while reducing costs. By providing real-time insights into network status, DT also improves connectivity and intelligent computing performance. Additionally, DT networks can leverage AI to dynamically manage tasks such as offloading, resource allocation, and network orchestration.

The advent of wireless networks and DT will enable precise and convenient passenger services, intelligent and efficient transport organization, panoramic station management, predictive maintenance, as well as comprehensive safety and incident management for intelligent railways \cite{10292659}. While still in its early stages, there have been ongoing efforts to explore the application of DTs in railway networks. To enhance the efficiency of constructing railway infrastructure DTs, Wang et al. introduced a self-attentive learning framework for point cloud semantic segmentation in HSR. This framework incorporates a local geometry embedding module and a prototype-guided regularization method \cite{10144473}. In another initiative \cite{10049521}, De Benedictis et al. presented an anomaly detection architecture leveraging DT and autonomic computing, which is designed to monitor, analyze, plan, and execute appropriate strategies. Validation of the proposed DT system was conducted using a reference operational scenario within the European Railway Traffic Management System, demonstrating its effectiveness. Furthermore, in \cite{9926529}, Ferdousi et al. provided a DT-based framework for railways aimed at estimating future states and pre-determining actions. By employing AI techniques, the proposed scheme enhances track health inspection and monitoring, thereby improving the operational efficiency of railway networks.

\begin{figure}[htbp]
	\centering
	\includegraphics[scale=0.19]{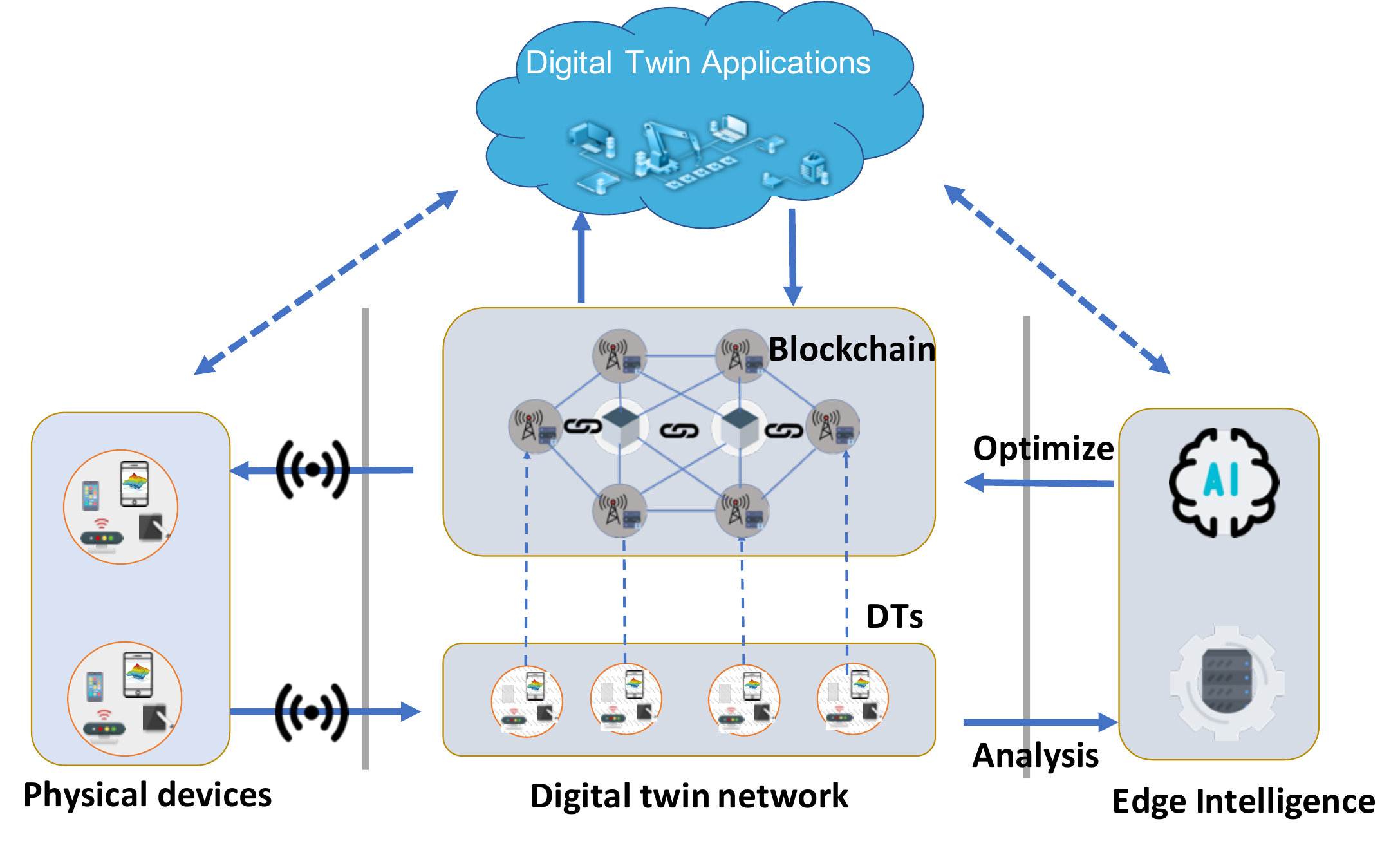}
	\caption{Digital twin networks}
	\label{fig:fig21}
\end{figure}

Building upon the aforementioned investigations and considering the structural characteristics of the intelligent transportation network, a railway DT network can be illustrated in Fig. \ref{fig:fig21}. The network is divided into three layers: the physical layer, the DT network layer, and the application layer. The physical layer includes trains, network entities, and IoT devices. In the DT network layer, each physical entity deploys its corresponding DT on the server, trained using AI algorithms. The DT network dynamically synchronizes real-time operational data and entity relationships within the railway system. At the application layer, edge intelligence enhances the DT by using AI algorithms for computation and analysis, enabling the generation of optimization decisions and feedback for both the DT network and the device layer.

DT has also attracted wide attention in related fields of rail transportation, including intelligent maintenance \cite{9690822}, smart stations \cite{9540153}, as well as operational fault detection and diagnosis, and environmental monitoring \cite{s21175757}. For example, the RailTwin framework proposed in \cite{9926529} integrates real-time system awareness, future state prediction, and AI-driven automation for intelligent decision-making. However, the intelligent railway system driven by DTs still faces the following challenges:
\begin{itemize}
\item \textit{DT modeling for dynamic railway systems with limited resources:} Building a DT network that comprehensively maps railway networks, encompassing numerous sensor terminals, IoT nodes, and a complex train operation system, poses significant design challenges. Effectively modeling DT for a highly dynamic railway system under limited resource conditions is a crucial aspect of DT network development. Solutions from various vertical areas can be leveraged to achieve accurate and efficient DT construction in railway networks. For instance, He et al. \cite{9887906} integrated DT and MEC technologies into a hierarchical federated learning framework within heterogeneous network scenarios to address the joint optimization problem of user association and resource allocation. In another approach \cite{9899364}, a DT-driven task offloading and IRS configuration framework was proposed to adapt to the time-varying nature of the physical operating environment in vehicular networks. The utilization of distributed AI and optimization algorithms is pivotal for enhancing the system performance and resource utility of DT networks.
\item \textit{Simulation errors between DT and physical railway entities:} Constructed based on historical and current operational data of entities, the DT model may unavoidably exhibit some discrepancies with the actual state of the physical entity due to losses in the data transmission and training processes. Further research is needed to minimize the errors between DT and physical entities in rail transportation and enhance the system fault tolerance. Sun et al. \cite{9311405} proposed a dynamic incentive scheme based on the Stackelberg game and federated learning, which adaptively selects the best client and training rounds to reduce the impact of bias. In the mobility scenario empowered by DT \cite{9830070}, a learning-based heterogeneous network selection scheme was proposed, optimizing the connection between the moving entity and its DT, thereby facilitating the data synchronization process.
\item \textit{Security and privacy for DT modeling:} 
The synchronization of the mapping between the physical entities and the DT counterparts necessitates real-time data transmissions, ensuring the convergence of physical and virtual states simultaneously. The data uploaded originate from various physical entities, such as track sensors, train terminals, and passenger terminals. However, data transmissions at the source through dynamic communication channels may lead to the potential leakage of sensitive information, thereby expanding the risk of potential attacks. On the other hand, malicious clients or attackers attempt to target either the physical railway system itself or its DT, posing a great threat to the entire system. 
Fan et al. \cite{9579446} proposed a DT-based MEC architecture for lane-changing scenarios, enhancing the perception and computation capabilities of connected and autonomous vehicles to ensure real-time safety. Liu et al. \cite{9447819} considered CSI in MEC servers selection and intelligent task offloading scheme, and developed a side-cooperative node selection scheme, which fully guarantees data security and provides high-quality communication links.
\end{itemize}

While the aforementioned research schemes partially address challenges in high-mobility scenarios, some of them may not be suitable for smart railway scenarios. The implementation of DTs is made difficult by the Doppler shift caused by high-speed train movement, as well as the time-varying nature of network topology and communication links. In addition, the wide geographical distribution of sensors and certain correlations among sensor data within a certain range pose significant challenges in terms of data controllability and security levels, which need to be addressed.

\subsection{Railway Digital Twin Empowered by Blockchain and Federated Intelligence}
\subsubsection{Digital twin system modeling}
\begin{figure}[htbp]
	\centering
	\includegraphics[scale=0.45]{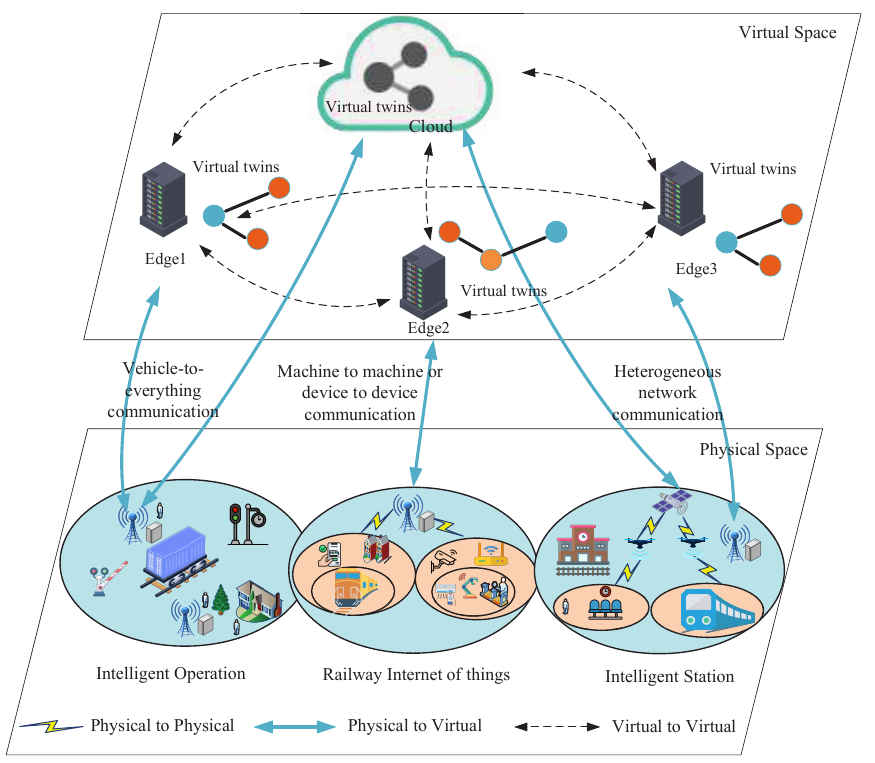}
	\caption{A hierarchical architecture of digital twin networks.}\label{fig:model4}
\end{figure}

To gain a deeper understanding of a railway DT network, we introduce a layered architecture consisting of a virtual space and a physical space, as shown in Fig. \ref{fig:model4}.

The physical plane encompasses several specific DT application scenarios, including intelligent transportation systems, communication networks, and IoT networks. In contrast, the virtual plane utilizes DT technology on both edge and cloud servers to create and manage virtual twins of physical objects. Physical devices such as trains, sensors, and smart terminals synchronize data in real-time with the virtual twins via wireless communication and receive feedback for instant control. MEC networks are expected to provide communication and computing power to fulfill key DT requirements including low latency, high reliability, and strong security. This enables seamless connectivity between the real and virtual worlds to interact more efficiently and securely. Due to the dynamic nature of railway computing and communication resources, associating DTs with their corresponding servers emerges as a crucial challenge in DT edge networks, necessitating a thorough exploration. Furthermore, given that federated learning in DT edge networks entails multiple data exchange communications, optimizing the allocation of limited communication resources can enhance the efficiency of DT networks. 

\begin{figure}[htb]
	\centering
	\includegraphics[scale=0.18]{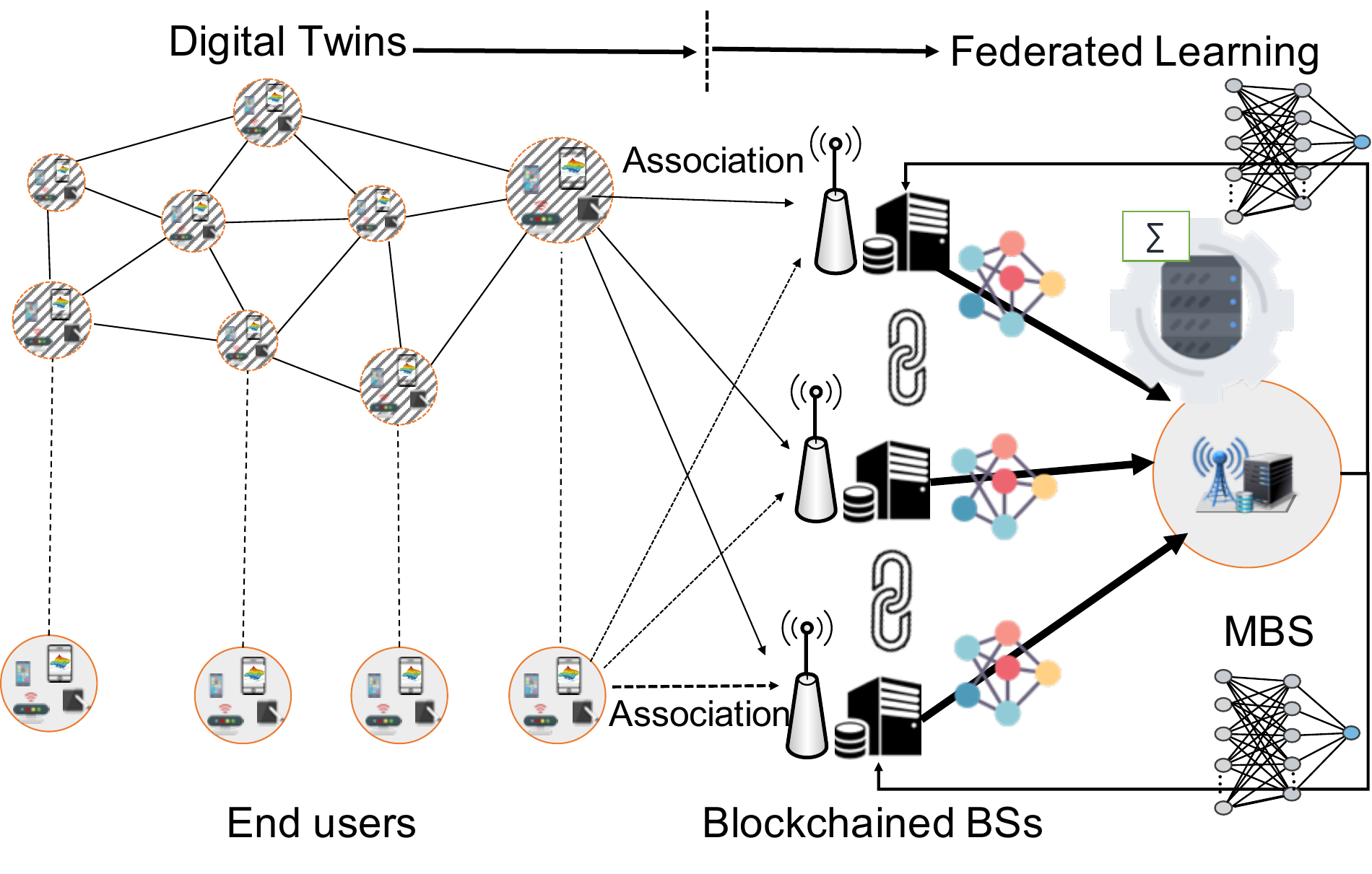}
	\caption{The proposed railway digital twin networks}
	\label{fig:DTWN}
\end{figure}

A model of a DT railway network empowered by blockchain and federated learning is depicted in Fig. \ref{fig:DTWN}. The system comprises end clients, BSs, and a macro base station (MBS). Both the BSs and the MBS are equipped with MEC servers. The end devices generate running data and synchronize their data with corresponding DTs running on the trackside BSs. 
In various application scenarios, end clients may communicate with each other to exchange operational information and share data. Consequently, DTs also establish a network based on connections among end clients. With analysis of the constructed DT network, operational states of physical devices can be gathered, enabling informed decisions to optimize and guide device operations.

In the proposed scheme, federated learning is employed to collaboratively execute the training and learning process for building DT models. Addressing the lack of mutual trust among end users and considering the inclusion of private data in DTs, the system incorporates a permissioned blockchain to enhance security and data privacy. The permissioned blockchain records data from DTs and manages client participation through permission controls. BSs are responsible for maintaining the blockchain and also act as clients of federated learning, while MBS serve as the federated learning server. Leveraging data from DTs, BSs train their local model and return the model parameters to MBS. This methodology ensures secure data management and collaborative model training in DT networks.


\subsubsection{Blockchain empowered DT}
To enhance security and privacy, a blockchain-enabled DT architecture is proposed to showcase the diverse integration of blockchain and DT technologies. BSs act as blockchain nodes, managing a permissioned blockchain to protect DTs from untrusted end users. DT parameters are recorded on the blockchain and updated as user states change. Additionally, local models generated at BSs are treated as transactions and verified by peer BSs to ensure model quality. This framework maintains three types of records: DT model records, DT data records, and training model records.

\begin{figure}[htb]
	\centering
	\includegraphics[scale=0.22]{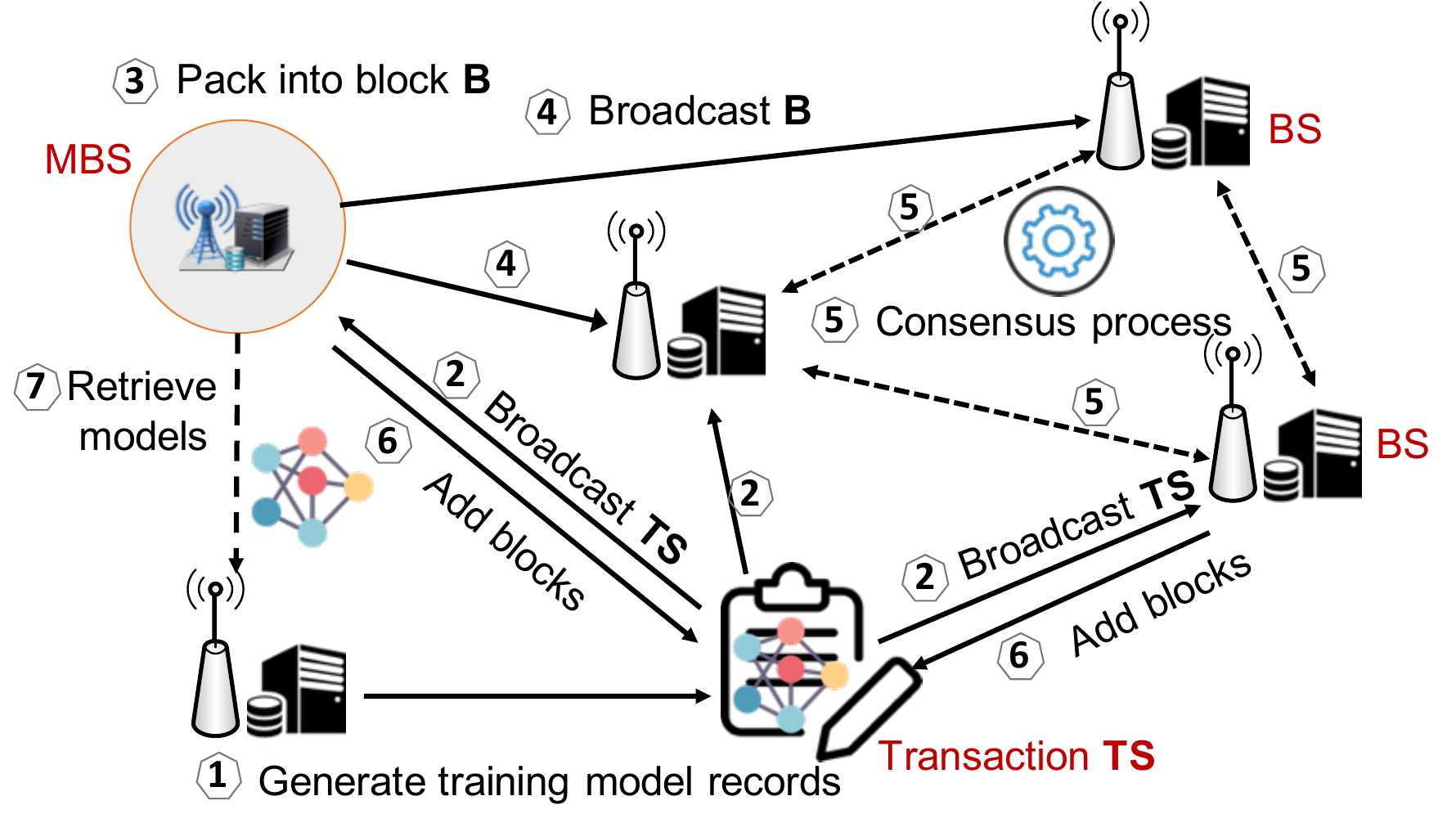}
	\caption{The blockchain empowered learning scheme \cite{li41}}
	\label{fig:blockchain}
\end{figure}

The overall blockchain framework for DTs is presented in Fig. \ref{fig:blockchain}. The process begins with BSs training local models on their datasets and subsequently uploading these trained models to MBS. These trained models are additionally documented as blockchain transactions and shared with other BSs for verification. Following this, other BSs gather these transactions and organize them into blocks, initiating the consensus process to authenticate the transactions within those blocks.
The consensus process is implemented using the delegated proof of stake (DPoS) protocol, where the stakes are the training coins. 


Following the training process, the coins of each BS are adjusted based on their performance. If a BS’s model passes verification by peer BSs and the MBS, it will be rewarded with coins; otherwise, it receives no compensation. A group of BSs is elected as block producers via a voting process where BSs use training coins to vote. The elected producers take turns packaging verified transactions into a block $B$ within a time interval $T$, and broadcast it for validation. In this framework, blockchain ensures local models are verified before global aggregation. To reduce verification overhead, block intervals are set as multiples of the local training period, requiring BSs to complete several training rounds before uploading models to the MBS.

\subsubsection{DT edge association}
In the proposed scheme, the mapping of end devices or users to DTs in BSs is a critical task, requiring significant computing and communication resources to maintain real-time data synchronization and model construction. Given the limited computation and communication resources, optimizing the allocation of these resources is essential to enhance resource utilization. Consequently, associating various physical devices with different BSs based on their computation capabilities and the communication channel state becomes a pivotal problem to address. As illustrated in Fig. \ref{fig:DTWN}, DTs are established and managed by their respective BSs. The distribution of training data and computation tasks for training occurs across different BSs, guided by the association between DTs and BSs, referred to as edge association. 

\begin{figure}[htb]
	\centering
	\includegraphics[scale=0.27]{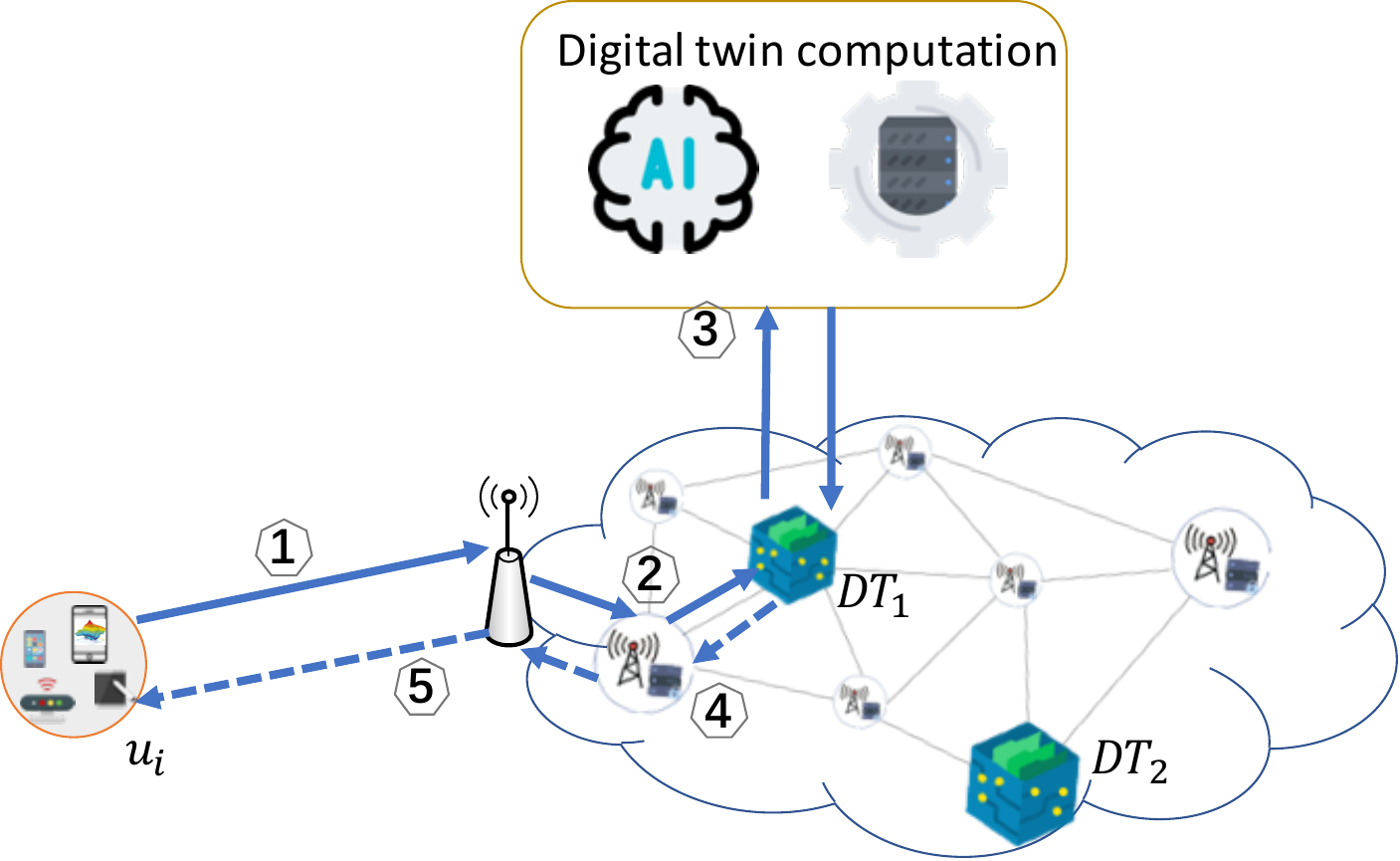}
	\caption{The processes of constructing a digital twin}
	\label{fig:latency}
\end{figure}

The process of edge association is determined by the client dataset $D_i$, BS computation capacity $f_j$, and transmission rate $R_{i,j}$ between client $u_i$ and BS $j$, modeled as $\Phi(i,j) = f(D_i, f_j, R_{i,j})$. The entire processes of constructing DTs are illustrated in Fig. \ref{fig:latency}, including local training, data transmission, and model aggregation.

The training time associated with the local training of BS $i$ hinges on the computing capability and data size of its DTs:
\begin{equation}
	\label{eq:local_training}
	T_i^{cmp} = \frac{\sum_{j=1}^{K_i} b_j D_{DT_j}}{f_i^C}f^C,
\end{equation}
where $f^C$ is the number of CPU cycles per data sample, $f_i^C$ is the CPU frequency of BS $i$, and $b_j$ is the batch size of $DT_j$. The process of aggregating local models from diverse DTs is executed by BSs. The aggregation time is:
\begin{equation}
	\label{eq:local_agg}
	T_i^{la} = \frac{\sum_{j=1}^{K_i} |w_j|}{f_i^C}f_b^C,
\end{equation}
where $|w_j|$ is the size of each local model, typically equal to the global model size $|w_g|$. After aggregation, BS $i$ broadcasts its model as transactions. The broadcast time, impacted by the BS network size $M$ and upload rate $R_i^U$, is:
\begin{equation}
	\label{eq:bs_broadcast}
	T_i^{pt} = \xi \log_2 M \frac{K_i |w_g|}{R_i^U}.
\end{equation}
Block validation is performed by elected producer BSs that collect transactions, form blocks, and verify them. The validation time is given by:
\begin{equation}
	\label{eq:block}
	T_{bp}^{bv} = \xi \log_2 M_p \frac{S_B}{R_i^D} + \max_i \frac{S_B f^v}{f_is},
\end{equation}
where $M_p$ is the number of block producers, and $S_B$ is the size of a block.

Given the small size of $|w_g|$ and the high processing power of BSs, the aggregation time $T_i^{la}$ is negligible. Thus, the total time per iteration is
\begin{equation}
	\begin{aligned}
		\label{eq:time}
		T &= \max_i\left\{\frac{\sum_{j=1}^{K_i} b_j D_{DT_j}}{f_i^C}f^C \right \} + \max_i\left\{ \xi \log_2 M \frac{K_i |w_g|}{R_i^U}\right \} \\ & +  \xi \log_2 M_p \frac{S_B}{R_i^D} + \max_i \frac{S_B f^v}{f_is}.
	\end{aligned}
\end{equation}
The goal of the DT edge association between railway clients and their DTs is to find the trade-off between learning accuracy and time cost of the learning process by leveraging integrated techniques, including optimization and emerging learning algorithms. The optimized edge association mechanism aims to ensure the efficiency and accuracy of DT systems.

The proposed frameworks play a crucial role in enhancing wireless communication quality. For example, these frameworks enable decentralized, secure, and efficient training of channel estimation models by aggregating local data while preserving user privacy. Additionally, DTs contribute significantly by simulating dynamic network environments in HSR scenarios, facilitating the development of more accurate and adaptive channel estimation methods.

\subsection {Lessons Learned}
The integration of intelligence and security technologies in 6G railways underscores the transformative synergy between AI-driven analytics and robust security measures, boosting efficiency, reliability, and resilience in high-speed rail operations. MEC, paired with AI, mitigates latency and computational challenges, enabling real-time decision-making and resource management in dynamic railway environments. Advanced technologies like blockchain, federated learning, and DTs further enhance secure spectrum sharing, predictive maintenance, and intelligent system optimization. However, challenges such as synchronization errors, resource limitations, and data security issues still remain. Future smart railway systems will require endogenous security architectures with zero-trust frameworks and proactive defenses to address evolving threats, ensuring adaptive, efficient, and highly-secure operations.

\section{Internet of Things Technologies}
\subsection{Massive Machine Communication for Smart HSR}


In recent years, the rapid development of IoT has led to an intensive attention to instant acquisition, analysis, and exchange of massive amounts of data, which puts forward new demands for next-generation wireless cellular networks. The next-generation wireless communication cellular networks must provide massive connections for massive machine-type communication (mMTC) to support various applications, such as environmental sensing, event detection, and monitoring. For instance, rail-side equipment can continuously monitor the condition of railway infrastructure over wide areas, reducing maintenance time, accurately identifying faults, and improving efficiency compared to traditional manual inspections.

Compared with traditional human-centered communications, mMTC has different characteristics. Firstly, the whole system needs to support the access of massive devices, ranging from $10^6$ to $10^7$ devices per square kilometer. Secondly, the access of devices is sporadic. For instance, rail-side cameras used to monitor the surrounding environment for railway safety only access the network and report to the command center if the surrounding environment is abnormal, and thus only a small percentage of devices have data to send at any given time. Thirdly, the transmitted packets are small, typically control messages of only several bits.

Yet, the traditional grant-based access scheme in 5G and earlier fails to support the increasing number of devices. Besides, the complex control signaling leads to high latency and more communication overheads and is inefficient for the transmissions of small packets. At the same time, it also causes high latency. To the end, the grant-free random access protocol that allows massive IoT devices to access wireless networks without a grant has become an attractive option, where active devices directly transmit their pilot and data to BS without the grant. The directional transmission raises the new challenge of activity detection at BS. Accurate user activity detection and channel estimation are the keys to successfully establishing communications between a device and a BS. To identify active users and estimate channels, each user can be assigned a unique pilot sequence. However, it is unwise to assign orthogonal pilot sequences as traditional access schemes due to limited resources \cite{8264818}. While the non-orthogonal pilots further increase the difficulty of user activity detection and channel estimation, the sporadic transmission provides the potential solution by constructing the user activity detection and channel estimation as a compressed sensing (CS) problem. As only a small number of users are active, the activity of all users can be represented as a sparse signal. Therefore, CS theory and related algorithms can be used to reconstruct this sparse signal to detect active users and estimate their channels. Solving this CS problem can alleviate the multi-user interference problem and achieve highly reliable activity detection and accurate channel estimation.

AI has also been used in mMTC scenarios to solve the key sparse signal recovery problem for massive random access. The AI-based methods can learn from data where specific structures are embedded, so they can better solve the sparse recovery problem. For the rail-side equipment, as their positions are fixed, the static channel could be learned by AI-based methods for better user activity detection and channel estimation. For the on-board equipment in the smart railway scenario, a train is running at a high speed and the surrounding environment is also changing rapidly. When the train enters the range of BS, the increase of on-board equipment within the range of BS leads to the increase of active users. The research shows that the process follows bimodal distribution. The AI-based methods can prevent the algorithm from falling into local optimality by learning such structure so as to avoid structural errors. Meanwhile, AI-based methods can also improve the convergence speed of the algorithm. Therefore, the AI-based methods can better use the bimodal distribution feature to solve the problem.

 Currently, grant-free random access receives considerable attention. In \cite{8264818}, the authors use sparse recovery algorithms like approximate message passing (AMP) and sparse Bayesian learning for activity detection and channel estimation.
 The smart railway scenario has many characteristics that differ from the general public network scenario, which can be utilized in AI-based methods to improve the spectrum efficiency and achieve higher accuracy and robustness. In AI-based methods, various networks are proposed to utilize different characteristics to enhance the performance of active user detection and data recovery \cite{9903376}. In \cite{9252937}, Zhang et al. propose a learned iterative soft-thresholding network (LISTA) with adaptive depth to fit the cases of a flexible number of active users. In \cite{9605579}, the checking information in the data recovery phase is utilized as the prior information and a modified LISTA using this information is proposed to improve the channel estimation and detection performance.

\begin{figure}[t]
	\centering
	\includegraphics[scale=0.28]{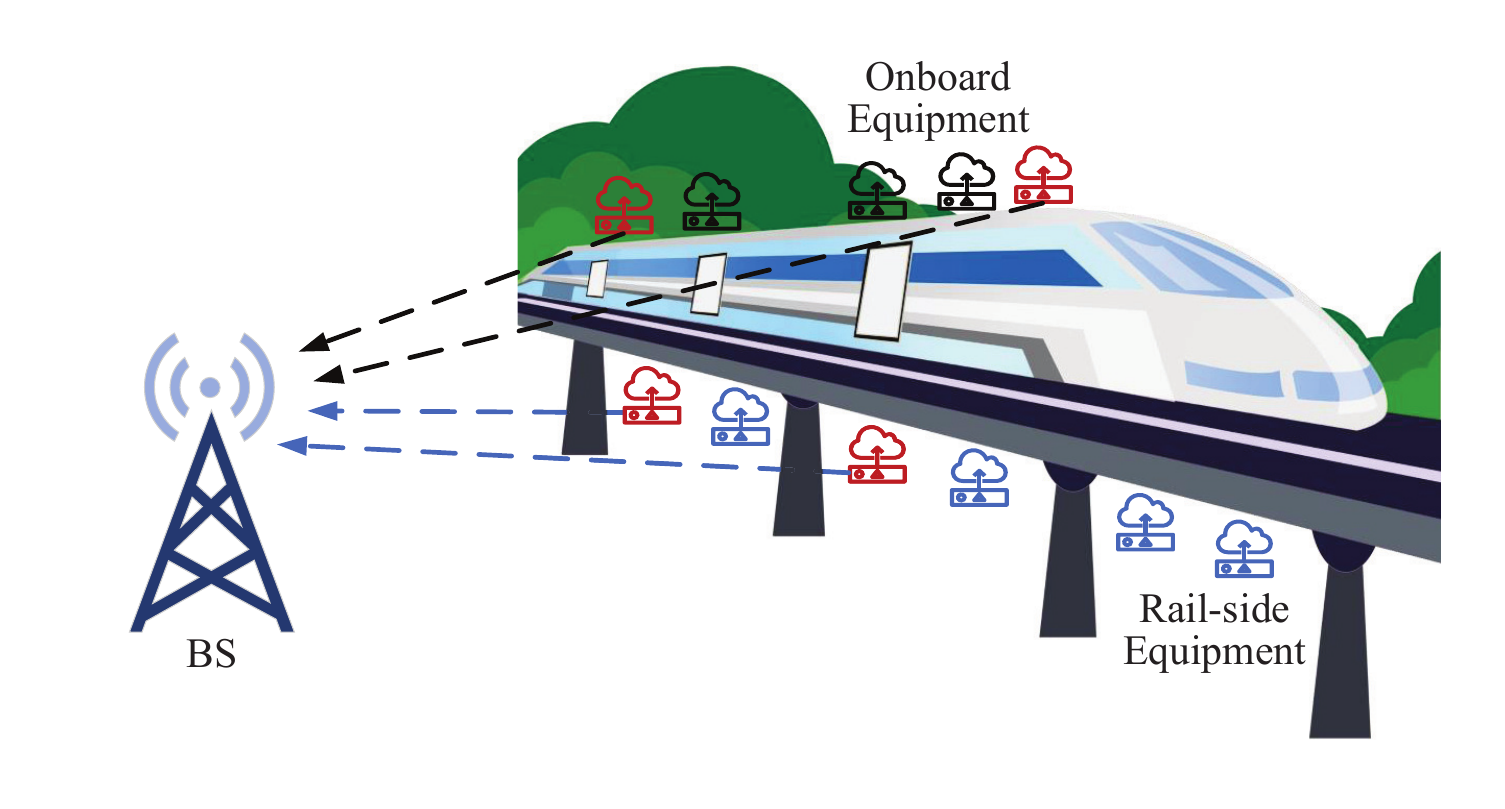}
	\caption{Railway access scenario when a train passing by.}
	\label{fig23}
\end{figure}

The smart railway scenario has many characteristics that are different from the general mMTC scenario.  In the general mMTC scenario, the number of active users is relatively stable. In the smart railway scenario, the potential access users include rail-side equipment and onboard equipment, as shown in Fig. \ref{fig23}. Changes in the number of potential access users caused by passing trains lead to dynamic changes in the number of active users. Besides, most related papers assume active users transmit small packets of the same size. Yet, in the smart railway scenario, various rail-side equipment is used to collect the maintenance status of trains to ensure safety and efficiency. Generally, the transmitted packets of different rail-side equipment could be different, which leads to data length diversity. These special characteristics of the smart railway scenario could be exploited to enhance massive wireless access in the smart railway scenario.

\subsubsection{Dynamic changes in the number of active users}
We consider the railway-dedicated mobile communication system where the number of rail-side equipment and onboard equipment are $N_r$ and $N_o$, respectively. In each coherence time, when the train passes by, the number of potential access users is $N_r+N_o$. With the same probability of being active, the number of active users will have two peaks depending on whether there is a train passing by, so it follows a bimodal distribution.

Due to the large number of potential access users, we consider a contention-based CS-based access scheme, where each potential access user randomly selects a pilot from the pilot pool. 
In each coherence time, active users send their chosen pilots to BS, which performs the pilot detection and estimates the corresponding channel. Then, active user detection is achieved by sending the user's ID in the data. All transmitted packets are assumed to be synchronized at BS. We adopt the block-fading channel model, wherein each coherence time, the channel follows independent quasi-static flat-fading.

We define $h_k$ as the channel gain corresponding to $\boldsymbol p_k$. Then, the received pilot signal $\boldsymbol y_p \in \mathbb {C}^{l_p}$ at BS is given by
\begin{equation}
	\label{equation1}
	\boldsymbol y_p = \sum_{k \in \mathcal K} h_k\boldsymbol p_k + \boldsymbol n_p = \boldsymbol {Ph} + \boldsymbol n_p,
\end{equation}
where $\mathcal K \subset \{1,2,..., N\}$, $\boldsymbol h \in \mathbb {C}^{N}$ and $\boldsymbol n_p \in \mathbb {C}^{l_p}$ denote the index set of the selected pilots, the channel coefficients corresponding to all pilots and the additive white Gaussian noise, respectively. When a pilot is not selected, its channel coefficient is 0, which makes $\boldsymbol h$ a sparse vector. The sparse linear inverse problems to obtain $\boldsymbol h$ can be formulated as 
\begin{equation}
	\label{equation_cvx}
	\min_{\boldsymbol h} \frac{1}{2}{\Vert \boldsymbol y_p - \boldsymbol {Ph} \Vert}_2^2 + \lambda{\Vert \boldsymbol h \Vert}_1,
\end{equation}
where $\lambda>0$ is the regularization coefficient. 

Since AI-based methods are data-driven, we consider the characteristics of data structures in the smart railway scenario. When the number of active users follows a bimodal distribution caused by passing trains, the deep unfolding network learned AMP (LAMP) \cite{7934066} can be used to learn the structural properties of the data, which in turn improves the active user detection and channel estimation in the smart railway scenario.

\subsubsection{Data length diversity}
The business of smart railway has different service requirements, which leads to data length diversity, e.g., the devices for intrusion detection only transmit 1 bit to indicate whether there is foreign body intrusion, while the devices for train condition monitoring transmit more bits to provide feedback on the condition of the train. Since most rail-side equipment and onboard equipment transmit small packets, we consider a grant-free CS-based access scheme to reduce the latency caused by signaling interactions between BS and users. Without loss of generality, the encoded data symbols and pilot symbols of user $k$ are spread by user-specific spreading sequence $s_k$. Then, the received signal to the receiver is
\begin{equation}
	\label{equation8}
	\mathbf Y = \sum_{k=1}^N h_k\mathbf s_k \mathbf x_k^\mathsf{T} + \mathbf N = \mathbf {SHV + N} = \mathbf {SX + N},
\end{equation}
where $\mathbf S $ and $\mathbf H$ denote the spread spectrum matrix and channel gains of $N$ potential users, respectively.

Since different devices transmit data of different lengths, the number of non-zero elements in the last column of data is the smallest and gradually increases forward. Considering that the smaller the sparsity, the better the estimation performance, BS can estimate the sparsity of $\mathbf X$ in a backward manner. After obtaining the estimated sparsity, BS can circularly perform activity detection, channel estimation, data recovery, and channel re-estimation utilizing the joint sparsity of pilot and data symbols, the error checking information and the modulation constellation information to improve the performance of joint active user detection, channel estimation, and data recovery. Exploiting data length diversity, the backward sparsity adaptive matching pursuit with checking and projecting (BSAMP-CP) algorithm is proposed in \cite{9268113}. A modified LAMP that incorporates a backward propagation algorithm is introduced \cite{10304065}, which exploits the three-level sparsity resulting from sporadic activity, symbol delays, and data length diversity.


\begin{figure}[t]
	\centering
	\includegraphics[scale=0.32]{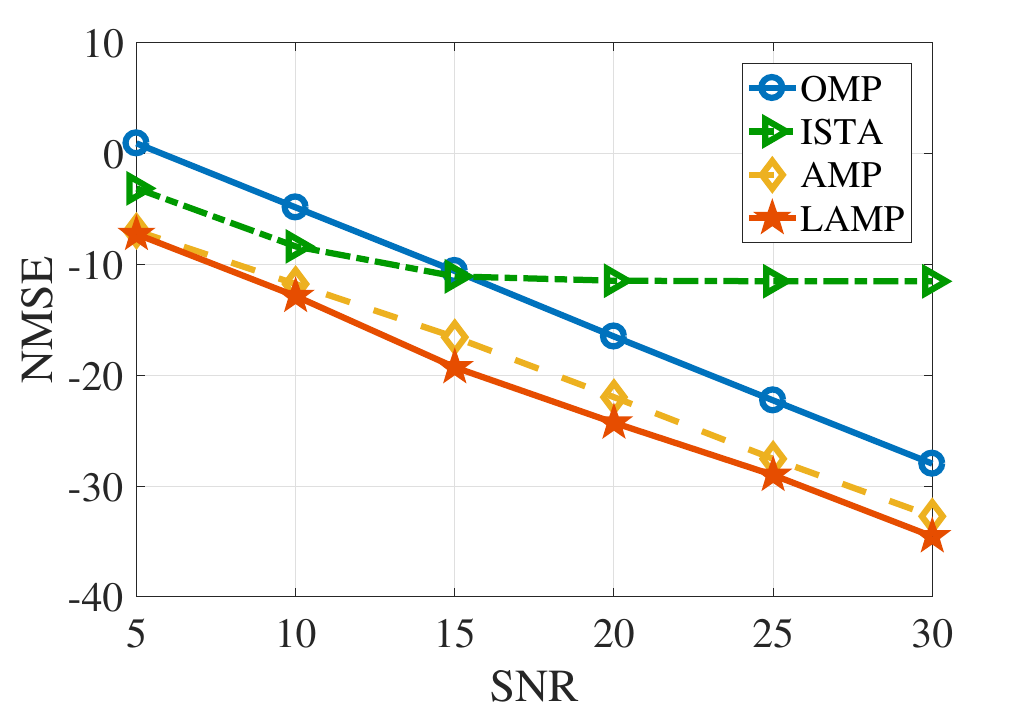}
	\caption{Comparison of channel estimation performance.}
	\label{fig25}
\end{figure}

\begin{figure}[t]
	\centering
	\includegraphics[scale=0.32]{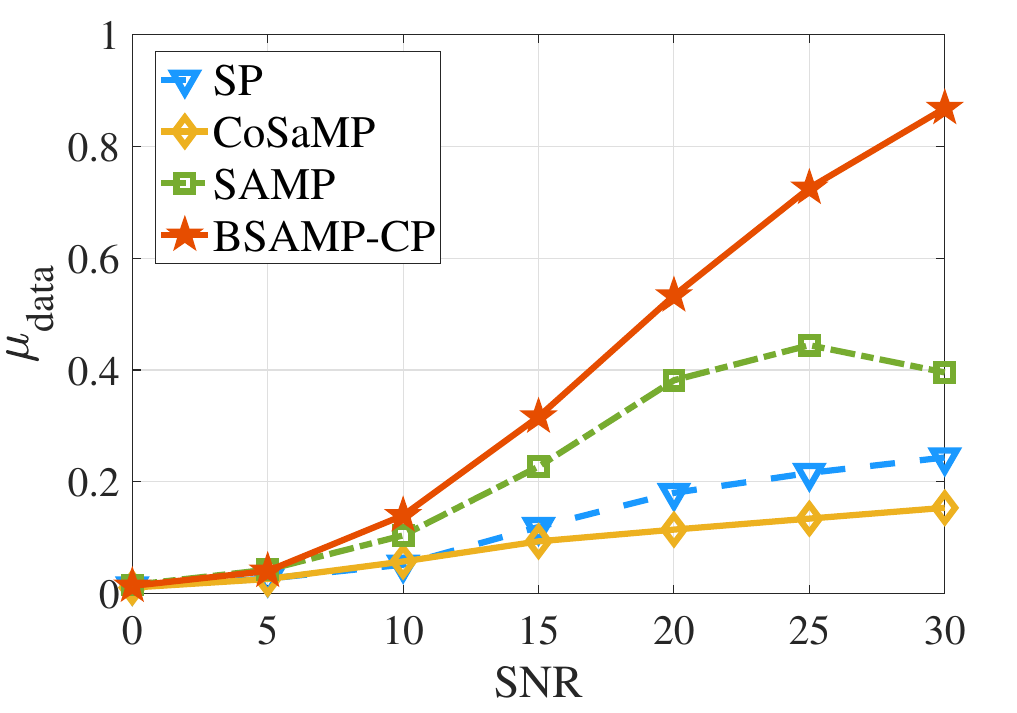}
	\caption{Comparison of data recovery ratio for varying SNRs.}
	\label{fig26}
\end{figure}

For the smart railway access scenario where the number of active users changes, we consider AI-enabled methods to solve the channel estimation problem. In the experiment, LAMP is trained with the sparsity following a bimodal distribution. The orthogonal matching pursuit (OMP) \cite{342465}, ISTA, and AMP are selected as the comparison solutions. We compare the channel estimation NMSE with the test set that follows the same distribution as the training set. 

Fig. \ref{fig25} shows the channel estimation performance under different SNRs. The performance of all methods improves with the increase of SNR, while the performance of ISTA gradually stabilizes. LAMP has obvious advantages over other traditional methods. Once the training process of the network is completed, LAMP does not need to iterate many times to solve the problem. Therefore, LAMP is more suitable for the smart railway scenario where the number of active users changes dynamically. In conclusion, unlike traditional methods, AI-enabled methods can effectively handle complex environments with interference and fading, delivering more robust and scalable solutions for next-generation networks.

Considering data length diversity in the smart railway scenarios, we compare the performance of SP, compressive sampling matching pursuit (CoSaMP), sparsity adaptive matching pursuit (SAMP), and BSAMP-CP. Fig. \ref{fig26} shows the performance of data recovery for varying SNRs. The data recovery ratio is defined as $\mu_{data} = \dfrac{\text {card}(\mathcal V)}{K}$, where $\mathcal V$ denotes the set of devices whose data are correctly recovered and $K$ denotes the number of active users. It can be seen that the data recovery performance of the proposed BSAMP-CP is superior to all other methods. The data recovery ratio of BSAMP-CP can even reach twice of that of SAMP when SNR is 30 dB. Therefore, BSAMP-CP is more suitable for recovering data of different sizes in smart railways.

\subsection{Integrated Sensing and Communications} 
The 6G network marks a shift from traditional communication systems to a unified infrastructure that integrates communication, sensing, computing, and storage. Unlike 5G, 6G aims to support advanced sensing modalities—including touch, taste, and even telepathy—while demanding ultra-low latency, ultra-high reliability, massive connectivity, and wide bandwidth. Traditional designs that treat sensing, communication, and computing separately are insufficient for emerging 6G applications like autonomous driving and environmental monitoring. There is a pressing need for an integrated approach that combines communication, computing, storage, and intelligence to effectively address these requirements \cite{9687468}. The network of integrated sensing, communication, and computing is referred to as the network that possesses the capabilities of physical-digital space perception, ubiquitous intelligent communication, and computing \cite{IMT2030}. In the realm of a perception-computing integrated network, the synergy and sharing of sensing, computing, and communication resources enable the seamless integration of multi-dimensional sensing, collaborative communication, and intelligent computing functions. This integrated approach holds significant potential for applications such as smart cities and smart transportation. Notably, the safety of high-speed trains necessitates the presence of both a highly reliable, low-latency communication network and a comprehensive, high-precision sensing network. Leveraging wireless communication networks and infrastructure, with integrated communication, sensing, computing, and storage as core technologies, and powered by edge AI, a comprehensive information collection and transmission system is built. Combined with cloud processing, this system enables seamless and holographic sensing for 6G rail transportation, supporting intelligent high-speed rail operations across all scenarios, frequency bands, and full coverage.

\begin{figure}[htb]
	\centering
	\includegraphics[scale=0.4]{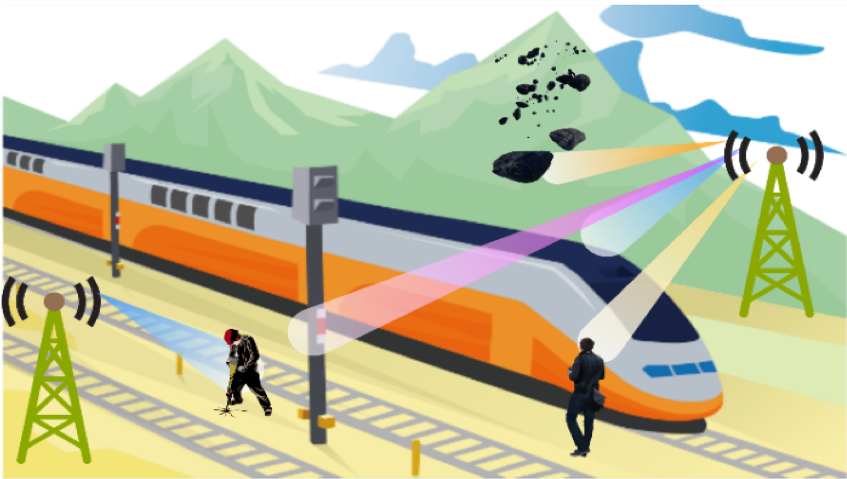}
	\caption{Integrated Sensing and Communication}
	\label{fig:fig27}
\end{figure}

The illustration of ISAC scenarios is shown in Fig. \ref{fig:fig27}, where sensing of rock rolling and human approaching is conducted while normal communications are carried out. In complex intelligent transportation scenarios, the information processing flow of communication and sensing services exhibits highly coupled characteristics in terms of temporal and spatial domains, functional impacts, and reliance on the large broadband spectrum and large aperture antennas. Accurate channel estimation further enhances ISAC in 6G railways by optimizing beamforming, positioning, and real-time tracking, ensuring efficient train-to-ground and train-to-train communication. It also improves sensing tasks such as obstacle detection and infrastructure monitoring by reducing interference and ensuring reliable signal processing. These ISAC solutions facilitate cooperation between sensing and communication systems, delivering high-reliability, low-latency communication and large-scale, high-precision sensing, while reducing hardware costs, minimizing spectrum usage, and improving energy efficiency for future intelligent high-speed railways.

However, the integrated design of sensing, communications, computing, storage, and intelligence (SCCSI) in railway scenarios still has three key challenges:

\subsubsection {Integrated Design of Sensing, Communication, Computing, and Intelligence for HSR Wireless Communications} The evolution of wireless communication demands a focus on efficient train-to-ground wireless communication, T2T wireless communication, and environmentally sustainable, low-carbon wireless communication and networks. In scenarios where a transceiver is in motion, continuous reconfiguration of communication link (i.e., beam direction) is essential to uphold optimal alignment and ensure the requisite throughput. 
Furthermore, a significant challenge lies in maintaining high communication throughput while concurrently engaging in synchronous environment sensing, precise positioning, and trajectory tracking \cite{9829746}. The potential for enhanced communication performance in the 6G ISAC system is emphasized in \cite{9799524}, suggesting improvements like more precise beamforming, quicker beam fault recovery, and reduced tracking channel state information overhead.

\subsubsection {Use Cases and Performance Indicators of Communication-Sensing-Computing Integration} 
Currently, communication, sensing, and computing exist largely in separate domains. While some research focuses on integrating communication and sensing, there is a need to develop new theories that unify all three into a cohesive system. Understanding the performance boundaries and trade-offs is crucial for such integrated systems. Unlike traditional wireless communication systems, 6G rail transportation requires maximizing the reuse of existing design patterns and modules, including hardware, algorithms, and protocols. A communication-centric approach is essential to address the unique needs of rail transportation and the limitations of wireless signal processing. This allows for the seamless integration of innovative waveform designs, new architectures, and advanced signal processing into wireless communication systems. Establishing a trade-off mechanism enables the optimization of communication and sensing performance, bridging the gap between the two domains. Additionally, rail transit environment monitoring demands near-real-time detection, requiring a comprehensive understanding of all elements and states. Challenges include limited coverage, network interruptions, and changes in topology and deployment during design and construction stages. Mu at al. \cite{9492131} proposed an integrating sensing and communication for precise beam tracking in vehicular networks, which enhances the estimation performance and maintains reliable communication in highly mobile scenarios. In intelligent reflecting surface-assisted ISAC systems, the limited signal processing capability of the passive IRS, combined with mutual interference between sensing and communication (SAC) signals, hampers effective channel estimation. To address this challenge, Liu et al. \cite{9997576} proposed a deep learning-based, three-stage approach to decouple the estimation problem. Two different convolutional neural network and two types of input-output pairs are designed to train the models and estimate the ISAC channels.

\subsubsection{Joint Optimization of Sensing, Communication and Computing Resources for Intelligent Railway} Current research on perception performance seldom considers the specific perception process but pays more attention to the optimization of communication and computing resources \cite{chen2023vehicle}. However, different types of tasks often require support for sensing, communication, and computing functions simultaneously, resulting in complex relationships such as coupling, collaboration, and even competition among the three modules. In \cite{9970330}, a specific case of human motion recognition based on wireless sensing is studied, and the joint communication-sensing computation resource allocation problem of FEEL is addressed. 
Li et al. \cite{10014666} attempts to combine over-the-air-computing with ISAC technology to improve sensing and transmission efficiency.
 
\subsubsection{Security and Privacy for the Integration of Intelligent Railway Communication, Sensing and Computing} The proliferation of sensors in the upcoming rail transit system is anticipated to be substantial. Smart railways will harness these sensor networks to gather information about train positions, the surrounding environment, and the railway infrastructure, thereby guaranteeing the secure operations of trains. However, the sensing and communication infrastructure is susceptible to numerous threats that can compromise the system's availability, integrity, and confidentiality. Moreover, the integrated design of sensing and communication may inadvertently facilitate eavesdropping attacks, as highlighted in \cite{9829746}.

To address these challenges, several strategies can be considered. First, railway scenarios have unique characteristics that require optimizing beam management with integrated sensing, reducing signaling overhead, and improving spectral efficiency and response speed. Research into AI-based, RIS-assisted communication, sensing, and computing is essential. By leveraging data from communication, sensing, and computation, and applying scene-specific algorithms, we can build a rail transportation database using big data techniques. This will overcome communication bottlenecks and improve wireless efficiency and the intelligence of high-speed rail systems. Furthermore, perception, computing, and communication are closely linked, especially with edge AI. Railway systems should deploy task-driven, integrated computing, storage, and intelligence using distributed edge AI to optimize resources. Performance metrics such as delay, energy efficiency, and classification accuracy can guide the optimization of latency under given accuracy requirements, for example, through an ISCC scheme. Lastly, security and privacy are critical. Data protection requires low-complexity, effective defense mechanisms, combining privacy technologies like zero-knowledge proof, federated learning, and secure multi-party computation.

\subsection {Lessons Learned}
The integration of IoT technologies in 6G smart railways underscores their critical role in enhancing operational efficiency, safety, and reliability. mMTC enables real-time monitoring and predictive maintenance, addressing challenges like sporadic device activity, small data packets, and high device density. Advanced solutions such as grant-free random access and AI-driven sparse signal recovery improve user activity detection, channel estimation, and resource utilization. Dynamic user activity and data length diversity in railway scenarios are effectively managed with innovative algorithms such as deep unfolding networks and backward sparsity adaptive matching pursuit, optimizing data recovery and allocation. Despite these advancements, challenges like synchronization errors, limited spectrum, and data security persist, highlighting the need for AI-powered optimization, efficient resource management, and robust privacy measures to ensure secure and intelligent IoT systems for next-generation railways.

\section{Future Research and Development}
\subsection{Cell-Free Network Architectures}
To address issues related to ultra-high mobility and reduce overhead in 6G Smart railways, cell-free networks can be developed and deployed. Distributed massive MIMO can improve coverage and reduce base station switching, leading to a lower probability of switching failure and link interruption. However, a fast-switching solution is still necessary to provide ultra-reliable and low-latency communication in high-speed rail scenarios.

Cell-free massive MIMO is a new framework that eliminates cell edges and is ideal for high-speed rail communication. While there have been studies on large-scale MIMO-OFDM in cellular environments, they overlook the issue of subcarrier interference and are based on static scenarios. Hence, there is a need to deploy massive MIMO-OFDM systems in high-speed rail communication.

A scalable cell-free massive MIMO-OFDM system can be designed based on the idea of dynamic cooperation clustering centered on vehicle antennas. Different transmission powers can be utilized to eliminate interference and compensate for performance loss due to high-speed movement. Various power control algorithms such as heuristic fractional power control algorithm, Max Min power control algorithm, and Max Sum rate control algorithm can be used according to different target requirements.

\subsection{Channel Modeling and Characteristics for All Frequency Bands and All Scenarios}
The vision of ubiquitous connection in the 6G era will require the use of a wide range of frequencies. Therefore, there is an urgent need for channel modeling and measurement for all frequency bands and all scenarios \cite{9237116} to ensure optimal communication performance. This includes the measurement, characterization, and modeling of cross-domain channels for emerging railway communication techniques such as THz communication, IoT, and ISAC. However, the challenges of sparse data and difficult registration under dynamic conditions need to be carefully addressed during this process. Additionally, beam alignment, beam adjustment, and quick link recovery are crucial challenges that must be addressed in THz communication.

AI-based methods are essential for cross-domain channel modeling, particularly for real-time channel clustering and tracking, communication scenario recognition, channel prediction, and final channel modeling. To achieve this, AI models such as convolutional neural networks, graph neural networks, and recurrent neural networks can be used for feature extraction, scene recognition, and channel modeling. Moreover, distributed and cooperative AI techniques are expected to play a crucial role in future 6G railway communications, with reduced resource requirements and improved learning performance.

\subsection{Endogenous Security for 6G Smart Railways}
The 6G network for railways will undergo comprehensive upgrades, shifting from a focus on Internet security to a focus on cyberspace security, with the scope of network security continuing to expand. The intelligent rail transit 6G network will integrate defense mechanisms at the system design stage to enhance its inherent ``immunity''. It will utilize AI to monitor system status in real-time and better assess potential security risks, combining resistance to malicious intrusion attacks with prediction of danger, thereby achieving intelligent endogenous security. 

Robust endogenous security architectures are critical to ensure the secure and reliable operation of railway systems in the smart railway. The endogenous security framework will promote engineering tasks such as next-generation identity security, reconstruction of network depth security, pan-terminal and IoT security, cloud data center security construction, and practical safe operation, thus integrating with the informatization development of the railway industry. 

To enhance the security, privacy, and reliability of railway systems, emerging technologies such as blockchain, private computing, and distributed collaboration can be integrated with railway communication architectures. In addition, the integration methods, collaborative mechanisms, and resource allocation strategies need to be explored to ensure the successful implementation of these technologies in the railway industry.

\subsection{Connected Cooperative Edge Intelligence}
With the vision of ``ubiquitous intelligent connection" in 6G railway networks, new techniques such as federated learning, distributed machine learning, and cooperative AI are needed to achieve connected edge intelligence. Conventional edge intelligence techniques, such as MEC, face challenges due to the large number of connected devices. 

However, efficient learning with limited computation and communication resources is a significant challenge that needs to be addressed. Two main solutions can be applied to tackle this problem. Firstly, new edge intelligence techniques and mechanisms can be developed to perform computation tasks in parallel with reduced resource consumption. Secondly, more effective resource allocation and utilization strategies are required to improve the utility of limited resources. Both aspects can be jointly considered in practical applications. For instance, emerging computing power networks can play an essential role in achieving connected edge intelligence by jointly utilizing terminal computing resources.

Efficient resource allocation and utilization are critical for connected edge intelligence in railway systems. The development of new techniques, such as network slicing and adaptive allocation, can effectively reduce resource consumption and improve the utilization of limited resources. Additionally, the joint optimization of spectrum resource allocation and computation offloading can further enhance the efficiency of edge intelligence. The integration of these techniques will enable the efficient and intelligent operation of railway systems.

\subsection{New URLLC Techniques}
To enable the safe and efficient operation of automatic trains in emerging super high-speed rail systems, reaching speeds over 1000 kilometers per hour, further research is required on the new URLLC frame structure under limited block code length. This will help achieve a balance between ultra-reliability and ultra-low latency, as they are critical for the successful functioning of such systems \cite{Youxiaohu3}.

Moreover, to support a wide range of latency and resource-sensitive applications, distributed AI, deep reinforcement learning, and optimization can be utilized to sense communication conditions and optimize communication system operations. These techniques can help achieve URLLC performance and provide a high-quality service experience.

\subsection{MEC Empowered Digital Twin Networks}
Real-time mapping analysis of railway networks, including sensing and processing of train operation status, monitoring, prediction, and decision-making on the states of train operation and track conditions, is of great importance for the safe operation of high-speed trains. By achieving real-time perception and close monitoring of the system and drivers through optimization, predictive analysis, fault-tolerant control, and automated processes, the cost of operation and maintenance can be reduced, while security and reliability can be improved. 

A 6G digital twin network can fundamentally change the digital characteristics of the historical and current behaviors and states of the train and driver, enabling feedback control and optimization of their performance. However, building a railway DT network with the assistance of MEC faces several challenges. Firstly, digital twin models need to be constructed for complex and changeable rail transit systems based on limited resources in near real-time. Additionally, the simulation error between DTs and physical entities in rail transit needs to be carefully addressed. Furthermore, due to high mobility of trains, the placement and transfer strategies of DT models with ultra-low latency need to be carefully considered. 

To address these challenges, it is important to explore methods for quickly and conveniently collecting massive sensing data through synchronization at a low cost. This will greatly improve the efficiency of optimal iterations for intelligent train automatic operation. Additionally, advanced distributed AI techniques can be used to efficiently construct DT networks and optimize network operations, enabling all-around network situational awareness.

\subsection{Generative AI Applications}

Generative AI can be leveraged to create realistic and diverse channel models for high-mobility environments, enhancing the robustness of channel estimation algorithms \cite{10490142}. It can also augment training datasets for machine learning-based 6G applications, addressing data scarcity and improving model generalization \cite{10398474}. By learning the normal patterns of network traffic and system dynamics, generative AI aids in detecting anomalies, bolstering network reliability and security. Furthermore, it supports lightweight model generation and compression for edge devices, enabling efficient deployment in resource-constrained environments. However, challenges such as computational complexity, privacy concerns, and the need for high-quality data must be carefully addressed to ensure successful integration.

\section{Conclusion}

This article provides a comprehensive overview of key 6G technologies for future smart railways. Firstly, we have presented an in-depth investigation into the development progress of 5G-R networks and discussed the benefits and drawbacks of 5G communication in HSR scenarios. Then, we have analyzed the business requirements and application challenges of 6G railway communication networks and provided the visions for 6G smart railways. Moreover, we have proposed an integrated network architecture that jointly considers factors such as wireless coverage, storage, computing, and communication. We have thoroughly reviewed the possible key enabling technologies of 6G railway communications and provided their application scenarios and deployment methods, correspondingly. Finally, we have identified several open research issues and promising directions. It is expected that with the application of new 6G networks and technologies, network coverage, broadcasting, communication, interaction, and safety will be greatly improved, thereby enhancing the travel experience and operation efficiency of future railway systems.
\ifCLASSOPTIONcaptionsoff
  \newpage
\fi
\bibliography{IEEEabrv, ref.bib}

\begin{thebibliography}{100}
\providecommand{\url}[1]{#1}
\csname url@samestyle\endcsname
\providecommand{\newblock}{\relax}
\providecommand{\bibinfo}[2]{#2}
\providecommand{\BIBentrySTDinterwordspacing}{\spaceskip=0pt\relax}
\providecommand{\BIBentryALTinterwordstretchfactor}{4}
\providecommand{\BIBentryALTinterwordspacing}{\spaceskip=\fontdimen2\font plus
\BIBentryALTinterwordstretchfactor\fontdimen3\font minus
  \fontdimen4\font\relax}
\providecommand{\BIBforeignlanguage}[2]{{%
\expandafter\ifx\csname l@#1\endcsname\relax
\typeout{** WARNING: IEEEtran.bst: No hyphenation pattern has been}%
\typeout{** loaded for the language `#1'. Using the pattern for}%
\typeout{** the default language instead.}%
\else
\language=\csname l@#1\endcsname
\fi
#2}}
\providecommand{\BIBdecl}{\relax}
\BIBdecl

\bibitem{9103348}
B.~Ai, A.~F. Molisch, M.~Rupp, and Z.-D. Zhong, ``5g key technologies for smart
  railways,'' \emph{Proceedings of the IEEE}, vol. 108, no.~6, pp. 856--893,
  2020.

\bibitem{3gpprailway}
``Use of 3gpp technologies by railways,''
  \url{https://www.3gpp.org/technologies/railways1}, accessed: 2025-03-25.

\bibitem{etsirailway}
``Rail communications (rt),''
  \url{https://www.etsi.org/technologies/rail-communications}, accessed:
  2025-03-25.

\bibitem{9454557}
J.~M. Pereira, ``5g for connected and automated mobility (cam) in europe:
  Targeting cross-border corridors,'' \emph{IEEE Network}, vol.~35, no.~3, pp.
  6--9, 2021.

\bibitem{10054381}
C.-X. Wang, X.~You, X.~Gao \emph{et~al.}, ``On the road to 6g: Visions,
  requirements, key technologies, and testbeds,'' \emph{IEEE Communications
  Surveys \& Tutorials}, vol.~25, no.~2, pp. 905--974, 2023.

\bibitem{8985528}
J.~Navarro-Ortiz, P.~Romero-Diaz, S.~Sendra \emph{et~al.}, ``A survey on 5g
  usage scenarios and traffic models,'' \emph{IEEE Communications Surveys \&
  Tutorials}, vol.~22, no.~2, pp. 905--929, 2020.

\bibitem{rail2050}
``Rail 2050 vision,''
  \url{https://errac.org/publications/rail-2050-vision-document/}, accessed:
  2023-05-30.

\bibitem{9482430}
M.~A. Uusitalo, M.~Ericson, B.~Richerzhagen \emph{et~al.}, ``Hexa-x the
  european 6g flagship project,'' in \emph{2021 Joint European Conference on
  Networks and Communications \& 6G Summit (EuCNC/6G Summit)}, 2021, pp.
  580--585.

\bibitem{6Genesis}
M.~Katz, M.~Matinmikko-Blue, and M.~Latva-aho, ``6genesis flagship program:
  Building the bridges towards 6g-enabled wireless smart society and
  ecosystem,'' 11 2018, pp. 1--9.

\bibitem{8766143}
Z.~Zhang, Y.~Xiao, Z.~Ma \emph{et~al.}, ``6g wireless networks: Vision,
  requirements, architecture, and key technologies,'' \emph{IEEE Vehicular
  Technology Magazine}, vol.~14, no.~3, pp. 28--41, 2019.

\bibitem{8418701}
R.~Chen, W.-X. Long, G.~Mao, and C.~Li, ``Development trends of mobile
  communication systems for railways,'' \emph{IEEE Communications Surveys and
  Tutorials}, vol.~20, no.~4, pp. 3131--3141, 2018.

\bibitem{10373408}
M.~Nold and F.~Corman, ``How will the railway look like in 2050? a survey of
  experts on technologies, challenges and opportunities for the railway
  system,'' \emph{IEEE Open Journal of Intelligent Transportation Systems},
  vol.~5, pp. 85--102, 2024.

\bibitem{9582617}
J.~Sheng, X.~Cai, Q.~Li, C.~Wu, B.~Ai, Y.~Wang, M.~Kadoch, and P.~Yu,
  ``Space-air-ground integrated network development and applications in
  high-speed railways: A survey,'' \emph{IEEE Transactions on Intelligent
  Transportation Systems}, vol.~23, no.~8, pp. 10\,066--10\,085, 2022.

\bibitem{9232926}
G.~Noh, B.~Hui, and I.~Kim, ``High speed train communications in 5g: Design
  elements to mitigate the impact of very high mobility,'' \emph{IEEE Wireless
  Communications}, vol.~27, no.~6, pp. 98--106, 2020.

\bibitem{10740531}
C.~Safitri, M.~S. Harjono, E.~S. Hasrito, and R.~Roestam, ``Comprehensive
  survey: Quality of service in railway communication using information-centric
  networking and light fidelity,'' \emph{IEEE Transactions on Intelligent
  Transportation Systems}, vol.~25, no.~12, pp. 19\,218--19\,251, 2024.

\bibitem{8026132}
O.~Jo, Y.-K. Kim, and J.~Kim, ``Internet of things for smart railway:
  Feasibility and applications,'' \emph{IEEE Internet of Things Journal},
  vol.~5, no.~2, pp. 482--490, 2018.

\bibitem{9651548}
X.~Shen, J.~Gao, W.~Wu, M.~Li, C.~Zhou, and W.~Zhuang, ``Holistic network
  virtualization and pervasive network intelligence for 6g,'' \emph{IEEE
  Communications Surveys \& Tutorials}, vol.~24, no.~1, pp. 1--30, 2022.

\bibitem{li1}
N.~Cheng, W.~Quan, W.~Shi, H.~Wu, Q.~Ye, H.~Zhou, W.~Zhuang, X.~Shen, and
  B.~Bai, ``A comprehensive simulation platform for space-air-ground integrated
  network,'' \emph{IEEE Wireless Communications}, vol.~27, no.~1, pp. 178--185,
  2020.

\bibitem{li2}
{E. Calvanese Strinati et al.}, ``6g: The next frontier: From holographic
  messaging to artificial intelligence using subterahertz and visible light
  communication,'' \emph{IEEE Vehicular Technology Magazine}, vol.~14, no.~3,
  pp. 42--50, Sep. 2019.

\bibitem{li7}
{S. Dang, O. Amin, et al.}, ``What should 6g be?'' \emph{Nature Electronics},
  vol.~3, no.~1, pp. 20--29, 2020.

\bibitem{li9}
{Y. Wang, Y. Xu, et al.}, ``Hybrid satellite-aerial-terrestrial networks in
  emergency scenarios: a survey,'' \emph{China Communications}, vol.~14, no.~7,
  pp. 1--13, 2017.

\bibitem{li10}
{G. Giambene, S. Kota, et al.}, ``Satellite-5g integration: A network
  perspective,'' \emph{IEEE Network}, vol.~32, no.~5, pp. 23--31, 2018.

\bibitem{li11}
{3GPP}, ``Solutions for nr to support non-terrestrial networks (ntn) (release
  16),'' \emph{TR 38.821 V1.0.0}, Dec. 2018.

\bibitem{li3}
{X. Wang et al.}, ``Holistic service-based architecture for space-air-ground
  integrated network for 5g-advanced and beyond,'' \emph{China Communications},
  vol.~19, no.~1, pp. 14--28, Jan. 2022.

\bibitem{li13}
{S. Chen, S. Sun, et al.}, ``System integration of terrestrial mobile
  communication and satellite communication---the trends, challenges and key
  technologies in b5g and 6g,'' \emph{China Communications}, vol.~17, no.~12,
  pp. 156--171, 2020.

\bibitem{li16}
{T. Li, H. Zhou, H. Luo, and S. Yu}, ``Service: A software defined framework
  for integrated space-terrestrial satellite communication,'' \emph{IEEE Trans.
  Mobile Comput.}, vol.~17, no.~3, pp. 703--716, Mar. 2018.

\bibitem{li17}
{S. Yao, J. Guan, Z. Yan, and K. Xu}, ``Si-stin: A smart identifier framework
  for space and terrestrial integrated network,'' \emph{IEEE Netw.}, vol.~33,
  no.~1, pp. 8--14, Feb. 2019.

\bibitem{Zhangping141}
P.~Zhang, K.~Niu, H.~Tian, G.~Nie, X.~Qin, q.~Qi, and J.~Zhang, ``Technology
  prospect of 6g mobile communications,'' \emph{Journal on Communications},
  vol.~40, no.~1, p. 141, 2019.

\bibitem{chen2023vehicle}
X.~Chen, Y.~Deng, H.~Ding, G.~Qu, H.~Zhang, P.~Li, and Y.~Fang, ``Vehicle as a
  service (vaas): Leverage vehicles to build service networks and capabilities
  for smart cities,'' \emph{arXiv preprint arXiv:2304.11397}, 2023.

\bibitem{song2016millimeter}
H.~Song, X.~Fang, and Y.~Fang, ``Millimeter-wave network architectures for
  future high-speed railway communications: Challenges and solutions,''
  \emph{IEEE Wireless Communications}, vol.~23, no.~6, pp. 114--122, 2016.

\bibitem{Youxiaohu3}
\BIBentryALTinterwordspacing
X.~You, H.~Yin, and H.~Wu, ``On 6g and wide-area iot,'' \emph{Chinese Journal
  on Internet of Things}, vol.~4, no.~1, p.~3, 2020. [Online]. Available:
  \url{http://www.infocomm-journal.com/wlw/CN/abstract/article_170019.shtml}
\BIBentrySTDinterwordspacing

\bibitem{9768113}
G.~Geraci, A.~Garcia-Rodriguez, M.~M. Azari, A.~Lozano, M.~Mezzavilla,
  S.~Chatzinotas, Y.~Chen, S.~Rangan, and M.~D. Renzo, ``What will the future
  of uav cellular communications be? a flight from 5g to 6g,'' \emph{IEEE
  Communications Surveys \& Tutorials}, vol.~24, no.~3, pp. 1304--1335, 2022.

\bibitem{9799524}
X.~Cheng, Z.~Huang, and L.~Bai, ``Channel nonstationarity and consistency for
  beyond 5g and 6g: A survey,'' \emph{IEEE Communications Surveys \&
  Tutorials}, vol.~24, no.~3, pp. 1634--1669, 2022.

\bibitem{9277535}
Y.~Yang, F.~Gao, C.~Xing, J.~An, and A.~Alkhateeb, ``Deep multimodal learning:
  Merging sensory data for massive mimo channel prediction,'' \emph{IEEE
  Journal on Selected Areas in Communications}, vol.~39, no.~7, pp. 1885--1898,
  2021.

\bibitem{yan2019machine}
L.~Yan, H.~Ding, L.~Zhang, J.~Liu, X.~Fang, Y.~Fang, M.~Xiao, and X.~Huang,
  ``Machine learning-based handovers for sub-6 ghz and mmwave integrated
  vehicular networks,'' \emph{IEEE Transactions on Wireless Communications},
  vol.~18, no.~10, pp. 4873--4885, 2019.

\bibitem{9585108}
M.~Guo and M.~C. Gursoy, ``Joint activity detection and channel estimation in
  cell-free massive mimo networks with massive connectivity,'' \emph{IEEE
  Transactions on Communications}, vol.~70, no.~1, pp. 317--331, 2022.

\bibitem{10186000}
J.~Xu, A.~Xu, L.~Chen, Y.~Chen, X.~Liang, and B.~Ai, ``Deep reinforcement
  learning for ris-aided secure mobile edge computing in industrial internet of
  things,'' \emph{IEEE Transactions on Industrial Informatics}, vol.~20, no.~2,
  pp. 2455--2464, 2024.

\bibitem{10238401}
Y.~Zhang, R.~He, B.~Ai, M.~Yang, R.~Chen, C.~Wang, Z.~Zhang, and Z.~Zhong,
  ``Generative adversarial networks based digital twin channel modeling for
  intelligent communication networks,'' \emph{China Communications}, vol.~20,
  no.~8, pp. 32--43, 2023.

\bibitem{li23}
{N. Kato et al.}, ``Optimizing space-air-ground integrated networks by
  artificial intelligence,'' \emph{IEEE Wireless Commun.}, vol.~26, no.~4, pp.
  140--147, Aug. 2019.

\bibitem{li24}
{C. Qiu, H. Yao, F. R. Yu, F. Xu, and C. Zhao}, ``Deep q-learning aided
  networking, caching, and computing resources allocation in softwaredefined
  satellite-terrestrial networks,'' \emph{IEEE Trans. Veh. Technol.}, vol.~68,
  no.~6, pp. 5871--5883, Jun. 2019.

\bibitem{li25}
{M. Jia, X. Zhang, J. Sun, X. Gu, and Q. Guo}, ``Intelligent resource
  management for satellite and terrestrial spectrum shared networking toward
  b5g,'' \emph{IEEE Wireless Commun.}, vol.~27, no.~1, pp. 54--61, Feb. 2020.

\bibitem{li26}
{Q. Liu, J. Yang, C. Zhuang, A. Barnawi, and B. A. Alzahrani}, ``Artificial
  intelligence based mobile tracking and antenna pointing in
  satelliteterrestrial network,'' \emph{IEEE Access}, vol.~7, pp.
  177\,497--177\,503, 2019.

\bibitem{li36}
{S. Moon, H. Kim, and I. Hwang}, ``Deep learning-based channel estimation and
  tracking for millimeter-wave vehicular communications,'' \emph{J. Commun.
  Netw.}, vol.~22, no.~3, pp. 177--184, June 2020.

\bibitem{li37}
{L. Liang, H. Ye, and G. Y. Li}, ``Toward intelligent vehicular networks: A
  machine learning framework,'' \emph{IEEE Internet Things J.}, vol.~6, no.~1,
  pp. 124--135, Feb. 2019.

\bibitem{li39}
{Y. Xu, L. Li, B.-H. Soong, and C. Li}, ``Fuzzy q-learning based vertical
  handoff control for vehicular heterogeneous wireless network,'' \emph{Proc.
  IEEE Int. Conf. Commun. (ICC)}, pp. 5653--5658, 2014.

\bibitem{li41}
{Y. Lu, X. Huang, K. Zhang, S. Maharjan, and Y. Zhang}, ``Low-latency federated
  learning and blockchain for edge association in digital twin empowered 6g
  networks,'' \emph{IEEE Trans. Ind. Informat.}, 2020.

\bibitem{li27}
{C. Stergiou, K. E. Psannis, B.-G. Kim, and B. Gupta}, ``Secure integration of
  iot and cloud computing,'' \emph{Future Gener. Comput. Syst.}, vol.~78,
  no.~3, pp. 964--975, Jan. 2018.

\bibitem{li52}
{H. Q. Ngo, A. Ashikhmin, H. Yang, E. G. Larsson, and T. L. Marzetta},
  ``Cell-free massive mimo versus small cells,'' \emph{IEEE Trans. Wireless
  Commun}, vol.~16, no.~3, pp. 1834--1850, Mar. 2017.

\bibitem{li54}
{S. Buzzi and C. D'Andrea}, ``Cell-free massive mimo: User-centric approach,''
  \emph{IEEE Wireless Commun. Lett.}, vol.~6, no.~6, pp. 706--709, Dec. 2017.

\bibitem{li55}
{E. Nayebi, A. Ashikhmin, T. L. Marzetta, H. Yang, and B. D. Rao}, ``Precoding
  and power optimization in cell-free massive mimo systems,'' \emph{IEEE Trans.
  Wireless Commun.}, vol.~16, no.~6, pp. 706--709, July 2017.

\bibitem{li51}
{X. Zhang, J. Wang, and H. V. Poor}, ``Statistical delay and error-rate bounded
  qos provisioning for murllc over 6g cf m-mimo mobile networks in the finite
  blocklength regime,'' \emph{IEEE J. Sel. Areas Commun.}, vol.~39, no.~3, pp.
  652--667, Mar. 2021.

\bibitem{li56}
{X. Zhang, J. Wang and H. V. Poor}, ``Statistical delay and error-rate bounded
  qos provisioning for swipt over cf m-mimo 6g mobile wireless networks using
  fbc,'' \emph{IEEE Journal of Selected Topics in Signal Processing}, vol.~15,
  no.~5, pp. 1272--1287, Aug. 2021.

\bibitem{li45}
{V. Petrov, M. Komarov, D. Moltchanov, J. M. Jornet, and Y. Koucheryavy},
  ``Interference and sinr in millimeter wave and terahertz communication
  systems with blocking and directional antennas,'' \emph{IEEE Trans. Wireless
  Commun.}, vol.~16, no.~3, pp. 1791--1808, Mar. 2017.

\bibitem{li48}
{H. Tataria, M. Shafi, A. F. Molisch, M. Dohler, H. Sj{\"o}land, and F.
  Tufvesson}, ``6g wireless systems: Vision, requirements, challenges,
  insights, and opportunities,'' \emph{Proc. IEEE}, vol. 109, no.~7, pp.
  1166--1199, July 2021.

\bibitem{li49}
{C. Chaccour, M. N. Soorki, W. Saad, M. Bennis, P. Popovski, and M. Debbah},
  ``Seven defining features of terahertz (thz) wireless systems: A fellowship
  of communication and sensing,'' \emph{IEEE Commun. Surveys Tuts.}, vol.~24,
  no.~2, pp. 967--993, 2nd Quart. 2022.

\bibitem{li50}
{C. Han and Y. Chen}, ``Propagation modeling for wireless communications in the
  terahertz band,'' \emph{IEEE Commun. Mag.}, vol.~56, no.~6, pp. 96--101, June
  2018.

\bibitem{10565301}
Z.~Cao, Y.~Qin, L.~Jia, Z.~Xie, Y.~Gao, Y.~Wang, P.~Li, and Z.~Yu, ``Railway
  intrusion detection based on machine vision: A survey, challenges, and
  perspectives,'' \emph{IEEE Transactions on Intelligent Transportation
  Systems}, vol.~25, no.~7, pp. 6427--6448, 2024.

\bibitem{li65}
{N. Neshenko, E. Bou-Harb, J. Crichigno, G. Kaddoum, and N. Ghani},
  ``Demystifying iot security: An exhaustive survey on iot vulnerabilities and
  a first empirical look on internet-scale iot exploitations,'' \emph{IEEE
  Commun. Surveys Tuts.}, vol.~21, no.~3, pp. 2702--2733, 3rd Quart. 2019.

\bibitem{li66}
{N. Xie, J. Chen, and L. Huang}, ``Physical-layer authentication using multiple
  channel-based features,'' \emph{IEEE Trans. Inf. Forensics Security},
  vol.~16, pp. 2356--2366, Jan. 2021.

\bibitem{li69}
{Y. Sun, J. Liu, J. Wang, Y. Cao, and N. Kato}, ``When machine learning meets
  privacy in 6g: A survey,'' \emph{IEEE Commun. Surveys Tuts.}, vol.~22, no.~4,
  pp. 2694--2724, 4th Quart. 2020.

\bibitem{li73}
{S. Yao, J. Guan, Y. Wu, K. Xu, and M. Xu}, ``Toward secure and lightweight
  access authentication in sagins,'' \emph{IEEE Wireless Commun.}, vol.~27,
  no.~6, pp. 75--81, Dec. 2020.

\bibitem{li74}
{T. Maksymyuka et al.}, ``Blockchain-empowered framework for decentralized
  network management in 6g,'' \emph{IEEE Commun. Mag.}, vol.~58, no.~9, pp.
  86--92, Sep. 2020.

\bibitem{9711524}
L.~U. Khan, W.~Saad, D.~Niyato, Z.~Han, and C.~S. Hong, ``Digital-twin-enabled
  6g: Vision, architectural trends, and future directions,'' \emph{IEEE
  Communications Magazine}, vol.~60, no.~1, pp. 74--80, 2022.

\bibitem{li59}
{Y. Lu, X. Huang, K. Zhang, S. Maharjan, and Y. Zhang},
  ``Communicationefficient federated learning and permissioned blockchain for
  digital twin edge networks,'' \emph{IEEE Internet Things}, 2020.

\bibitem{li61}
{W. Sun, H. Zhang, R. Wang, and Y. Zhang}, ``Reducing offloading latency for
  digital twin edge networks in 6g,'' \emph{IEEE Trans. Veh. Technol.},
  vol.~69, no.~10, pp. 12\,240--12\,251, Oct. 2020.

\bibitem{li63}
Y.~Lu, X.~Huang, K.~Zhang, S.~Maharjan, and Y.~Zhang, ``Communication-efficient
  federated learning and permissioned blockchain for digital twin edge
  networks,'' \emph{IEEE Internet of Things Journal}, vol.~8, no.~4, pp.
  2276--2288, 2021.

\bibitem{li64}
{Y. Lu, S. Maharjan and Y. Zhang}, ``Adaptive edge association for wireless
  digital twin networks in 6g,'' \emph{IEEE Internet of Things Journal},
  vol.~8, no.~22, pp. 16\,219--16\,230, Nov. 2021.

\bibitem{li60}
{W. Sun, N. Xu, L. Wang, H. Zhang, and Y. Zhang}, ``Dynamic digital twin and
  federated learning with incentives for air-ground networks,'' \emph{IEEE
  Trans. Netw. Sci. Eng.}, vol.~9, no.~1, pp. 321--333, Jan./Feb. 2022.

\bibitem{9984956}
E.~Mozo, P.~Unterhuber, A.~A. Gómez, S.~Sand, and M.~Mendicute, ``Measurement
  based tapped delay line model for train-to-train communications,'' \emph{IEEE
  Transactions on Vehicular Technology}, pp. 1--13, 2022.

\bibitem{9474457}
P.~Unterhuber, M.~Walter, U.-C. Fiebig, and T.~Kürner, ``Stochastic channel
  parameters for train-to-train communications,'' \emph{IEEE Open Journal of
  Antennas and Propagation}, vol.~2, pp. 778--792, 2021.

\bibitem{9127790}
D.~Yu, G.~Yue, N.~Wei, L.~Yang, H.~Tan, D.~Liang, and Y.~Gong, ``Empirical
  study on directional millimeter-wave propagation in railway communications
  between train and trackside,'' \emph{IEEE Journal on Selected Areas in
  Communications}, vol.~38, no.~12, pp. 2931--2945, 2020.

\bibitem{6378492}
R.~He, Z.~Zhong, B.~Ai, G.~Wang, J.~Ding, and A.~F. Molisch, ``Measurements and
  analysis of propagation channels in high-speed railway viaducts,'' \emph{IEEE
  Transactions on Wireless Communications}, vol.~12, no.~2, pp. 794--805, 2013.

\bibitem{8438326}
D.~He, B.~Ai, K.~Guan, L.~Wang, Z.~Zhong, and T.~K{\"u}rner, ``The design and
  applications of high-performance ray-tracing simulation platform for 5g and
  beyond wireless communications: A tutorial,'' \emph{IEEE Communications
  Surveys \& Tutorials}, vol.~21, no.~1, pp. 10--27, 2019.

\bibitem{9786750}
C.-X. Wang, Z.~Lv, X.~Gao, X.~You, Y.~Hao, and H.~Haas, ``Pervasive wireless
  channel modeling theory and applications to 6g gbsms for all frequency bands
  and all scenarios,'' \emph{IEEE Transactions on Vehicular Technology},
  vol.~71, no.~9, pp. 9159--9173, 2022.

\bibitem{9135487}
P.~Unterhuber, I.~Rashdan, M.~Walter, and T.~Kürner, ``Path loss models and
  large scale fading statistics for c-band train-to-train communication,'' in
  \emph{2020 14th European Conference on Antennas and Propagation (EuCAP)},
  2020, pp. 1--5.

\bibitem{7794737}
T.~Hattori and T.~Kudo, ``Propagation experiment on millimeter wave for
  high-speed rail trains,'' in \emph{2016 IEEE 27th Annual International
  Symposium on Personal, Indoor, and Mobile Radio Communications (PIMRC)},
  2016, pp. 1--6.

\bibitem{8392735}
B.~Zhang, Z.~Zhong, R.~He, G.~Dahman, J.~Ding, S.~Lin, B.~Ai, and M.~Yang,
  ``Measurement-based markov modeling for multi-link channels in railway
  communication systems,'' \emph{IEEE Transactions on Intelligent
  Transportation Systems}, vol.~20, no.~3, pp. 985--999, 2019.

\bibitem{6987277}
B.~Chen, Z.~Zhong, B.~Ai, and D.~G. Michelson, ``A geometry-based stochastic
  channel model for high-speed railway cutting scenarios,'' \emph{IEEE Antennas
  and Wireless Propagation Letters}, vol.~14, pp. 851--854, 2015.

\bibitem{9933809}
X.~Zhang, R.~He, M.~Yang, B.~Ai, S.~Wang, W.~Li, W.~Sun, L.~Li, P.~Huang, and
  Y.~Xue, ``Measurements and modeling of large-scale channel characteristics in
  subway tunnels at 1.8 and 5.8 ghz,'' \emph{IEEE Antennas and Wireless
  Propagation Letters}, vol.~22, no.~3, pp. 561--565, 2023.

\bibitem{9378519}
M.~Berbineau, R.~Behaegel, J.~M. Garcia-Loygorri, R.~Torrego, R.~D’Errico,
  A.~Sabra, Y.~Yan, and J.~Soler, ``Channel models for performance evaluation
  of wireless systems in railway environments,'' \emph{IEEE Access}, vol.~9,
  pp. 45\,903--45\,918, 2021.

\bibitem{9397335}
K.~Guan, D.~He, B.~Ai, Y.~Chen, C.~Han, B.~Peng, Z.~Zhong, and T.~Kürner,
  ``Channel characterization and capacity analysis for thz communication
  enabled smart rail mobility,'' \emph{IEEE Transactions on Vehicular
  Technology}, vol.~70, no.~5, pp. 4065--4080, 2021.

\bibitem{8114238}
R.~He, B.~Ai, G.~L. Stüber, G.~Wang, and Z.~Zhong, ``Geometrical-based
  modeling for millimeter-wave mimo mobile-to-mobile channels,'' \emph{IEEE
  Transactions on Vehicular Technology}, vol.~67, no.~4, pp. 2848--2863, 2018.

\bibitem{9316444}
B.~Ai, R.~He, H.~Zhang, M.~Yang, Z.~Ma, G.~Sun, and Z.~Zhong, ``Feeder
  communication for integrated networks,'' \emph{IEEE Wireless Communications},
  vol.~27, no.~6, pp. 20--27, 2020.

\bibitem{9130836}
J.~Yu, X.~Liu, Y.~Gao, and X.~Shen, ``3d channel tracking for uav-satellite
  communications in space-air-ground integrated networks,'' \emph{IEEE Journal
  on Selected Areas in Communications}, vol.~38, no.~12, pp. 2810--2823, 2020.

\bibitem{9631953}
Y.~Wang, Z.~Su, J.~Ni, N.~Zhang, and X.~Shen, ``Blockchain-empowered
  space-air-ground integrated networks: Opportunities, challenges, and
  solutions,'' \emph{IEEE Communications Surveys \& Tutorials}, vol.~24, no.~1,
  pp. 160--209, 2022.

\bibitem{9120643}
X.~Wang, H.~Liy, W.~Yao, T.~Lany, and Q.~Wu, ``Content delivery for high-speed
  railway via integrated terrestrial-satellite networks,'' in \emph{2020 IEEE
  Wireless Communications and Networking Conference (WCNC)}, 2020, pp. 1--6.

\bibitem{9275613}
M.~Giordani and M.~Zorzi, ``Non-terrestrial networks in the 6g era: Challenges
  and opportunities,'' \emph{IEEE Network}, vol.~35, no.~2, pp. 244--251, 2021.

\bibitem{7470933}
Y.~Zeng, R.~Zhang, and T.~J. Lim, ``Wireless communications with unmanned
  aerial vehicles: opportunities and challenges,'' \emph{IEEE Communications
  Magazine}, vol.~54, no.~5, pp. 36--42, 2016.

\bibitem{1623307}
C.~Loo, ``A statistical model for a land mobile satellite link,'' \emph{IEEE
  Transactions on Vehicular Technology}, vol.~34, no.~3, pp. 122--127, 1985.

\bibitem{corazza1994statistical}
G.~E. Corazza and F.~Vatalaro, ``A statistical model for land mobile satellite
  channels and its application to nongeostationary orbit systems,'' \emph{IEEE
  Transactions on vehicular technology}, vol.~43, no.~3, pp. 738--742, 1994.

\bibitem{289418}
E.~Lutz, D.~Cygan, M.~Dippold, F.~Dolainsky, and W.~Papke, ``The land mobile
  satellite communication channel-recording, statistics, and channel model,''
  \emph{IEEE Transactions on Vehicular Technology}, vol.~40, no.~2, pp.
  375--386, 1991.

\bibitem{8840846}
L.~Bai, C.-X. Wang, G.~Goussetis, S.~Wu, Q.~Zhu, W.~Zhou, and E.-H.~M. Aggoune,
  ``Channel modeling for satellite communication channels at q-band in high
  latitude,'' \emph{IEEE Access}, vol.~7, pp. 137\,691--137\,703, 2019.

\bibitem{arapoglou2012railway}
P.-D. Arapoglou, K.~P. Liolis, and A.~D. Panagopoulos, ``Railway satellite
  channel at ku band and above: Composite dynamic modeling for the design of
  fade mitigation techniques,'' \emph{International Journal of Satellite
  Communications and Networking}, vol.~30, no.~1, pp. 1--17, 2012.

\bibitem{8739589}
S.~Rougerie and J.~Israel, ``Mobile satellite propagation channels at ka band
  for railway and highway environnement,'' in \emph{2019 13th European
  Conference on Antennas and Propagation (EuCAP)}, 2019, pp. 1--5.

\bibitem{9909265}
Q.~Huang, H.~An, K.~Guan, Y.~Li, D.~Fei, F.~Zhu, and H.~Wang,
  ``Measurement-based tapped delay line channel modeling for inter-uav
  communications with typical uav attitudes,'' in \emph{2022 IEEE 5th
  International Conference on Electronic Information and Communication
  Technology (ICEICT)}, 2022, pp. 421--426.

\bibitem{7501562}
R.~Sun and D.~W. Matolak, ``Air–ground channel characterization for unmanned
  aircraft systems part ii: Hilly and mountainous settings,'' \emph{IEEE
  Transactions on Vehicular Technology}, vol.~66, no.~3, pp. 1913--1925, 2017.

\bibitem{7835273}
D.~W. Matolak and R.~Sun, ``Air–ground channel characterization for unmanned
  aircraft systems—part iii: The suburban and near-urban environments,''
  \emph{IEEE Transactions on Vehicular Technology}, vol.~66, no.~8, pp.
  6607--6618, 2017.

\bibitem{8725553}
H.~Chang, J.~Bian, C.-X. Wang, Z.~Bai, W.~Zhou, and e.-H.~M. Aggoune, ``A 3d
  non-stationary wideband gbsm for low-altitude uav-to-ground v2v mimo
  channels,'' \emph{IEEE Access}, vol.~7, pp. 70\,719--70\,732, 2019.

\bibitem{8569308}
W.~Zeng, J.~Zhang, K.~P. Peppas, B.~Ar, and Z.~Zhong, ``Uav-aided wireless
  information and power transmission for high-speed train communications,'' in
  \emph{2018 21st International Conference on Intelligent Transportation
  Systems (ITSC)}, 2018, pp. 3409--3414.

\bibitem{8885447}
H.~S. Khallaf and M.~Uysal, ``Uav-based fso communications for high speed train
  backhauling,'' in \emph{2019 IEEE Wireless Communications and Networking
  Conference (WCNC)}, 2019, pp. 1--6.

\bibitem{9193934}
R.~K. Saha, ``Licensed countrywide full-spectrum allocation: A new paradigm for
  millimeter-wave mobile systems in 5g/6g era,'' \emph{IEEE Access}, vol.~8,
  pp. 166\,612--166\,629, 2020.

\bibitem{6858019}
R.~He, B.~Ai, Z.~Zhong, A.~F. Molisch, R.~Chen, and Y.~Yang, ``A
  measurement-based stochastic model for high-speed railway channels,''
  \emph{IEEE Transactions on Intelligent Transportation Systems}, vol.~16,
  no.~3, pp. 1120--1135, 2015.

\bibitem{8568640}
M.~Soliman, Y.~Dawoud, E.~Staudinger, S.~Sand, A.~Schuetz, and A.~Dekorsy,
  ``Influences of train wagon vibrations on the mmwave wagon-to-wagon
  channel,'' in \emph{12th European Conference on Antennas and Propagation
  (EuCAP 2018)}, 2018, pp. 1--5.

\bibitem{9411172}
J.-J. Park, J.~Lee, K.-W. Kim, and M.-D. Kim, ``Large- and small-scale fading
  characteristics of mmwave hst propagation channel based on 28-ghz
  measurements,'' in \emph{2021 15th European Conference on Antennas and
  Propagation (EuCAP)}, 2021, pp. 1--5.

\bibitem{8641489}
Y.~Liu, C.-X. Wang, C.~F. Lopez, G.~Goussetis, Y.~Yang, and G.~K.
  Karagiannidis, ``3d non-stationary wideband tunnel channel models for 5g
  high-speed train wireless communications,'' \emph{IEEE Transactions on
  Intelligent Transportation Systems}, vol.~21, no.~1, pp. 259--272, 2020.

\bibitem{9303579}
H.~B.~H. Dutty, M.~M. Mowla, and M.~A. Mou, ``A statistical mmwave channel
  modeling for railway communications backhaul in 5g networks,'' in \emph{2019
  3rd International Conference on Electrical, Computer \& Telecommunication
  Engineering (ICECTE)}, 2019, pp. 121--124.

\bibitem{zhao2019channel}
X.~Zhao, Z.~Wang, S.~Geng, Y.~Zhang, F.~Du, and Z.~Fu, ``Channel sounding,
  modelling, and characterisation in a large waiting hall of a high-speed
  railway station at 28 ghz,'' \emph{IET Microwaves, Antennas \& Propagation},
  vol.~13, no.~15, pp. 2619--2624, 2019.

\bibitem{8438925}
J.~Kim, M.~Schmieder, M.~Peter, H.~Chung, S.-W. Choi, I.~Kim, and Y.~Han, ``A
  comprehensive study on mmwave-based mobile hotspot network system for
  high-speed train communications,'' \emph{IEEE Transactions on Vehicular
  Technology}, vol.~68, no.~3, pp. 2087--2101, 2019.

\bibitem{8597759}
G.~Adriandi and C.~Apriono, ``Short range propagation simulation of modified
  ground microstrip antenna design for near field communication application at
  frequency of 0.35 thz,'' in \emph{2018 Progress in Electromagnetics Research
  Symposium (PIERS-Toyama)}, 2018, pp. 962--966.

\bibitem{10333669}
Y.~Dai, J.~Wu, J.~Zhao, B.~Gong, and Y.~Lu, ``Intelligent reflecting surfaces
  aided task offloading in digital twin edge networks,'' in \emph{2023 IEEE
  98th Vehicular Technology Conference (VTC2023-Fall)}, 2023, pp. 1--5.

\bibitem{9613722}
Y.~Wang, G.~Wang, R.~Xu, R.~He, B.~Ai, and H.~Xiao, ``Joint channel estimation
  and data detection for intelligent transparent surface (its) aided wireless
  communications on railways,'' in \emph{2021 13th International Conference on
  Wireless Communications and Signal Processing (WCSP)}, 2021, pp. 1--5.

\bibitem{9690475}
J.~Zhang, H.~Liu, Q.~Wu, Y.~Jin, Y.~Chen, B.~Ai, S.~Jin, and T.~J. Cui,
  ``Ris-aided next-generation high-speed train communications: Challenges,
  solutions, and future directions,'' \emph{IEEE Wireless Communications},
  vol.~28, no.~6, pp. 145--151, 2021.

\bibitem{9769365}
R.~Zhou, X.~Chen, W.~Tang, X.~Li, S.~Jin, E.~Basar, Q.~Cheng, and T.~J. Cui,
  ``Modeling and measurements for multi-path mitigation with reconfigurable
  intelligent surfaces,'' in \emph{2022 16th European Conference on Antennas
  and Propagation (EuCAP)}, 2022, pp. 1--5.

\bibitem{9206044}
W.~Tang, M.~Z. Chen, X.~Chen, J.~Y. Dai, Y.~Han, M.~Di~Renzo, Y.~Zeng, S.~Jin,
  Q.~Cheng, and T.~J. Cui, ``Wireless communications with reconfigurable
  intelligent surface: Path loss modeling and experimental measurement,''
  \emph{IEEE Transactions on Wireless Communications}, vol.~20, no.~1, pp.
  421--439, 2021.

\bibitem{9880837}
B.~Gao, J.~Li, Z.~Yu, J.~Sang, M.~Zhou, J.~Lan, W.~Tang, X.~Li, and S.~Jin,
  ``Propagation characteristics of ris-assisted wireless channels in corridors:
  Measurements and analysis,'' in \emph{2022 IEEE/CIC International Conference
  on Communications in China (ICCC)}, 2022, pp. 550--554.

\bibitem{9541182}
E.~Basar, I.~Yildirim, and F.~Kilinc, ``Indoor and outdoor physical channel
  modeling and efficient positioning for reconfigurable intelligent surfaces in
  mmwave bands,'' \emph{IEEE Transactions on Communications}, vol.~69, no.~12,
  pp. 8600--8611, 2021.

\bibitem{9775205}
G.~Sun, R.~He, B.~Ai, Z.~Ma, P.~Li, Y.~Niu, J.~Ding, D.~Fei, and Z.~Zhong, ``A
  3d wideband channel model for ris-assisted mimo communications,'' \emph{IEEE
  Transactions on Vehicular Technology}, vol.~71, no.~8, pp. 8016--8029, 2022.

\bibitem{9690144}
Y.~Dai, Y.~L. Guan, K.~K. Leung, and Y.~Zhang, ``Reconfigurable intelligent
  surface for low-latency edge computing in 6g,'' \emph{IEEE Wireless
  Communications}, vol.~28, no.~6, pp. 72--79, 2021.

\bibitem{submitted-1}
Y.~{Yuan}, R.~{He}, B.~{Ai} \emph{et~al.}, ``A 3d geometry-based reconfigurable
  intelligent surfaces-assisted mmwave channel model for high-speed train
  communications,'' \emph{IEEE Transactions On Vehicular Technology}, 2022.

\bibitem{9496190}
Y.~Ma, G.~Ma, N.~Wang, Z.~Zhong, and B.~Ai, ``Otfs-tsma for massive internet of
  things in high-speed railway,'' \emph{IEEE Transactions on Wireless
  Communications}, vol.~21, no.~1, pp. 519--531, 2022.

\bibitem{9714331}
H.~A.~H. Alobaidy, M.~Jit~Singh, M.~Behjati, R.~Nordin, and N.~F. Abdullah,
  ``Wireless transmissions, propagation and channel modelling for iot
  technologies: Applications and challenges,'' \emph{IEEE Access}, vol.~10, pp.
  24\,095--24\,131, 2022.

\bibitem{8340813}
F.~Jameel, Z.~Hamid, F.~Jabeen, S.~Zeadally, and M.~A. Javed, ``A survey of
  device-to-device communications: Research issues and challenges,'' \emph{IEEE
  Communications Surveys \& Tutorials}, vol.~20, no.~3, pp. 2133--2168, 2018.

\bibitem{9383771}
G.~Caso, {\"O}.~Alay, L.~De~Nardis, A.~Brunstrom, M.~Neri, and M.-G.
  Di~Benedetto, ``Empirical models for nb-iot path loss in an urban scenario,''
  \emph{IEEE Internet of Things Journal}, vol.~8, no.~17, pp. 13\,774--13\,788,
  2021.

\bibitem{8287901}
E.~Bedeer, J.~Pugh, C.~Brown, and H.~Yanikomeroglu, ``Measurement-based path
  loss and delay spread propagation models in vhf/uhf bands for iot
  communications,'' in \emph{2017 IEEE 86th Vehicular Technology Conference
  (VTC-Fall)}, 2017, pp. 1--5.

\bibitem{6953146}
S.~L. Cotton, ``Human body shadowing in cellular device-to-device
  communications: Channel modeling using the shadowed $\kappa-\mu$ fading
  model,'' \emph{IEEE Journal on Selected Areas in Communications}, vol.~33,
  no.~1, pp. 111--119, 2015.

\bibitem{7500097}
S.~Kim and A.~Zajić, ``Statistical modeling and simulation of short-range
  device-to-device communication channels at sub-thz frequencies,'' \emph{IEEE
  Transactions on Wireless Communications}, vol.~15, no.~9, pp. 6423--6433,
  2016.

\bibitem{7400962}
S.~Hur, S.~Baek, B.~Kim, Y.~Chang, A.~F. Molisch, T.~S. Rappaport, K.~Haneda,
  and J.~Park, ``Proposal on millimeter-wave channel modeling for 5g cellular
  system,'' \emph{IEEE Journal of Selected Topics in Signal Processing},
  vol.~10, no.~3, pp. 454--469, 2016.

\bibitem{8792137}
T.~Nishio, H.~Okamoto, K.~Nakashima, Y.~Koda, K.~Yamamoto, M.~Morikura,
  Y.~Asai, and R.~Miyatake, ``Proactive received power prediction using machine
  learning and depth images for mmwave networks,'' \emph{IEEE Journal on
  Selected Areas in Communications}, vol.~37, no.~11, pp. 2413--2427, 2019.

\bibitem{8608834}
J.~Pascual-Garcia, J.-M. Molina-Garcia-Pardo, M.-T. Martinez-Ingles, J.-V.
  Rodriguez, and L.~Juan-Llacer, ``Wireless channel simulation using
  geometrical models extrated from point clouds,'' in \emph{2018 IEEE
  International Symposium on Antennas and Propagation \& USNC/URSI National
  Radio Science Meeting}, 2018, pp. 83--84.

\bibitem{8345613}
U.~T. Virk, S.~L.~H. Nguyen, K.~Haneda, and J.-F. Wagen, ``On-site permittivity
  estimation at 60 ghz through reflecting surface identification in the point
  cloud,'' \emph{IEEE Transactions on Antennas and Propagation}, vol.~66,
  no.~7, pp. 3599--3609, 2018.

\bibitem{9325887}
P.~Koivumäki, G.~Steinböck, and K.~Haneda, ``Impacts of point cloud modeling
  on the accuracy of ray-based multipath propagation simulations,'' \emph{IEEE
  Transactions on Antennas and Propagation}, vol.~69, no.~8, pp. 4737--4747,
  2021.

\bibitem{6194400}
J.~Gozalvez, M.~Sepulcre, and R.~Bauza, ``Ieee 802.11p vehicle to
  infrastructure communications in urban environments,'' \emph{IEEE
  Communications Magazine}, vol.~50, no.~5, pp. 176--183, 2012.

\bibitem{7780459}
K.~He, X.~Zhang, S.~Ren, and J.~Sun, ``Deep residual learning for image
  recognition,'' in \emph{2016 IEEE Conference on Computer Vision and Pattern
  Recognition (CVPR)}, 2016, pp. 770--778.

\bibitem{10556753}
E.~Shi, J.~Zhang, H.~Du, B.~Ai, C.~Yuen, D.~Niyato, K.~B. Letaief, and X.~Shen,
  ``{RIS}-aided cell-free massive {MIMO} systems for {6G}: Fundamentals, system
  design, and applications,'' \emph{Proceedings of the IEEE}, vol. 112, no.~4,
  pp. 331--364, Apr. 2024.

\bibitem{9765773}
Y.~Zhang, J.~Zhang, D.~W.~K. Ng, H.~Xiao, and B.~Ai, ``Performance analysis of
  reconfigurable intelligent surface assisted systems under channel aging,''
  \emph{Intelligent and Converged Networks}, vol.~3, no.~1, pp. 74--85, 2022.

\bibitem{chapala2022reconfigurable}
V.~K. Chapala and S.~Zafaruddin, ``Reconfigurable intelligent surface for
  vehicular communications: Exact performance analysis with phase noise and
  mobility,'' \emph{arXiv preprint arXiv:2209.10528}, 2022.

\bibitem{9384497}
Z.~Mao, M.~Peng, and X.~Liu, ``Channel estimation for reconfigurable
  intelligent surface assisted wireless communication systems in mobility
  scenarios,'' \emph{China Communications}, vol.~18, no.~3, pp. 29--38, 2021.

\bibitem{9292080}
S.~Sun and H.~Yan, ``Channel estimation for reconfigurable intelligent
  surface-assisted wireless communications considering doppler effect,''
  \emph{IEEE Wireless Communications Letters}, vol.~10, no.~4, pp. 790--794,
  2021.

\bibitem{9875062}
C.~Xu, J.~An, T.~Bai, S.~Sugiura, R.~G. Maunder, Z.~Wang, L.-L. Yang, and
  L.~Hanzo, ``Channel estimation for reconfigurable intelligent surface
  assisted high-mobility wireless systems,'' \emph{IEEE Transactions on
  Vehicular Technology}, vol.~72, no.~1, pp. 718--734, 2023.

\bibitem{9651545}
J.~Xu and B.~Ai, ``When mmwave high-speed railway networks meet reconfigurable
  intelligent surface: A deep reinforcement learning method,'' \emph{IEEE
  Wireless Communications Letters}, vol.~11, no.~3, pp. 533--537, 2022.

\bibitem{9110889}
B.~Di, H.~Zhang, L.~Song, Y.~Li, Z.~Han, and H.~V. Poor, ``Hybrid beamforming
  for reconfigurable intelligent surface based multi-user communications:
  Achievable rates with limited discrete phase shifts,'' \emph{IEEE Journal on
  Selected Areas in Communications}, vol.~38, no.~8, pp. 1809--1822, 2020.

\bibitem{fraga2017towards}
P.~Fraga-Lamas, T.~M. Fern{\'a}ndez-Caram{\'e}s, and L.~Castedo, ``Towards the
  internet of smart trains: A review on industrial iot-connected railways,''
  \emph{Sensors}, vol.~17, no.~6, p. 1457, 2017.

\bibitem{li2019railway}
P.~Li, S.~Shao, R.~Xue, and X.~Zhang, ``Railway digitalization and intelligent
  railway development in other countries,'' \emph{Chin. Railways}, vol.~4,
  no.~2, pp. 25--31, 2019.

\bibitem{hadani2017orthogonal}
R.~Hadani, S.~Rakib, M.~Tsatsanis, A.~Monk, A.~J. Goldsmith, A.~F. Molisch, and
  R.~Calderbank, ``Orthogonal time frequency space modulation,'' in \emph{2017
  IEEE Wireless Communications and Networking Conference (WCNC)}.\hskip 1em
  plus 0.5em minus 0.4em\relax IEEE, 2017, pp. 1--6.

\bibitem{9829188}
H.~Lin and J.~Yuan, ``Orthogonal delay-doppler division multiplexing
  modulation,'' \emph{IEEE Transactions on Wireless Communications}, vol.~21,
  no.~12, pp. 11\,024--11\,037, 2022.

\bibitem{10806729}
Y.~Ma, A.~Shafie, J.~Yuan, G.~Ma, Z.~Zhong, and B.~Ai, ``Orthogonal
  delay-doppler division multiplexing modulation with tomlinson-harashima
  precoding,'' \emph{IEEE Transactions on Communications}, pp. 1--1, 2024.

\bibitem{8671740}
P.~Raviteja, K.~T. Phan, and Y.~Hong, ``Embedded pilot-aided channel estimation
  for otfs in delay–doppler channels,'' \emph{IEEE Transactions on Vehicular
  Technology}, vol.~68, no.~5, pp. 4906--4917, 2019.

\bibitem{7842433}
Z.~Ding, Y.~Liu, J.~Choi, Q.~Sun, M.~Elkashlan, I.~Chih-Lin, and H.~V. Poor,
  ``Application of non-orthogonal multiple access in lte and 5g networks,''
  \emph{IEEE Communications Magazine}, vol.~55, no.~2, pp. 185--191, 2017.

\bibitem{wang2019massive}
F.~Wang and G.~Ma, \emph{Massive Machine Type Communications: Multiple Access
  Schemes}.\hskip 1em plus 0.5em minus 0.4em\relax Springer, 2019.

\bibitem{9411900}
K.~Deka, A.~Thomas, and S.~Sharma, ``Otfs-scma: A code-domain noma approach for
  orthogonal time frequency space modulation,'' \emph{IEEE Transactions on
  Communications}, vol.~69, no.~8, pp. 5043--5058, 2021.

\bibitem{9928043}
X.~Zhou, K.~Ying, Z.~Gao, Y.~Wu, Z.~Xiao, S.~Chatzinotas, J.~Yuan, and
  B.~Ottersten, ``Active terminal identification, channel estimation, and
  signal detection for grant-free noma-otfs in leo satellite
  internet-of-things,'' \emph{IEEE Transactions on Wireless Communications},
  vol.~22, no.~4, pp. 2847--2866, 2023.

\bibitem{8352626}
G.~Ma, B.~Ai, F.~Wang, X.~Chen, Z.~Zhong, Z.~Zhao, and H.~Guan, ``Coded tandem
  spreading multiple access for massive machine-type communications,''
  \emph{IEEE Wireless Communications}, vol.~25, no.~2, pp. 75--81, 2018.

\bibitem{1386525}
A.~Poon, R.~Brodersen, and D.~Tse, ``Degrees of freedom in multiple-antenna
  channels: a signal space approach,'' \emph{IEEE Transactions on Information
  Theory}, vol.~51, no.~2, pp. 523--536, 2005.

\bibitem{10056866}
Y.~Ma, G.~Ma, B.~Ai, D.~Fei, N.~Wang, Z.~Zhong, and J.~Yuan, ``Characteristics
  of channel spreading function and performance of otfs in high-speed
  railway,'' \emph{IEEE Transactions on Wireless Communications}, vol.~22,
  no.~10, pp. 7038--7054, 2023.

\bibitem{8888372}
Y.~Niu, J.~Ding, D.~Fei, Z.~Zhong, and Y.~Liu, ``Doppler effect on high-speed
  railway at 465 mhz,'' in \emph{2019 IEEE International Symposium on Antennas
  and Propagation and USNC-URSI Radio Science Meeting}, 2019, pp. 2117--2118.

\bibitem{8786203}
Z.~Ding, R.~Schober, P.~Fan, and H.~Vincent~Poor, ``Otfs-noma: An efficient
  approach for exploiting heterogenous user mobility profiles,'' \emph{IEEE
  Transactions on Communications}, vol.~67, no.~11, pp. 7950--7965, 2019.

\bibitem{9181410}
M.~Li, S.~Zhang, F.~Gao, P.~Fan, and O.~A. Dobre, ``A new path division
  multiple access for the massive mimo-otfs networks,'' \emph{IEEE Journal on
  Selected Areas in Communications}, vol.~39, no.~4, pp. 903--918, 2021.

\bibitem{10274133}
Y.~Ma, G.~Ma, B.~Ai, J.~Liu, N.~Wang, and Z.~Zhong, ``Otfcs-modulated waveform
  design for joint grant-free random access and positioning in c-v2x,''
  \emph{IEEE Journal on Selected Areas in Communications}, vol.~42, no.~1, pp.
  103--119, 2024.

\bibitem{8634036}
B.~Wei, Z.~Li, L.~Liu, and J.~Wang, ``Field distribution characteristics of
  leaky-wave system in the vacuum tube for high-speed rail,'' in \emph{2018
  12th International Symposium on Antennas, Propagation and EM Theory (ISAPE)},
  2018, pp. 1--3.

\bibitem{latva2020key}
M.~Latva-Aho, K.~Lepp{\"a}nen \emph{et~al.}, ``Key drivers and research
  challenges for 6g ubiquitous wireless intelligence,'' 2019.

\bibitem{9269930}
K.~Rikkinen, P.~Kyosti, M.~E. Leinonen, M.~Berg, and A.~Parssinen, ``Thz radio
  communication: Link budget analysis toward 6g,'' \emph{IEEE Communications
  Magazine}, vol.~58, no.~11, pp. 22--27, 2020.

\bibitem{9267779}
H.~Jiang, M.~Mukherjee, J.~Zhou, and J.~Lloret, ``Channel modeling and
  characteristics for 6g wireless communications,'' \emph{IEEE Network},
  vol.~35, no.~1, pp. 296--303, 2021.

\bibitem{9254121}
Z.~Lv, L.~Qiao, and I.~You, ``6g-enabled network in box for internet of
  connected vehicles,'' \emph{IEEE Transactions on Intelligent Transportation
  Systems}, vol.~22, no.~8, pp. 5275--5282, 2021.

\bibitem{9833304}
Y.~Ma, Y.~Niu, Z.~Han, B.~Ai, K.~Li, Z.~Zhong, and N.~Wang, ``Robust
  transmission scheduling for uav-assisted millimeter-wave train-ground
  communication system,'' \emph{IEEE Transactions on Vehicular Technology},
  vol.~71, no.~11, pp. 11\,741--11\,755, 2022.

\bibitem{9779990}
Y.~Wang, Y.~Niu, H.~Wu, S.~Mao, B.~Ai, Z.~Zhong, and N.~Wang, ``Scheduling of
  uav-assisted millimeter wave communications for high-speed railway,''
  \emph{IEEE Transactions on Vehicular Technology}, vol.~71, no.~8, pp.
  8756--8767, 2022.

\bibitem{zhang2011key}
Y.~Zhang, D.~Oster, M.~Kumada, J.~Yu, and S.~Li, ``Key vacuum technology issues
  to be solved in evacuated tube transportation,'' \emph{Journal of Modern
  Transportation}, vol.~19, pp. 110--113, 2011.

\bibitem{4455844}
R.~Piesiewicz, T.~Kleine-Ostmann, N.~Krumbholz, D.~Mittleman, M.~Koch,
  J.~Schoebel, and T.~Kurner, ``Short-range ultra-broadband terahertz
  communications: Concepts and perspectives,'' \emph{IEEE Antennas and
  Propagation Magazine}, vol.~49, no.~6, pp. 24--39, 2007.

\bibitem{8598887}
Y.~Niu, W.~Ding, H.~Wu, Y.~Li, X.~Chen, B.~Ai, and Z.~Zhong, ``Relay-assisted
  and qos aware scheduling to overcome blockage in mmwave backhaul networks,''
  \emph{IEEE Transactions on Vehicular Technology}, vol.~68, no.~2, pp.
  1733--1744, 2019.

\bibitem{8344113}
W.~Ding, Y.~Niu, H.~Wu, Y.~Li, and Z.~Zhong, ``Qos-aware full-duplex concurrent
  scheduling for millimeter wave wireless backhaul networks,'' \emph{IEEE
  Access}, vol.~6, pp. 25\,313--25\,322, 2018.

\bibitem{6364219}
J.~Qiao, L.~X. Cai, X.~Shen, and J.~W. Mark, ``Stdma-based scheduling algorithm
  for concurrent transmissions in directional millimeter wave networks,'' in
  \emph{2012 IEEE International Conference on Communications (ICC)}, 2012, pp.
  5221--5225.

\bibitem{9261169}
M.~Jankowski, D.~Gündüz, and K.~Mikolajczyk, ``Wireless image retrieval at
  the edge,'' \emph{IEEE Journal on Selected Areas in Communications}, vol.~39,
  no.~1, pp. 89--100, 2021.

\bibitem{10772628}
L.~Guo, W.~Chen, Y.~Sun, and B.~Ai, ``Digital-sc: Digital semantic
  communication with adaptive network split and learned non-linear
  quantization,'' \emph{IEEE Transactions on Cognitive Communications and
  Networking}, pp. 1--1, 2024.

\bibitem{xu2023deep}
J.~Xu, T.-Y. Tung, B.~Ai, W.~Chen, Y.~Sun, and D.~G{\"u}nd{\"u}z, ``Deep joint
  source-channel coding for semantic communications,'' \emph{IEEE
  communications Magazine}, vol.~61, no.~11, pp. 42--48, 2023.

\bibitem{10431795}
G.~Zhang, Q.~Hu, Z.~Qin, Y.~Cai, G.~Yu, and X.~Tao, ``A unified multi-task
  semantic communication system for multimodal data,'' \emph{IEEE Transactions
  on Communications}, vol.~72, no.~7, pp. 4101--4116, 2024.

\bibitem{xie2022task}
H.~Xie, Z.~Qin, X.~Tao, and K.~B. Letaief, ``Task-oriented multi-user semantic
  communications,'' \emph{IEEE Journal on Selected Areas in Communications},
  vol.~40, no.~9, pp. 2584--2597, 2022.

\bibitem{9954153}
J.~Xu, B.~Ai, N.~Wang, and W.~Chen, ``Deep joint source-channel coding for csi
  feedback: An end-to-end approach,'' \emph{IEEE Journal on Selected Areas in
  Communications}, vol.~41, no.~1, pp. 260--273, 2023.

\bibitem{10660530}
Y.~Guo, W.~Chen, J.~Xu, L.~Li, and B.~Ai, ``Deep joint csi feedback and
  multiuser precoding for mimo ofdm systems,'' \emph{IEEE Transactions on
  Vehicular Technology}, vol.~74, no.~1, pp. 1730--1735, 2025.

\bibitem{9596610}
D.~Xu, T.~Li, Y.~Li, X.~Su, S.~Tarkoma, T.~Jiang, J.~Crowcroft, and P.~Hui,
  ``Edge intelligence: Empowering intelligence to the edge of network,''
  \emph{Proceedings of the IEEE}, vol. 109, no.~11, pp. 1778--1837, 2021.

\bibitem{2021Edge}
S.~M. Asad, A.~Tahir, R.~N.~B. Rais, S.~Ansari, A.~I. Abubakar, S.~Hussain,
  Q.~H. Abbasi, and M.~A. Imran, ``Edge intelligence in private mobile networks
  for next-generation railway systems,'' \emph{Frontiers in Communications and
  Networks}, vol.~2, p. 769299, 2021.

\bibitem{9922178}
C.~Chen, L.~Zhu, and H.~Zhao, ``5g-enabled edge intelligence for autonomous
  train control: A practical perspective,'' in \emph{2022 IEEE 25th
  International Conference on Intelligent Transportation Systems (ITSC)}, 2022,
  pp. 1790--1794.

\bibitem{9354351}
D.~Yang, E.~Cui, H.~Wang, and H.~Zhang, ``Eh-edge--an energy harvesting-driven
  edge iot platform for online failure prediction of rail transit vehicles: A
  case study of a cloud, edge, and end device collaborative computing
  paradigm,'' \emph{IEEE Vehicular Technology Magazine}, vol.~16, no.~2, pp.
  95--103, 2021.

\bibitem{9899356}
J.~Xu, B.~Ai, L.~Chen, Y.~Cui, and N.~Wang, ``Deep reinforcement learning for
  computation and communication resource allocation in multiaccess mec assisted
  railway iot networks,'' \emph{IEEE Transactions on Intelligent Transportation
  Systems}, vol.~23, no.~12, pp. 23\,797--23\,808, 2022.

\bibitem{10061573}
J.~Zhao, L.~He, D.~Zhang, and X.~Gao, ``A tp-ddpg algorithm based on cache
  assistance for task offloading in urban rail transit,'' \emph{IEEE
  Transactions on Vehicular Technology}, pp. 1--11, 2023.

\bibitem{9506824}
J.~Xu, Z.~Wei, Z.~Lyu, L.~Shi, and J.~Han, ``Throughput maximization of
  offloading tasks in multi-access edge computing networks for high-speed
  railways,'' \emph{IEEE Transactions on Vehicular Technology}, vol.~70, no.~9,
  pp. 9525--9539, 2021.

\bibitem{8926409}
H.~Chen, B.~Jiang, W.~Chen, and Z.~Li, ``Edge computing-aided framework of
  fault detection for traction control systems in high-speed trains,''
  \emph{IEEE Transactions on Vehicular Technology}, vol.~69, no.~2, pp.
  1309--1318, 2020.

\bibitem{10258056}
Z.~Li, H.~Wu, Y.~Lu, B.~Ai, Z.~Zhong, and Y.~Zhang, ``Matching game for
  multi-task federated learning in internet of vehicles,'' \emph{IEEE
  Transactions on Vehicular Technology}, vol.~73, no.~2, pp. 1623--1636, 2024.

\bibitem{10437116}
Z.~Li, H.~Wu, Y.~Dai, and Y.~Lu, ``Pafl: Parameter-authentication federated
  learning for internet of vehicles,'' in \emph{GLOBECOM 2023 - 2023 IEEE
  Global Communications Conference}, 2023, pp. 1241--1246.

\bibitem{5634099}
E.~J.~C. Kelkboom, J.~Breebaart, T.~A.~M. Kevenaar, I.~Buhan, and R.~N.~J.
  Veldhuis, ``Preventing the decodability attack based cross-matching in a
  fuzzy commitment scheme,'' \emph{IEEE Transactions on Information Forensics
  and Security}, vol.~6, no.~1, pp. 107--121, 2011.

\bibitem{dang2020should}
S.~Dang, O.~Amin, B.~Shihada, and M.-S. Alouini, ``What should 6g be?''
  \emph{Nature Electronics}, vol.~3, no.~1, pp. 20--29, 2020.

\bibitem{zhang2023endogenous}
S.~Zhang, Z.~Yao, H.~Liao, Z.~Zhou, Y.~Chen, and Z.~You, ``Endogenous
  security-aware resource management for digital twin and 6g edge intelligence
  integrated smart park,'' \emph{China Communications}, vol.~20, no.~2, pp.
  46--60, 2023.

\bibitem{chang20226g}
L.~Chang, Z.~Zhang, P.~Li, S.~Xi, W.~Guo, Y.~Shen, Z.~Xiong, J.~Kang,
  D.~Niyato, X.~Qiao \emph{et~al.}, ``6g-enabled edge ai for metaverse:
  Challenges, methods, and future research directions,'' \emph{Journal of
  Communications and Information Networks}, vol.~7, no.~2, pp. 107--121, 2022.

\bibitem{ramezanpour2023security}
K.~Ramezanpour, J.~Jagannath, and A.~Jagannath, ``Security and privacy
  vulnerabilities of 5g/6g and wifi 6: Survey and research directions from a
  coexistence perspective,'' \emph{Computer Networks}, vol. 221, p. 109515,
  2023.

\bibitem{10075050}
S.~Soderi, D.~Masti, and Y.~Z. Lun, ``Railway cyber-security in the era of
  interconnected systems: A survey,'' \emph{IEEE Transactions on Intelligent
  Transportation Systems}, vol.~24, no.~7, pp. 6764--6779, 2023.

\bibitem{10234591}
Y.~Wang, W.~Zhang, X.~Wang, M.~K. Khan, and P.~Fan, ``Security enhanced
  authentication protocol for space-ground integrated railway networks,''
  \emph{IEEE Transactions on Intelligent Transportation Systems}, vol.~25,
  no.~1, pp. 370--385, 2024.

\bibitem{9484083}
L.~Zhu, H.~Liang, H.~Wang, B.~Ning, and T.~Tang, ``Joint security and train
  control design in blockchain-empowered cbtc system,'' \emph{IEEE Internet of
  Things Journal}, vol.~9, no.~11, pp. 8119--8129, 2022.

\bibitem{10036973}
N.~Qin, J.~Du, Y.~Zhang, D.~Huang, and B.~Wu, ``Fault diagnosis of
  multi-railway high-speed train bogies by improved federated learning,''
  \emph{IEEE Transactions on Vehicular Technology}, vol.~72, no.~6, pp.
  7184--7194, 2023.

\bibitem{dai2021review}
D.~Dai and S.~Boroomand, ``A review of artificial intelligence to enhance the
  security of big data systems: state-of-art, methodologies, applications, and
  challenges,'' \emph{Archives of Computational Methods in Engineering}, pp.
  1--19, 2021.

\bibitem{10078088}
J.~Pei, S.~Li, Z.~Yu, L.~Ho, W.~Liu, and L.~Wang, ``Federated learning
  encounters 6g wireless communication in the scenario of internet of things,''
  \emph{IEEE Communications Standards Magazine}, vol.~7, no.~1, pp. 94--100,
  2023.

\bibitem{chen2022zero}
X.~Chen, W.~Feng, N.~Ge, and Y.~Zhang, ``Zero trust architecture for 6g
  security,'' \emph{IEEE Network}, vol.~38, no.~4, pp. 224--232, 2024.

\bibitem{loven2019edgeai}
L.~Lov{\'e}n, T.~Lepp{\"a}nen, E.~Peltonen, J.~Partala, E.~Harjula,
  P.~Porambage, M.~Ylianttila, and J.~Riekki, ``Edgeai: A vision for
  distributed, edge-native artificial intelligence in future 6g networks,''
  \emph{The 1st 6G wireless summit}, pp. 1--2, 2019.

\bibitem{8998330}
Y.~Dai, D.~Xu, K.~Zhang, S.~Maharjan, and Y.~Zhang, ``Deep reinforcement
  learning and permissioned blockchain for content caching in vehicular edge
  computing and networks,'' \emph{IEEE Transactions on Vehicular Technology},
  vol.~69, no.~4, pp. 4312--4324, 2020.

\bibitem{10177944}
H.~Liang, L.~Zhu, and F.~R. Yu, ``Collaborative edge intelligence service
  provision in blockchain empowered urban rail transit systems,'' \emph{IEEE
  Internet of Things Journal}, vol.~11, no.~2, pp. 2211--2223, 2024.

\bibitem{zhu2019applications}
Q.~Zhu, S.~W. Loke, R.~Trujillo-Rasua, F.~Jiang, and Y.~Xiang, ``Applications
  of distributed ledger technologies to the internet of things: A survey,''
  \emph{ACM computing surveys (CSUR)}, vol.~52, no.~6, pp. 1--34, 2019.

\bibitem{feng2023blockchain}
Y.~Feng, Z.~Zhong, X.~Sun, L.~Wang, Y.~Lu, and Y.~Zhu, ``Blockchain enabled
  zero trust based authentication scheme for railway communication networks,''
  \emph{Journal of Cloud Computing}, vol.~12, no.~1, pp. 1--21, 2023.

\bibitem{10091151}
S.~Hong, L.~Xu, J.~Huang, H.~Li, H.~Hu, and G.~Gu, ``Sysflow: Toward a
  programmable zero trust framework for system security,'' \emph{IEEE
  Transactions on Information Forensics and Security}, vol.~18, pp. 2794--2809,
  2023.

\bibitem{9745481}
Y.~Dai and Y.~Zhang, ``Adaptive digital twin for vehicular edge computing and
  networks,'' \emph{Journal of Communications and Information Networks},
  vol.~7, no.~1, pp. 48--59, 2022.

\bibitem{10292659}
S.~Ghaboura, R.~Ferdousi, F.~Laamarti, C.~Yang, and A.~E. Saddik, ``Digital
  twin for railway: A comprehensive survey,'' \emph{IEEE Access}, vol.~11, pp.
  120\,237--120\,257, 2023.

\bibitem{10144473}
Z.~Wang, Y.~Geng, L.~Jia, Y.~Qin, Y.~Chai, L.~Tong, and K.~Liu,
  ``Self-attentive local aggregation learning with prototype guided
  regularization for point cloud semantic segmentation of high-speed
  railways,'' \emph{IEEE Transactions on Intelligent Transportation Systems},
  vol.~24, no.~10, pp. 11\,157--11\,170, 2023.

\bibitem{10049521}
A.~De~Benedictis, F.~Flammini, N.~Mazzocca, A.~Somma, and F.~Vitale, ``Digital
  twins for anomaly detection in the industrial internet of things: Conceptual
  architecture and proof-of-concept,'' \emph{IEEE Transactions on Industrial
  Informatics}, vol.~19, no.~12, pp. 11\,553--11\,563, 2023.

\bibitem{9926529}
R.~Ferdousi, F.~Laamarti, C.~Yang, and A.~El~Saddik, ``Railtwin: A digital twin
  framework for railway,'' in \emph{2022 IEEE 18th International Conference on
  Automation Science and Engineering (CASE)}, 2022, pp. 1767--1772.

\bibitem{9690822}
E.~Dimitrova and S.~Tomov, ``Digital twins: An advanced technology for railways
  maintenance transformation,'' in \emph{2021 13th Electrical Engineering
  Faculty Conference (BulEF)}, 2021, pp. 1--5.

\bibitem{9540153}
R.~Chen, C.~Jin, Y.~Zhang, J.~Dai, and X.~Lv, ``Digital twin for equipment
  management of intelligent railway station,'' in \emph{2021 IEEE 1st
  International Conference on Digital Twins and Parallel Intelligence (DTPI)},
  2021, pp. 374--377.

\bibitem{s21175757}
A.~Kampczyk and K.~Dybeł, ``The fundamental approach of the digital twin
  application in railway turnouts with innovative monitoring of weather
  conditions,'' \emph{Sensors}, vol.~21, no.~17, 2021.

\bibitem{9887906}
Y.~He, M.~Yang, Z.~He, and M.~Guizani, ``Resource allocation based on digital
  twin-enabled federated learning framework in heterogeneous cellular
  network,'' \emph{IEEE Transactions on Vehicular Technology}, vol.~72, no.~1,
  pp. 1149--1158, 2023.

\bibitem{9899364}
X.~Yuan, J.~Chen, N.~Zhang, J.~Ni, F.~R. Yu, and V.~C.~M. Leung, ``Digital
  twin-driven vehicular task offloading and irs configuration in the internet
  of vehicles,'' \emph{IEEE Transactions on Intelligent Transportation
  Systems}, vol.~23, no.~12, pp. 24\,290--24\,304, 2022.

\bibitem{9311405}
W.~Sun, N.~Xu, L.~Wang, H.~Zhang, and Y.~Zhang, ``Dynamic digital twin and
  federated learning with incentives for air-ground networks,'' \emph{IEEE
  Transactions on Network Science and Engineering}, vol.~9, no.~1, pp.
  321--333, 2022.

\bibitem{9830070}
J.~Zheng, T.~H. Luan, Y.~Hui, Z.~Yin, N.~Cheng, L.~Gao, and L.~X. Cai,
  ``Digital twin empowered heterogeneous network selection in vehicular
  networks with knowledge transfer,'' \emph{IEEE Transactions on Vehicular
  Technology}, vol.~71, no.~11, pp. 12\,154--12\,168, 2022.

\bibitem{9579446}
B.~Fan, Y.~Wu, Z.~He, Y.~Chen, T.~Q. Quek, and C.-Z. Xu, ``Digital twin
  empowered mobile edge computing for intelligent vehicular lane-changing,''
  \emph{IEEE Network}, vol.~35, no.~6, pp. 194--201, 2021.

\bibitem{9447819}
T.~Liu, L.~Tang, W.~Wang, Q.~Chen, and X.~Zeng, ``Digital-twin-assisted task
  offloading based on edge collaboration in the digital twin edge network,''
  \emph{IEEE Internet of Things Journal}, vol.~9, no.~2, pp. 1427--1444, 2022.

\bibitem{8264818}
Z.~Chen, F.~Sohrabi, and W.~Yu, ``Sparse activity detection for massive
  connectivity,'' \emph{IEEE Transactions on Signal Processing}, vol.~66,
  no.~7, pp. 1890--1904, 2018.

\bibitem{9903376}
Y.~Bai, W.~Chen, F.~Sun, B.~Ai, and P.~Popovski, ``Data-driven compressed
  sensing for massive wireless access,'' \emph{IEEE Communications Magazine},
  vol.~60, no.~11, pp. 28--34, 2022.

\bibitem{9252937}
W.~{Chen}, B.~{Zhang}, S.~{Jin}, B.~{Ai}, and Z.~{Zhong}, ``Solving sparse
  linear inverse problems in communication systems: A deep learning approach
  with adaptive depth,'' \emph{IEEE Journal on Selected Areas in
  Communications}, vol.~39, no.~1, pp. 4--17, 2021.

\bibitem{9605579}
Y.~{Bai}, W.~{Chen}, B.~{Ai}, Z.~{Zhong}, and W.~{Ian}, ``Prior information
  aided deep learning method for grant-free {NOMA} in {mMTC},'' \emph{IEEE
  Journal on Selected Areas in Communications}, vol.~40, no.~1, pp. 112--126,
  2022.

\bibitem{7934066}
M.~Borgerding, P.~Schniter, and S.~Rangan, ``Amp-inspired deep networks for
  sparse linear inverse problems,'' \emph{IEEE Transactions on Signal
  Processing}, vol.~65, no.~16, pp. 4293--4308, 2017.

\bibitem{9268113}
H.~Xiao, W.~Chen, J.~Fang, B.~Ai, and I.~J. Wassell, ``A grant-free method for
  massive machine-type communication with backward activity level estimation,''
  \emph{IEEE Transactions on Signal Processing}, vol.~68, pp. 6665--6680, 2020.

\bibitem{10304065}
Y.~Bai, W.~Chen, B.~Ai, and P.~Popovski, ``Deep learning for asynchronous
  massive access with data frame length diversity,'' \emph{IEEE Transactions on
  Wireless Communications}, vol.~23, no.~6, pp. 5529--5540, 2024.

\bibitem{342465}
Y.~Pati, R.~Rezaiifar, and P.~Krishnaprasad, ``Orthogonal matching pursuit:
  recursive function approximation with applications to wavelet
  decomposition,'' in \emph{Proceedings of 27th Asilomar Conference on Signals,
  Systems and Computers}, 1993, pp. 40--44 vol.1.

\bibitem{9687468}
Z.~Feng, Z.~Wei, X.~Chen, H.~Yang, Q.~Zhang, and P.~Zhang, ``Joint
  communication, sensing, and computation enabled 6g intelligent machine
  system,'' \emph{IEEE Network}, vol.~35, no.~6, pp. 34--42, 2021.

\bibitem{IMT2030}
\BIBentryALTinterwordspacing
I.-.~P. Group, ``6g overall vision and potential key technology white paper,''
  Tech. Rep., Jul. 2021. [Online]. Available:
  \url{http://www.caict.ac.cn/kxyj/qwfb/ztbg/202106/P020210604552572072895.pdf}
\BIBentrySTDinterwordspacing

\bibitem{9829746}
J.~Wang, N.~Varshney, C.~Gentile, S.~Blandino, J.~Chuang, and N.~Golmie,
  ``Integrated sensing and communication: Enabling techniques, applications,
  tools and data sets, standardization, and future directions,'' \emph{IEEE
  Internet of Things Journal}, vol.~9, no.~23, pp. 23\,416--23\,440, 2022.

\bibitem{9492131}
J.~Mu, Y.~Gong, F.~Zhang, Y.~Cui, F.~Zheng, and X.~Jing, ``Integrated sensing
  and communication-enabled predictive beamforming with deep learning in
  vehicular networks,'' \emph{IEEE Communications Letters}, vol.~25, no.~10,
  pp. 3301--3304, 2021.

\bibitem{9997576}
Y.~Liu, I.~Al-Nahhal, O.~A. Dobre, and F.~Wang, ``Deep-learning channel
  estimation for irs-assisted integrated sensing and communication system,''
  \emph{IEEE Transactions on Vehicular Technology}, pp. 1--14, 2022.

\bibitem{9970330}
P.~Liu, G.~Zhu, S.~Wang, W.~Jiang, W.~Luo, H.~V. Poor, and S.~Cui, ``Toward
  ambient intelligence: Federated edge learning with task-oriented sensing,
  computation, and communication integration,'' \emph{IEEE Journal of Selected
  Topics in Signal Processing}, vol.~17, no.~1, pp. 158--172, 2023.

\bibitem{10014666}
X.~Li, F.~Liu, Z.~Zhou, G.~Zhu, S.~Wang, K.~Huang, and Y.~Gong, ``Integrated
  sensing, communication, and computation over-the-air: Mimo beamforming
  design,'' \emph{IEEE Transactions on Wireless Communications}, pp. 1--1,
  2023.

\bibitem{9237116}
C.-X. Wang, J.~Huang, H.~Wang, X.~Gao, X.~You, and Y.~Hao, ``6g wireless
  channel measurements and models: Trends and challenges,'' \emph{IEEE
  Vehicular Technology Magazine}, vol.~15, no.~4, pp. 22--32, 2020.

\bibitem{10490142}
N.~Van~Huynh, J.~Wang, H.~Du, D.~T. Hoang, D.~Niyato, D.~N. Nguyen, D.~I. Kim,
  and K.~B. Letaief, ``Generative ai for physical layer communications: A
  survey,'' \emph{IEEE Transactions on Cognitive Communications and
  Networking}, vol.~10, no.~3, pp. 706--728, 2024.

\bibitem{10398474}
M.~Xu, H.~Du, D.~Niyato \emph{et~al.}, ``Unleashing the power of edge-cloud
  generative ai in mobile networks: A survey of aigc services,'' \emph{IEEE
  Communications Surveys \& Tutorials}, vol.~26, no.~2, pp. 1127--1170, 2024.

\end{thebibliography}

\clearpage
\end{document}